\documentclass[
    ,final            
    ,numberedheadings 
  ]
  {aipproc}

\layoutstyle{8x11single}

\DeclareRobustCommand\openone{\leavevmode\hbox{\small1\normalsize\kern-.33em1}}%

\begin{document}

\title{Diagonalization- and Numerical Renormalization-Group-Based Methods
for Interacting Quantum Systems
}

\author{Reinhard M.\ Noack}{
  address={Fachbereich Physik, Philipps-Universit\"at Marburg,
  D-35032 Marburg, Germany}
}

\author{Salvatore R.\ Manmana}{
  address={Institut f\"ur Theoretische Physik III, Universit\"at Stuttgart,
  Pfaffenwaldring 57, D-70550 Stuttgart, Germany}
  ,altaddress={Fachbereich Physik, Philipps-Universit\"at Marburg,
  D-35032 Marburg, Germany} 
}

\begin{abstract}
In these lecture notes, we present a pedagogical review of a number of 
related {\it numerically exact} 
approaches to quantum many-body problems.
In particular, we focus on methods based on the exact diagonalization
of the Hamiltonian matrix and on methods extending exact
diagonalization using renormalization group ideas, i.e., Wilson's
Numerical Renormalization Group (NRG) and
White's Density Matrix Renormalization Group (DMRG). 
These methods are standard tools for the investigation of a variety of
interacting quantum systems, especially low-dimensional quantum
lattice models.
We also survey extensions to the methods to calculate properties such
as dynamical quantities and behavior at finite temperature, and
discuss generalizations of the DMRG method to a wider variety of
systems, such as classical models and quantum chemical problems.
Finally, we briefly review some recent developments for obtaining a
more general formulation of the DMRG in the context of matrix product
states as well as recent progress in calculating the time evolution of
quantum systems using the DMRG and  
the relationship of the foundations of the method with quantum 
information theory.
\end{abstract}


\maketitle

\section{Introduction}

\subsection{Strongly Correlated Quantum Systems}

Many problems in condensed matter physics can be described effectively
within a single-particle picture \cite{ashcroft}. 
However, in systems where interaction leads to strong correlation
effects between the system's component parts, mapping to an
effective single particle model can lead to an
erroneous description of the system's behavior.  
In such cases, it is necessary to treat the full many body problem. 
As we will see later, it is generally impossible to treat the
full {\sl ab initio} problem, even numerically. 
However, in the investigation of condensed matter systems one
is often interested in the low-energy properties of a system. 
In this article, we will discuss {\it numerically exact} methods to
treat effective models that describe these low-energy properties.  

One possibility to numerically investigate a physical system
is to discretize the underlying differential equation or to 
map the model onto or formulate it directly on a lattice. 
It is then necessary to describe model properties on the lattice sites
(e.g., is it a spin system? Are the particles fermions or bosons?) and
to specify the 
topology of the lattice, its dimension and the boundary conditions one
needs to apply (e.g., a one-dimensional chain or ring, a two-dimensional
square lattice, a honeycomb lattice, etc.).    
The system is then composed of $N$ quantum 
mechanical subsystems, where each subsystem is located on a
site $j$ and is described by a (usually finite) number of basis states 
$|\alpha_j^{(\ell)} \rangle$, $\ell = 1, \ldots,s_{\ell}$, which depend
on the model and may vary from site to site.
This basis will be referred to as the 
{\it site basis} in the following and is usually ``put in by hand'' in
order to model a physical situation or system.

One often is interested in general properties of the physical models
that go beyond their behavior on finite lattices.
In these cases, it is necessary to investigate the behavior in
particular limits, e.g., by investigating the model's behavior when
$s_{\ell} \to \infty$ (the continuum or thermodynamic limit) or $N \to
\infty$ (the thermodynamic limit). 
Insight can also be gained by considering the limit of a continuous
quantum field by taking $\ell \to x/a$, where $a$ is the lattice
constant. 
In the following, we will restrict ourselves to the description of
quantum systems on a finite lattice with $N$ sites, as, e.g., realized
in solid state systems or in recent experiments on optical lattices
\cite{Bloch_Mott}.  
Given the site-basis $\{ | \alpha_j^{(\ell)} \rangle \}$, a
possible configuration  $| \alpha^{(\ell)}_1, \alpha^{(\ell)}_2,
\ldots \alpha^{(\ell)}_N \rangle$ of the total system is given by the
direct product of the component basis,  
\begin{equation}
| \alpha^{(\ell)}_1, \alpha^{(\ell)}_2, \ldots, \alpha^{(\ell)}_N \rangle
  \equiv | \alpha^{(\ell)}_1 \rangle \otimes | \alpha^{(\ell)}_2 \rangle
  \otimes \ldots \otimes | \alpha^{(\ell)}_N \rangle.
\end{equation}
In the following, we will omit the state-index $\ell$.
The set of all of these states constitutes a complete basis and 
hence can be used to represent 
an arbitrary state 
\begin{equation}
|\Psi\rangle = \sum\limits_{\{\alpha_j\}}
 \psi(\alpha_1,\alpha_2,\ldots\alpha_N) \, |\alpha_1, \alpha_2 \ldots
 \alpha_N \rangle 
\end{equation}
where the total number of possible combinations and therefore the 
dimension of the resulting Hamiltonian matrix or state vector is
$\prod_{j=1}^N s_j$.   
This exponentially increasing dimension of the basis of the system is the
key problem to overcome when treating quantum many-body systems. 
In principle, a quantum computer would be needed to provide the full
solution of an arbitrary quantum many-body problem 
\cite{feynman1,feynman2,feynman3}.
However, a variety of numerical approaches exist to investigate
such systems {\it exactly} on ordinary computers. 
In this article, we restrict ourselves to a set of related techniques;
specifically, we will discus exact diagonalization and
numerical renormalization group (NRG) methods, including both Wilson's
original NRG \cite{wilson_rg} and the density matrix renormalization
group (DMRG) \cite{dmrg_prl,dmrg_prb}.

The authors are participants in the ALPS collaboration, whose aim it is to
develop and provide free, efficient, and flexible software for the
simulation of quantum many body systems \cite{alps}.
In the ALPS project, implementations of a number of different
numerical methods for many-body systems are available -- including
exact diagonalization and DMRG.
A variety of models on different lattices can be treated; the programs
themselves can be used as a ``black box'' to carry out calculations or
can be a helpful starting point for developing one's own implementation.
The reader is encouraged to visit the URL and to try out the
programs.\\ 

The behavior of a quantum mechanical system is, in general, governed
by the time-dependent 
\begin{equation}
\label{eqn:timedependent_schroedinger}
i\hbar \frac{\partial}{\partial t}| \Psi(t) \rangle = H |
\Psi(t) \rangle
\end{equation}
or time-independent Schr\"odinger equation, 
\begin{equation}
H | \Psi \rangle = E | \Psi \rangle \; .
\end{equation}
One approach to treating quantum mechanical systems numerically is to 
solve (at least partially) the eigenvalue
problem formed by the system's Hamiltonian.
Note that Hamiltonians without interaction, i.e., consisting of a sum
of single-particle terms, can be solved by treating the 
single-particle problem -- there is no need to work in the full many
body basis.
The many-body wave function can then be constructed out of the
single-particle solutions.
However, for some systems, e.g., systems with strong interactions, it
is not possible to reduce the problem to a single particle problem
without introducing substantial and, in many cases, uncontrolled
approximations. 
The approaches discussed in this article are suitable for such systems
and have been applied
successfully (mostly for one-dimensional systems) to investigate
the low-energy behavior.

A typical Hamiltonian for a quantum lattice system can contain
a variety of terms which can connect an arbitrary number of subsystems 
with each other. 
A general form for quantum lattice Hamiltonians is given by 
\begin{equation}
H = \sum_l H^{(1)}_l + \sum_{l,m} H^{(2)}_{l m} + \ldots
+ \sum_{l,m,n,p} H^{(4)}_{l m n p} + \ldots \; .
\end{equation}
The local or site Hamiltonian $ H^{(1)}_l$
describes general, local model properties of the system and may
contain additional terms such as a local chemical potential or
coupling to a local external field. 
The two-site terms $H^{(2)}_{l m}$ connect pairs of subsystems.
Such terms can connect two sites or, in the case of 
impurity systems or multi-band systems may also contain
hybridization or couplings between orbitals on the same site.
The coupling associated with  $H^{(2)}_{l m}$ is often short-ranged
(between nearest neighbors or next-nearest neighbors only) in models
of interest.
Hamiltonians containing only terms of the type  $H^{(1)}_l$
separate into single-particle problems.
In some cases, a change of the single-particle basis can transform 
terms of the type $H^{(2)}_{l m}$ into terms of the type
$H^{(1)}_l$, e.g., the diagonalization of a hopping term by
Fourier transformation.
However, two-site terms in general lead to an interacting many-body
system.
Four-site terms
$H^{(4)}_{l m n p}$ generally cannot be reduced to a sum over
single-site terms;
as demonstrated in the following, they always couple at
least two particles and therefore preclude separation into a 
single particle basis.   
The interaction terms of the strongly correlated quantum systems
treated here
will contain terms of the type $H^{(2)}_{l m}$ and sometimes
$H^{(4)}_{l m n p}$.

\subsubsection{Example Hamiltonians}

\begin{figure}
\includegraphics[width=\textwidth]{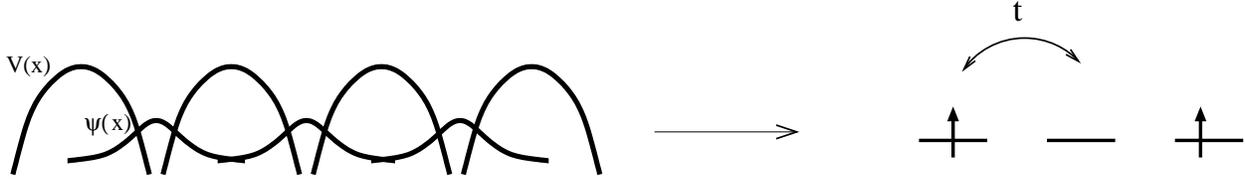}
\caption{Left-hand side: sketch of the pseudo-potential for a chain of
atoms and the 
corresponding localized Wannier 
orbital centered on the positions of the atoms, which are assumed to be
fixed (Born-Oppenheimer approximation). 
Right-hand side: sketch of the effective lattice model. 
Since the overlap between the 
Wannier-orbitals falls off very rapidly with distance, it is often
sufficient to include only the nearest-neighbor hopping $t$, which
leads to the simple tight-binding-model \cite{ashcroft}.}
\label{fig::wannier_orbitals}
\end{figure}

Roughly speaking, one can classify the models of interest as fermionic
or bosonic
lattice models, as pure spin-models, and as impurity models. 
To model a regular solid, one usually works within the
Born-Oppenheimer approximation, i.e., one assumes the atomic nuclei
to be moving slowly enough so that they can be considered to be fixed
on the relevant time scale for the electronic problem. 
The simplest fermionic lattice model is the {\it tight-binding} model
\cite{ashcroft}, which treats electrons as being localized to
Wannier-orbitals centered on a regular array of sites. 
Such a model would be appropriate for unfilled $d$ or $f$ orbitals
in transition metals, for example.
In the simplest case, one restricts the description to only a single
band, although {\it multi-band} models can also be considered.
Due to the finite overlap between nearby orbitals, the particles can
``hop'', from one site to another, i.e., tunnel between the
corresponding Wannier orbitals, with an amplitude $t$.
The mapping of unfilled strongly localized orbitals to the
tight-binding model is sketched in Fig.~\ref{fig::wannier_orbitals}. 

Since the overlap between strongly localized orbitals typically
decreases exponentially with separation, 
in most cases one considers hopping only between nearest-neighbor
sites, or at most hopping to next-nearest neighbor sites. 
The Hamiltonian for the simple tight-binding chain with
nearest-neighbour hopping in second quantization is given by
\begin{equation}
\label{eqn:tb}
H^{\rm tb} = - \sum\limits_{\langle i,j \rangle} t_{ij} \,
\left( c^\dagger_{j+1} c_j + c^\dagger_j c_{j+1}  \right) \; .
\label{eqn:tight-binding}
\end{equation}
Here and throughout this article, $\langle i,j \rangle$ will denote 
pairs of nearest-neighbor sites, and we will choose units so that
$\hbar = 1$.
The quantity $t_{ij}$ is the tunneling matrix element between the
sites $i$ and $j$ 
and corresponds to the kinetic energy of the particle ``hopping''
between these two sites.
The operators $c_j^{\dagger}$ ($c_j^{\phantom{\dagger}}$) create
(annihilate)  fermions in the subsystem labeled by $j$, and obey the
usual anticommutation relations, $\{c_i^\dagger,c_j\} = \delta_{ij}$
and $\{c_i^\dagger , c_j^\dagger\} = \{c_i , c_j \} = 0$. 
Without additional interaction terms, this model can be reduced to a
single-particle problem, which will be treated in 
Sec.~\ref{sec::nrg_noninteracting}. 

When spin is included, Pauli's exclusion principle allows up to two
fermions with opposite spin to occupy a Wannier orbital.
The possible states on a site $j$ are then $ \{ | \alpha_j^{(\ell)}
\rangle \} = \{ | 0_j \rangle, \, | \uparrow_j \rangle, \, |
\downarrow_j \rangle, \, | (\uparrow\downarrow)_j \rangle \}$.   
This is an appropriate site basis for modeling systems with
Coulomb interaction, e.g., between electrons, 
\begin{equation}
 H^{\rm C}_{j m} = \frac{e^2}{|{\bf r}_j - {\bf r}_m|}.
\end{equation}
One prominent example is the {\it Hubbard model}, 
\begin{equation}
H = -t \sum\limits_{\langle j,m\rangle,\sigma} 
\left( c^\dagger_{j,\sigma} \,
c^{\phantom{\dagger}}_{m,\sigma} 
+ c^\dagger_{m,\sigma} \, c^{\phantom{\dagger}}_{j,\sigma} \right) + 
U \sum\limits_j n_{j,\uparrow}n_{j,\downarrow} \; ,
\label{eqn:hubbard}
\end{equation}
with $\sigma$ labeling the spin and $n_{j,\sigma} =
c_{j,\sigma}^\dagger c_{j,\sigma}$ the local particle number
operator. 
Since screening effects are strong in strongly localized $d$ and
$f$ orbitals, it is usual to reduce the Coulomb interaction to the
on-site term (the {\it Hubbard term}), or sometimes
the nearest-neighbor term $V \sum n_i n_{i+r}$,
in which case the model is called the extended Hubbard model.  
For such models, the dimension of the Hilbert space grows as $4^N$,
where $N$ is the number of lattice sites. 
Since the interaction terms couple two particles, it is not possible
to reformulate the problem in a single particle basis -- an exact
treatment must treat the full many-body
basis $| \alpha_1, \alpha_2, \ldots, \alpha_N \rangle$.  
The eigenstates of such systems are many-body states, and the excited
states cannot, in general, be mapped to single-particle excitations. 
Within this formalism, it is also possible to model disorder,
using, for example, the {\it local Anderson Hamiltonian}
\begin{equation}
H^{\rm A}_j =  \sum_\sigma \lambda_j \; n_{j,\sigma} \; ,
\end{equation}
where $\lambda_j$ varies randomly from site to site according to a
given probability distribution.\\

\begin{figure}
\label{fig::austausch}
\includegraphics[height=2cm]{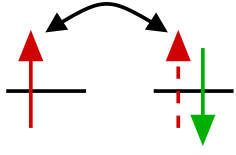}
\caption{The second-order virtual exchange process in the Hubbard
model which leads to the antiferromagnetic Heisenberg coupling, see
Ref.~\cite{auerbach}.}
\end{figure}

Another important class of models are spin models, where ${\bf S}_i$
describes localized quantum mechanical spins ($S=1/2, 1, 3/2,
\ldots$). 
The possible states on a site $j$ are then (omitting the site index) 
$ \{ | \alpha^{(\ell)} \rangle \} = \{ | -S \rangle, \; | -S+1
\rangle, \ldots, | S \rangle \} $, i.e., such models possess $(2S+1)^N$
degrees of freedom. 
A prominent example is the {\it Heisenberg Hamiltonian}, 
\begin{equation}
H^{\rm Heis} = \mathbf{J} \sum\limits_{\langle j,m \rangle}\; {\bf S}_j
\cdot {\bf S}_m  =  J^z \sum\limits_{\langle j,m \rangle} \; S^z_j
S^z_m + {\tiny\frac{1}{2}} J^{xy} \sum\limits_{\langle j,m \rangle}
\left ( S^+_j S^-_m + S^-_j S^+_m \right ).
\label{eqn:heisenberg}
\end{equation}
The $S=1/2$ Heisenberg model can be derived as the strong coupling
limit of the Hubbard model ($U/t \to \infty$) at half filling, i.e.,
average particle density $\langle n \rangle =1$.
The corresponding antiferromagnetic
exchange coupling in second-order perturbation theory 
(see Fig.~\ref{fig::austausch}) is $J=4t^2/U$
\cite{auerbach}. 

Several variations of this model are known. 
For example, when $J^{xy} = 0$, it is termed Heisenberg model with an
Ising anisotropy, or, more generally, with XY anisotropy, if $J^z \ne
J^{xy}$. 
Additional terms such as $H^{n}_j = D (S^z_j)^2$ (``single-ion
anisotropy''),
or a biquadratic coupling, $H^{bq}_{j m} = J_2 \;
\left ({\bf S}_j \cdot {\bf S}_m \right)^2$ can be introduced.  
When doped below half-filling, the strong-coupling limit of the Hubbard 
model is described by the {\it $t$--$J$ model}
\begin{equation}
H^{tJ} = \mathcal{P} \; H^{tb} \; \mathcal{P}
+ J \; \sum\limits_{\langle j,m \rangle} \left ({\bf S}_j \cdot {\bf
S}_m \; - \; \frac{1}{4} n_j \; n_m \right ),
\end{equation}
where $\mathcal{P}$ projects onto the space of singly occupied sites,
see, e.g., Ref.~\cite{auerbach}, leading to a site-basis with three
states per site. 
This Hamiltonian on a two-dimensional square lattice has been proposed
as an effective model for the CuO$_2$ planes in the 
high-$T_c$ superconductors.

A third important class of models are Hamiltonians describing systems
containing impurities.
A prominent example is the {\it Anderson impurity model}, which
models the hybridization of the localized $d$ (or $f$) orbital of an
impurity located at a site $j$ with conduction-band orbitals (in the
simplest case) located at that site, while the electrons on the
impurity interact with each other via a Hubbard term. 
The localized part of the Hamiltonian is then
\begin{equation}
H^{AI}_{j} \; = \; \sum\limits_{\sigma}\varepsilon_d \,
n^d_{j, \sigma} \; + \; V \sum\limits_{\sigma} \left (
d^\dagger_{j, \sigma} \, c^{\phantom{\dagger}}_{j, \sigma} + {\rm H.c.}
\right ) \; + \; U n^d_{j, \uparrow} n^d_{j, \downarrow} \;,
\end{equation}
while the conduction band is modeled as a tight-binding band
(\ref{eqn:tight-binding}).
Here $d^{(\dagger)}_{j,\sigma}$ is the annihilation (creation)
operator for a fermion with spin $\sigma$ on the impurity site, and
$n^d_{j,\sigma} = d_{j,\sigma}^\dagger d_{j,\sigma}$. 
This problem has been treated with one or two impurities, leading to the
the single or two impurity Anderson model, (SIAM or TIAM,
respectively) or as a system with impurities located at every lattice
site, leading to the periodic Anderson model (PAM). 
   
Perhaps the most well-known example of an impurity system is the {\it Kondo
model}, which describes the localized impurity level as a quantum
mechanical spin ${\bf S}$, leading
to the on-site Hamiltonian
\begin{equation}
H^{K}_j = 
\frac{J_K}{2} \; {\bf S}_j \cdot \left(
	c^\dag_{j,\alpha} 
	\sigma^{\phantom{\dag}}_{\alpha,\beta}
	c^{\phantom{\dag}}_{j,\beta} \right)
\end{equation}
where $\sigma^{\phantom{\dag}}_{\alpha,\beta}$ is a vector whose
components are the Pauli matrices.
This model can be derived as the limit of the symmetric SIAM at strong
coupling, $U \gg (t,V)$ \cite{schrieffer-wolff}. 

\subsubsection{Lattices}
\begin{figure}
\includegraphics[width=\textwidth]{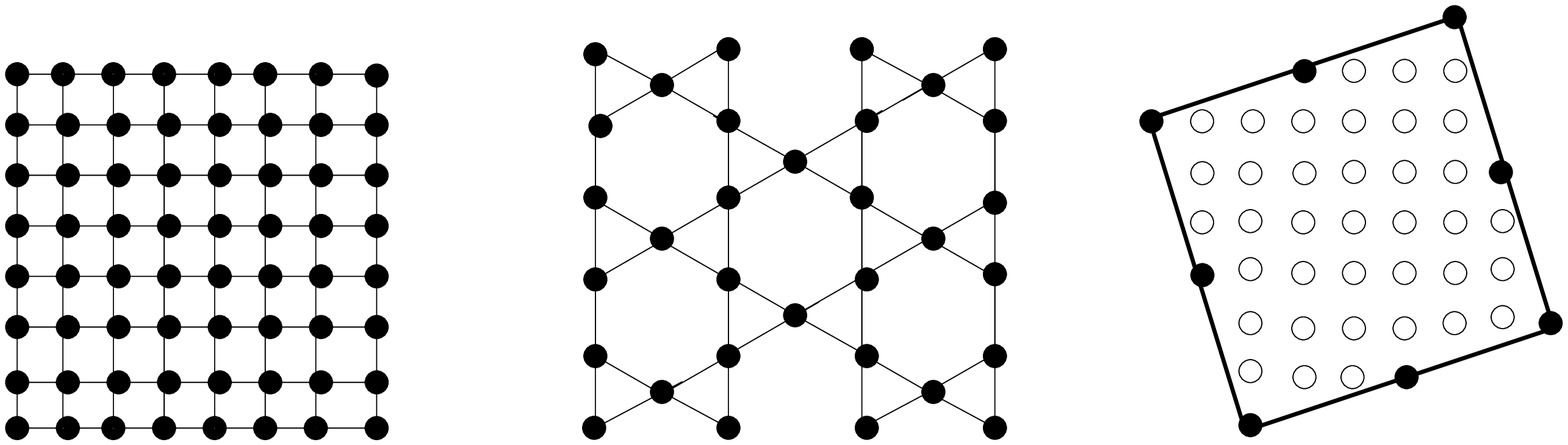} 
\caption{Examples of two-dimensional lattices. On the left and in the
middle the square and the Kagom\'e lattices, respectively, are
shown. The sketch on the right shows how a lattice can be tilted along
the symmetry axis -- in this way, it is possible to treat larger
system sizes by exploiting possible additional translation
symmetries.} 
\label{fig::lattices}
\end{figure}

The lattice structure of a regular solid is described by a Bravais
lattice, which is built up from a unit cell
and a set of translation vectors which create the regular
pattern.  
Since the memory of a computer is finite, it is often necessary to
restrict the treatment to finite lattices.
In this case, boundary conditions must be specified.
For open boundary conditions (OBC, sometimes they are also called {\it
fixed BC}) the intersite terms of the Hamiltonian are simply set to
zero at the boundaries. 
This leads to a lattice which is not translationally invariant.
Translational invariance can be restored on a finite lattice by
applying  periodic boundary
conditions (PBC, also sometimes called Born-von-K\'arm\'an BC), in
which the edges of the lattice are connected together in a ring
geometry for one-dimensional systems or a torus for two-dimensional systems.
It is also possible to introduce a phase at the boundaries, leading to
antiperiodic or twisted boundary conditions.
In Fig.~\ref{fig::lattices} some examples of
two-dimensional lattices are shown.

Lattices possess symmetries which can be exploited in
calculations. 
The translation symmetries, parameterized by
multiples of the Bravais lattice vector, can be used
for boundary conditions that do not break translational invariance,
i.e., (A)PBC or twisted boundary conditions.  
Other useful and simple symmetries are symmetries under a discrete
lattice rotation, 
e.g., rotation by $\pi/2$ for a square lattice (group $C_{4v}$), or
reflections about a symmetry axis.

In order to have more sizes available for two-dimensional
translationally invariant lattices, it is helpful
to work in geometries where the clusters are tilted with respect to some
symmetry axis, as depicted in Fig.~\ref{fig::lattices}. 
For the two dimensional square lattices, the spanning vectors for such
tilted lattices are given by $F_1 = (n,m)$, $F_2 = (-m,n)$ and the
total number of sites is then $N = n^2 + m^2$.
The lattice periodicity then appears for translations at appropriate
values of $(n,m)$.
Reflection and discrete rotation symmetries are still present in
these lattices, although they become somewhat more complicated.
For methods which treat the entire Hilbert space, such as exact
diagonalization, it can be important to use these symmetries to reduce
as much as possible the 
dimension 
of the basis that must be treated numerically.

\section{Exact Diagonalization}

The expression ``Exact Diagonalization'' is used to describe a number
of different approaches which yield numerically exact results for a
finite lattice system by 
directly diagonalizing the matrix representation of the system's
Hamiltonian in an appropriate many-particle basis. 
%
The simplest, and the most time- and memory- consuming approach is the
{\it complete diagonalization} of the matrix which enables one to
calculate {\it all} desired properties.  
However, as shown in the introduction, the dimension of the basis for
a strongly interacting quantum system grows exponentially with the
system size, so that it is impossible to treat systems with more than
a few sites.
If only properties of low- or high-lying eigenstates are required,
(in the investigation of condensed matter systems one is often
interested in the low-energy properties), it is possible
to reach substantially larger system sizes using {\it iterative
diagonalization} procedures, which also yield results to almost machine
precision in most cases.
The iterative diagonalization methods allow for the calculation of
ground state properties and (with some extra effort) some low-lying
excited states are also accessible. 
In addition, it is possible to calculate dynamical 
properties (e.g., spectral functions, time-evolution) as well as
behavior at finite temperature.  
Nearly every system and observable can be calculated in principle,
although the convergence properties may depend on the system under
investigation.
In the following, we describe the most useful methods and their
variants, which should allow for the investigation of most systems of
interest. 

We also discuss the use of the symmetries of a system, which can be
important in reaching the largest possible system sizes, because the
dimension of the matrices to be diagonalized is then significantly
reduced. 
An additional advantage is that the results obtained are resolved
according to the quantum numbers associated with the symmetries. 
Useful introductions and discussion of the use of symmetries for
specific example Hamiltonians can be found in
Refs.~\cite{ED_firstwithsymmetries,ED_golinelli,ED_introduction}.  
A more general mathematical description in terms of group theory is
presented in Ref.~\cite{ED_review}.   

With present-day computers, system sizes can be treated which are large
enough to provide insight into the physics of many systems of
interest.
It is possible to treat $S=1/2$ spin models with up to about $N=40$
sites; the maximum sizes reached thus far are, $N=40$ sites on a
square lattice, $N=39$ sites on a triangular
lattice, and $N=42$ sites on a star lattice geometry.
The $t$-$J$ model on a checkerboard and on a square lattice with 2 holes
have been treated on up to $N=32$ sites.
Hubbard models at half filling on a square lattice with up to $N=20$
sites have been diagonalized; the same size was also reached for
quantum dot structures.
Models with phonon degrees of freedom such as the Holstein model, are
much harder to treat because the phonons are bosons, which, in
principle, have an infinite number of degrees of freedom.
By truncating the number of phonon states, 
it was possible to treat a chain with $N=14$ + phonon pseudo-sites. 
For these calculations, it was necessary to store between $1.5-30 \cdot
10^9$ basis states \cite{laeuchli}.


\subsection{Representation of Many Body States}
\label{sec::representation_manybodystates}
In order to represent the many-body basis on a computer, it is
convenient to map the states of the site basis to a single bit or to a
set of bits. 
Thus, every basis state $| \alpha^{(\ell)}_1, \alpha^{(\ell)}_2, \ldots
\alpha^{(\ell)}_N \rangle$ 
can be mapped to a 
sequence of bits.
In C or C++, the most efficient way to handle a bit sequence is to
work directly on the bit representation of a (long) unsigned integer
variable using bit-operators. 

There is usually a natural bit representation for a particular choice
of the site basis,
The terms of the Hamiltonian then can be implemented as sequences of
bit operations on the 
chosen bit set.
For example, a basis state of a Heisenberg spin-1/2 lattice may be
mapped to the bit sequence
\[
|\uparrow_1\downarrow_2\downarrow_3\ldots\uparrow_{N-1}\uparrow_N
 \rangle \longrightarrow 1_10_20_3\ldots1_{N-1}1_N \; .
\]
In C or C++, a spin flip operation can be implemented compactly and
efficiently as a sequence of boolean operations that invert a
particular bit.

For Hubbard-like models, one natural choice 
for the site basis to the
mapping is
\[
|N_l^{\uparrow} N_l^{\downarrow} \rangle \longrightarrow
 N_l^{\uparrow} N_l^{\downarrow} 
\]
with $N_l^{\sigma} = \{0,1\}$, i.e., one needs two bits for the mapping
of a single site. 
Alternately,
a mapping $|N_l^{e} S_l^{z} \rangle \longrightarrow N_l^{e} S_l^{z}$
may also be used (mapping $S^z_\ell=\{\pm1/2\}$ to the bits $\{0, 1\}$). 

Other models can be implemented similarly; 
for bosonic
systems, however, it is necessary to introduce a cut-off in the number of
particles per site so that the site basis has a finite number of states.
Typically,
one only keeps a very small number of on-site bosons, 
which may depend on the desired system sizes and the parameter
values. 
A minor complication occurs when the number of bits needed to
represent a state exceeds the number of bits in an integer; then more
than one integer variable must be used
for the description of a single basis state.

In order to minimize the dimension of the Hamiltonian matrix, it is
essential to exploit symmetries of the system, as described
in more detail in Ref.\ \cite{ED_review}.  
Given a symmetry group $\mathcal{G}$ with generators $g_p$, 
if
\begin{equation}
[H,g_p] = 0, 
\end{equation}
the Hilbert space can be 
partitioned into sectors corresponding to irreducible
representations of the symmetry group   
so that the Hamiltonian matrix
$\mathbf{H}$ becomes block diagonal.   
The solution of the eigenvalue problem for each block then yields the
portion of the spectrum of the system associated with a particular
conserved quantum number.


It is possible to exploit continuous symmetries, such as the conservation
of the particle number or the conservation or the $z$ projection of
the spin, $S^z$. 
For these Abelian U(1)-symmetries, the conserved quantum numbers
correspond to the sum of all bits representing a basis state.
Therefore, 
all possible basis states 
preserving this symmetry 
can be
obtained by calculating all possible permutations of the bits
representing a suitable basis state.
Another important symmetry is the conservation of the total spin,
which is an SU(2) symmetry. 
This symmetry is more difficult to implement than U(1) symmetries
because it is non-Abelian.
However, the $Z_2$ symmetry corresponding to spin inversion can be
used instead and is easily implemented.
Space group symmetries include translational
invariance, which is an Abelian symmetry, or point group symmetries
such as reflections or rotations, which are non-Abelian in general.
%
A set of representative basis states can be formed from symmetrized
linear combinations of the original basis states generated by applying the
appropriate generators \cite{ED_review}. 

As an example of how symmetries can reduce the dimension
of the largest block of the Hamiltonian that must be diagonalized, 
we show the size of the sector containing the ground state
in Table~\ref{tab::use_symmetries} for a $S=$1/2 Heisenberg model
on a $\sqrt{40} \times \sqrt{40}$ cluster. 
In this case, the dimension of the sector to be diagonalized is
reduced by a factor of more than 2500. 
\begin{table}
\label{tab::use_symmetries}
\begin{tabular}{|ll|}
\hline
full Hilbert space:           & dim$ = 2^{40} = 10^{12}$ \\         
constrain to $S_z=0$:         & dim$ = 138\times10^9$ \\
using spin inversion:         & dim$ = 69\times 10^9$ \\
utilizing all 40 translations:& dim$ = 1.7\times 10^9$ \\
using all 4 rotations:        & dim$ = 430,909,650$ \\
\hline
\end{tabular}
\caption{The reduction of the dimension of the sector of the
Hamiltonian containing the ground state for the
$S=1/2$ Heisenberg model on a tilted square lattice with $\sqrt{40}
\times \sqrt{40}$ sites 
(from Ref.~\cite{laeuchli}).}
\end{table}

After considering the symmetries, a set of 
basis states $\{ | n \rangle \}$ is stored in an appropriate form such
as a set of bit strings or linear combinations of bit strings.
The multiplication is then carried out as 
\begin{equation}
H |\psi\rangle = \sum_n \langle n | \psi \rangle 
H | n \rangle
\end{equation}
where the set of coefficients $\langle n|\psi\rangle$ is a stored as
vector of real (or complex) numbers with dimension equal to that of
the targeted block of $H$.  
When the Hamiltonian is applied to a basis state 
$| n \rangle$, the result is a linear combination of basis states 
\begin{equation}
H | n \rangle =  \sum_{n'} 
\langle n^\prime | H | n \rangle | n^\prime \rangle \; ,
\end{equation}
where there are typically only few nonzero terms for a short-ranged
Hamiltonian.
When the states $|n\rangle$ can be represented as bit strings, it is
usually easy to identify the bit string corresponding to the
$\{|n^\prime\rangle\}$ and to determine the coefficients 
$\langle n^\prime | H | n \rangle$.
However, the bit strings then have to be mapped back to an index into
the vector of basis coefficients in order to calculate the 
coefficients $\langle n|\psi\rangle$.
A simple way to implement the needed bookkeeping is to index 
all the basis states $| n \rangle$ by an integer corresponding to
the bit string and to store this index in an additional vector. 
In this way, it is easy to identify which element of the coefficient
vector has to be modified. 
This implementation is simple and fast, but the index vector is
of the dimension of the full bit string used to encode the states and
is thus not memory efficient.  
More memory-efficient implementations are possible; for example, the
bit strings could be stored in a lookup-table along with the associated
indices and the lookup could be performed using a hash table
\cite{numerical_recipes}. 

\subsection{Complete Diagonalization}
\label{sec:ed_complete}
The diagonalization of real symmetric or complex Hermitian
matrices is a problem often encountered in numerical methods in
physics.
For example, as we will see later, in one step of the DMRG
procedure it is necessary to diagonalize the density matrix,
represented as a dense, real, symmetric matrix.
A number of software libraries provide complete diagonalization
routines which take a matrix as input and return all of the
eigenvalues and eigenvectors as output. 
Among the most prominent are the NAG library\cite{nag}, the
routines published in {\it Numerical Recipes}
\cite{numerical_recipes,quarteroni} 
as well as the
LAPACK \cite{lapack} library, which in combination with 
the BLAS \cite{blas,atlas,libgoto} provides a very efficient 
non-commercial implementation of linear algebra tools.   
Such routines could, in principle, also be used to diagonalize the
Hamiltonian matrix of a finite quantum lattice system directly.
However, they
are, in general, substantially less efficient than the iterative
methods described in Sec.\ \ref{sec:iterative_diag} for finding the
low-lying states of the sparse matrices found for such systems.


The approach normally used \cite{numerical_recipes} is first to
transform the matrix to tridiagonal form using a sequence of
Householder transformations and then to diagonalize the resulting
tridiagonal matrix $\mathbf{T}$ using the QL or QR algorithm, which
carries out a factorization $\mathbf{T}=\mathbf{QL}$ with $\mathbf{Q}$
an orthogonal and $\mathbf{L}$ a lower triangular matrix.   
The computational cost of this combined approach scales as 
$\frac{2}{3}n^3$ if only the eigenvalues are obtained, and  $\approx 3
n^3$ when the eigenvectors are also calculated, where $n$ is the
dimension of the matrix.

The limitations of using such complete diagonalization algorithms to
diagonalize the Hamiltonian matrix are obvious: the entire matrix has
to be stored and diagonalized.  
Since the dimension of the Hamiltonian matrix grows exponentially with
system size even when all 
symmetries are taken into account, the largest attainable lattice
sizes for strongly correlated quantum systems are generally very small. 
For a Hubbard chain, one may reach less than 10 sites on a
supercomputer, while with the iterative diagonalization procedures
(presented in Sec.\ \ref{sec:iterative_diag}) 
around 20 sites can be reached when most of the symmetries of the
system are taken into account.
As we will see later, one-dimensional systems containing more than 
1000 sites can be treated on a standard desktop PC using the DMRG, a
method carrying out an iterative diagonalization in a reduced Hilbert
space.
A complete diagonalization of the Hamiltonian matrix is nevertheless
useful for testing purposes, and if a significant fraction of all
eigenstates is needed on small systems.

\subsection{Iterative Diagonalization: the Lanczos and the Davidson
  Algorithm}  

\label{sec:iterative_diag}

If only the ground state and the low-lying excited
states of a system are required, powerful iterative diagonalization
procedures exist which can handle matrix representations of 
the Hamiltonian with a dimension substantially (up to four or five orders
of magnitude for short-range quantum lattice models) larger than
complete diagonalization.
These methods can be extended to investigate dynamical
properties, time evolution, and the finite-temperature behavior of the
system.
In addition, they form a key part of the DMRG algorithm, which carries
out an iterative diagonalization in an optimal, self-consistently
generated reduced basis for a system.

The basic common idea of the different iterative diagonalization
algorithms is to project the matrix to be treated $H$ onto a subspace
of dimension $M \ll N$ (where $N$ is the dimension of the Hilbert
space in which the diagonalization is carried out) which is cleverly
chosen so that 
the extremal eigenstates within the subspace converge very quickly
with $M$ to the extremal eigenstates of the system.
The main approach used in physics is the {\it Lanczos} method, while
in quantum-chemistry the {\it Davidson} or its generalization the 
{\it Jacobi-Davidson} algorithm is more widely used.

A number of standard textbooks 
treat 
numerical methods for matrix computations in detail. 
In this section, we will discuss only the basic Lanczos and Davidson
methods for handling hermitian eigenvalue problems for pedagogical
reasons.
For further details and for the applicability of the methods to
non-hermitian and to generalized eigenvalue problems, we refer the
reader to the following sources: 
Refs. \cite{parlett,van_loan,watkins} discuss numerical methods for
matrix computations in general and also include iterative
diagonalization algorithms, including
the more modern Jacobi-Davidson algorithm (at least in the later editions).
A nice overview and a compact representation of the algorithms can be
found in Ref. \cite{templatized_evproblems}.
The mathematical theory of the Lanczos algorithm has been worked out
in Ref. \cite{lanczos_book}.

The basic idea behind iterative diagonalization procedures is
illustrated by the very simple {\it power method}.
In this approach, the eigenpair with the extremal eigenvalue is 
obtained by repeatedly applying the Hamiltonian to a random initial
state $|v_0\rangle$, 
\begin{equation}
|v_n\rangle = H^n |v_0\rangle \; .
\end{equation}
Expanding in the eigenbasis
$H|i\rangle = \lambda_i |i\rangle$ yields
\begin{eqnarray*}
|v_n\rangle &=& \sum\limits_i \, \langle i | v_0 \rangle \, H^n | i \rangle \\
&=& \sum\limits_i \, \langle i | v_0 \rangle \, \lambda_i^n| i \rangle
\; .
\end{eqnarray*}
It is clear that the state with the eigenvalue with the largest
absolute value will have the highest weight after many iterations $n$,
provided that $|v_0\rangle$ has a finite overlap with this state.
The convergence behavior is determined by the spacing between the
extremal eigenvalue and the next one; 
the contribution of the state with the eigenvalue with the
next-highest magnitude will not be negligible if the 
difference between the magnitudes of these eigenenergies is not
sufficiently large. 
Since the convergence of the power method is generally much poorer than
other methods we will discuss below, it is generally not used in practice.
However, the approach is very simple to implement and is very memory
efficient because only the two vectors $|v_n\rangle$ and 
$|v_{n-1}\rangle$ must be stored in memory.

The subspace generated by the sequence of steps in the power method,
\begin{equation}
\{|v_0\rangle, H |v_0\rangle, H^2 |v_0\rangle,...,H^n|v_0\rangle\}
\label{eqn:power_sequence}
\end{equation}
is called {\it the $n^{\rm th}$ Krylov space} and is the starting point
for the other procedures.  

\subsubsection{The Lanczos Method}
\label{lanczos_method}

In the Lanczos method \cite{lanczos}, the Hamiltonian is projected
onto the Krylov subspace using a basis generated by 
orthonormalizing the sequence of vectors (\ref{eqn:power_sequence}) to
each other as they are generated.
This results in a basis in which the
matrix representation of the Hamiltonian becomes {\it tridiagonal}. 
The basic algorithm is as follows:
\begin{itemize}
\item[0)] Choose an initial state $|u_0\rangle$ which can be represented 
in the system's many-body basis $|n\rangle$, and which has finite
overlap with the groundstate of the Hamiltonian.
This can be done by taking
$\{| n \rangle \}$ as a vector with random entries.
\item[1)] Generate the states of the Lanczos basis using the recursion
relation  
\begin{eqnarray}
|u_{n+1}\rangle &=& H \; |u_{n}\rangle -a_n|u_{n}\rangle - b_n^2 \;
  |u_{n-1}\rangle \nonumber \\
  \mbox{where} \quad a_n &=& \frac{\langle u_n |H|u_n \rangle}{\langle
  u_n |u_n \rangle} \nonumber \\
   \mbox{and} \quad b_n^2 &=& \frac{\langle u_n |u_n \rangle} {\langle u_{n-1} |u_{n-1} \rangle}
\end{eqnarray}
with $b_{0} \equiv 0$ and $|u_{-1} \rangle \equiv 0$.
Note that the Lanczos vectors in this formulation of the recursion
are not normalized.
\item[2)] Check if the stopping criterion
$\langle u_{n+1}| u_{n+1}\rangle < \varepsilon$ is fulfilled. \\
If yes: carry out step 4) and then halt. \\
If no: continue.
\item[3)] Repeat starting with 1) until $n=M$ (the maximum dimension).
\item[4)] If the stopping criterion is fulfilled, diagonalize the
resulting tridiagonal matrix
\begin{equation}
\mathbf{T}_n=
\left(
\begin{array}{ccccc}
a_0 & b_1    &          &            &          \\
b_1  & a_1   & b_2  & \mathbf{0} &          \\
         & b_2    & a_2 & \ddots     &          \\
         & \mathbf{0} & \ddots   & \ddots     & b_n  \\
         &            &          & b_n    & a_n \\
\end{array}
 \right)
\end{equation}
using the QL algorithm.
Note that the off-diagonal elements have the value $b_i$ and not
$b_i^2$ for normalized Lanczos vectors $|u_i \rangle$.
The diagonalization yields eigenvalues $E_0, \, \ldots, \,
E_n$, and eigenstates $|\psi_0\rangle, \, \ldots, \, |\psi_n \rangle$
which are represented in the Lanczos basis.   
\item[5)] In order to avoid storing all of the basis states 
$|u_0 \rangle, \ldots, | u_n\rangle$, the procedure can be repeated
(starting with the same initial vector $|u_0\rangle$) to
calculate the eigenstates represented in the
original many-body basis.
This is necessary in order to be able to calculate properties
dependent on the wavefunction, such as quantum mechanical observables. 
One obtains the coefficients $\alpha_n$ of the eigenstate in the
original many-body basis $|n \rangle$ by carrying out the basis
transformation  
\begin{eqnarray}
|\psi_m \rangle &=& \sum\limits_i c_i \, | u_i \rangle
 \nonumber \\
&=& \sum\limits_n \sum\limits_i c_i \, \langle n | u_i \rangle | n
 \rangle \; ,
\end{eqnarray}
with $\alpha_n \equiv  \sum\limits_i c_i \, \langle n | u_i \rangle$. 
\end{itemize}

The algorithm is memory efficient, since only the 3 vectors
$|u_{n-1}\rangle, \, |u_n\rangle$, and $|u_{n+1}\rangle$ need to be
stored at once. 
Note that it is also possible to formulate the algorithm in a way such
that only 2 vectors need to be stored at the cost of complicating the
algorithm moderately;
the memory efficiency is then the same as the power method.
As is typically the case in iterative diagonalization procedures, the 
most time-consuming step is carrying out the multiplication 
$H | u_n \rangle$, 
which should be implemented as efficiently as possible, either as
described in  
Sec.\ \ref{sec::representation_manybodystates}, or using other sparse-matrix
multiplication routines.
The time needed to perform the other steps of the algorithm
is generally negligible when $N$ is realistically large.
Therefore, optimizing the routine performing
$H | \psi \rangle$ is crucial to the efficiency of the implementation.

The Lanczos procedure results in a variational approximation to the
extremal eigenvalue which usually attains quite high accuracy 
after a number of iterations much smaller than the
dimension of the Hilbert space.
Typically,  100 recursion steps or less are
sufficient to attain convergence to almost machine precision for
ground state properties \cite{dagotto}.
As is evident because the Lanczos method is based on the power method, 
convergence to extremal eigenvalues occurs first; additional iterations are
then necessary to obtain converged excited states.
The algorithm is generally considered to be a standard method which can
be robustly applied to a wide spectrum of systems.
Nevertheless, there are two technical problems which require some care
to be taken.
First, the convergence of excited states can be irregular; in
particular apparent convergence to a particular value can occur for
some range of iterative steps, followed by a relatively abrupt change
to another, substantionally lower value.
It is therefore important to carry out sufficient iterations for the
higher excited states.
Generally, the number of iterations required to obtain convergence
becomes larger for higher excited states.

The second technical problem is the appearance of so-called ``ghost''
eigenvalues, i.e., spurious eigenvalues which cannot be mapped to
eigenvalues of the original Hamiltonian. 
The origin of such ``ghost'' eigenvalues can be traced to the loss of
the orthogonality of the Lanczos 
vectors $|u_n\rangle$ due to finite machine precision.   
This is an intrinsic limitation of the algorithm.
For this reason, the much more stable Householder
algorithm \cite{numerical_recipes} rather than the Lanczos procedure is
generally used to transform an entire matrix to tridiagonal
form. 
However, because of the good convergence to the ground state
and the algorithm's memory efficiency, the Lanczos method is
nevertheless widely
used for the investigation of quantum lattice systems with
short-ranged coupling. 

With some extra effort, it is also possible to overcome the loss of
orthogonality which is a basic drawback of the algorithm.
The most straightforward solution is to {\it reorthogonalize} the
Lanczos vectors relative to each other using a modified Gram-Schmidt
procedure.  
However, this requires {\it all} vectors $|u_n\rangle$ to be stored in
memory, so that the advantage of memory efficiency is lost.
However, an appropriately chosen partial reorthogonalization is also
sufficient; 
for details see Ref. \cite{van_loan,watkins,templatized_evproblems}. 
Cullum and Willoughby \cite{lanczos_book,cullum_willoughby_trick}
developed a method to eliminate ghost states without
reorthogonalizing the Lanczos vectors. 
In their approach, the eigenvalues of the resulting tridiagonal
matrix $\mathbf{T}_n$ are compared to the ones of a similar matrix 
$\tilde{\mathbf{T}}_n$, which can be obtained by deleting the first row and
column of $\mathbf{T}_n$. 
This gives a heuristic criterion for the elimination of spurious
eigenvalues:
since the ghost eigenvalues are generated by roundoff errors, they do not
depend on the initial state $|v_0 \rangle$ and will be the same for
both matrices. 
After sufficiently many iterations, ghost eigenvalues will converge
towards true eigenvalues of the original matrix $\mathbf{H}$. 
Thus, every multiple eigenvalue of $\mathbf{T}_n$ is not a ghost and
every unique eigenvalue which is not an eigenvalue of
$\tilde{\mathbf{T}}_n$ is a true eigenvalue of $H$. 
This approach is as memory-efficient as the original Lanczos algorithm
and approximately as fast, but generally yields the wrong multiplicity
for the eigenvalues.


A variant of the algorithm which is easy to implement is the  
{\it ``modified''} Lanczos method \cite{dagotto}. 
In this approach, the recursion procedure is terminated after two
steps, i.e., only two Lanczos vectors are considered and the
resulting $2\times 2$ matrix is diagonalized.
The resultant eigenvector is taken as the starting point for a new
$2\times 2$ Lanzcos procedure; this process is then repeated
until convergence (i.e., the change in $|\lambda_0|$
is sufficiently small) is achieved.
The modified Lanczos method has only limited usefulness: the convergence is
only marginally better than the power method and excited states are
rather difficult to obtain. 
However, the idea of restarting the Lanczos procedure is often used in
practical implementations, i.e., after $\sim 10$ to $\sim 100$ Lanczos
iterations, the resulting tridiagonal matrix is diagonalized and the
extremal eigenstate is used as starting vector for a new Lanczos
procedure.
Further important variants of the algorithm are the 
{\it implicitly restarted Lanczos} method and the 
{\it Band-} or {\it Block-Lanczos} method.
The latter puts the ``modified'' Lanczos variant on a more
systematic footing and deals with the problem that it is not {\it a priori}
known how many iterations are needed for convergence by limiting the
number of steps and then restarting the Lanczos-procedure with an
appropriate, better initial vector obtained by taking into account the
outcome of the previous Lanczos run.
The Block Lanczos approach is useful when degeneracies are
present in the desired eigenvalues and all eigenstates
and eigenvalues of the corresponding subspace must be obtained.
Rather than starting with a single initial state vector $|u_0 \rangle$, a set 
of $p$ orthonormal state vectors $|u_0^{(1)}\rangle, |u_0^{(2)}\rangle,
\ldots, |u_0^{(p)}\rangle$ is used to initialize the Lanczos procedure.
For an eigenstate with degeneracy $g_d$,  $p \geq g_d$
should be chosen in order to fully resolve the degenerate
subspace. 
The Lanczos procedure is then performed on this subspace and a
block-tridiagonal matrix is obtained.
For details and for other variants of the Lanczos procedure, see
e.g., Ref.~\cite{templatized_evproblems}.
\begin{figure}
\includegraphics[width=0.5\textwidth,angle=270]{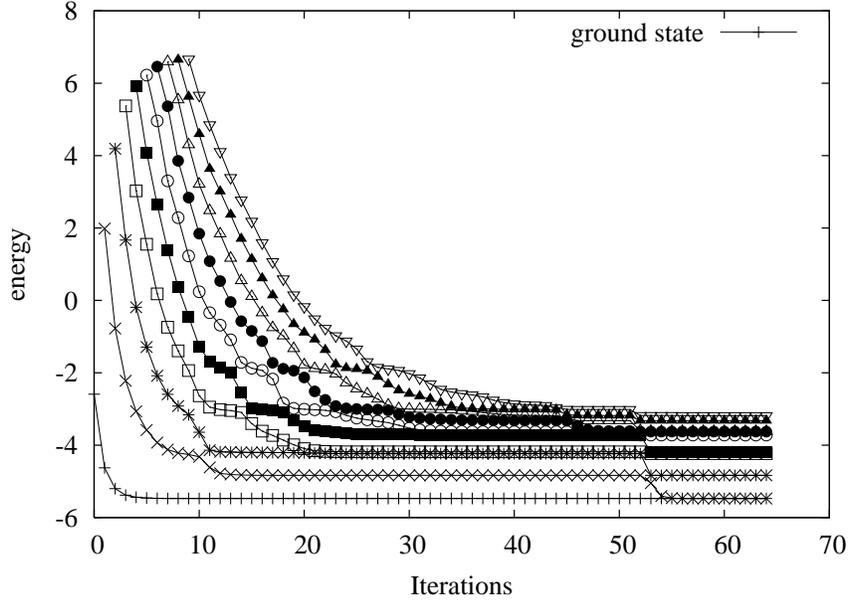}
\caption{Typical convergence behavior of the Lanczos algorithm. In
this example, convergence to the ground-state is already reached after
approximately 10 iterations, while additional iterations are required
to reach convergence for the excited states. 
``Ghost'' eigenvalues appear and converge to a real
eigenvalue as additional iterations are performed,
leading to erroneous multiplicity at the completion of the calculation.}
\end{figure}

The generalization of the Lanczos method to non-hermitian operators is
the {\it Arnoldi} method, see, e.g., Ref.~\cite{templatized_evproblems}. 
For such problems, a Krylov-space approach similar to the
Lanczos procedure is used to reduce a general matrix to upper Hessenberg
form. 

A number of software packages are available which provide
implementations of iterative diagonalization routines, such as ARPACK 
\cite{arpack} or the IETL \cite{ietl}, a part of the
ALPS-library \cite{alps}.
In order to use this software, a routine
which performs the multiplication $H | \psi \rangle$ must be defined
by the user; an effient implementation is crucial to the overall
efficiency of the algorithm.

There have been many investigations using the Lanczos procedure.
Examples for recent work using ED are investigations treating
frustrated quantum magnets, Ref.~\cite{lauchli:236404}, two-leg ladder
systems, Ref.~\cite{Roux}, transport through molecules and
nanodevices~\cite{aligia:076801}, and atomic gases on optical
lattices~\cite{Honecker}. 


\subsubsection{Davidson and Jacobi-Davidson}

The common idea of iterative diagonalization methods is to 
project the matrix to be diagonalized onto a subspace much smaller
than the complete Hilbert space.
The subspace, spanned by a set of
orthonormal states  $\left\{| u_k \rangle\right\}$, is then expanded
in a stepwise manner so that the approximation to the extremal
eigenstates improves.
At particular points in or at the end of the procedure, the
representation of the Hamiltonian matrix in this subspace is
diagonalized. 
The resulting extremal eigenvalue $\tilde{\lambda}_k$  is 
called the {\it Ritz-value} and the corresponding eigenvector the 
{\it Ritz-vector} $|\tilde{\psi}_k \rangle$.  
According to the Ritz variational principle, $\tilde{\lambda}_k$ is always
an upper bound to the real ground state energy, see,
e.g., Ref.~\cite{cohentannoudji}. 
The error in applying $H$ to the eigenvector $|\tilde{\psi}_k \rangle$
associated with the Ritz-value $\tilde{\lambda}_k$,  
is approximated by the 
{\it residual vector}   
\begin{equation}
| r_k \rangle = H |\tilde{\psi}_k \rangle -  \tilde{\lambda}_k
  |\tilde{\psi}_k \rangle \; .
\end{equation}
(This expression would be exact if $\tilde{\lambda}_k$ were replaced
by the exact eigenvalue $\lambda_k$.)
In the Lanczos procedure, the recursion is formulated so that the
subspace is expanded by the component of the residual vector
$| r_k \rangle$ orthogonal to the subspace.

In 1975, Davidson formulated an alternate iterative algorithm 
in which the subspace is expanded in the following 
way \cite{davidson_original}.
The exact correction to the Ritz-vector is given by 
\[
| z \rangle = | \psi \rangle - | \tilde{\psi}_k \rangle
\]
so that
\begin{equation}
   \left(\mathbf{H} - \lambda_k
   \mathbf{1}\right) | z \rangle = - \left(\mathbf{H} - \lambda_k
   \mathbf{1}\right) | \tilde{\psi}_k \rangle \; .
\end{equation}
Thus, solving
\begin{equation}
\left(\mathbf{H} - \lambda_k \mathbf{1}\right) | z
\rangle = - | r_k \rangle
\end{equation}
would lead to the exact correction to $ | \tilde{\psi}_k \rangle $.  
This amounts to {\it inverse iteration} (see, e.g.,
Ref.~\cite{numerical_recipes}).   
However, the exact eigenvalue $\lambda_k$ is not known, and the
numerical solution of this linear system is of comparable numerical
difficulty to the entire iterative diagonalization. 
Davidson's idea was to approximate the correction vector $|z\rangle$ by 
\begin{equation}
| \tilde{z}\rangle = - (\mathbf{D} - \tilde{\lambda}_k
  \mathbf{1})^{-1} | r_k \rangle  \; ,
\end{equation}
where the diagonal matrix $\mathbf{D}$ contains the diagonal
elements of $\mathbf{H}$.  
This is a good approximation if $\mathbf{H}$ is diagonally dominant.  
If $\mathbf{D}$ were replaced by the unit matrix $\mathbf{1}$,
the Davidson algorithm would be equivalent to the Block Lanczos
procedure.
Therefore, if the diagonal elements of $\mathbf{H}$ were all the same,
both methods would have the same performance.
For many problems, however, this variation of the diagonal elements is
important and the Davidson algorithm converges more rapidly.

In its original formulation, the Davidson algorithm follows the
procedure \cite{davidson_original}:
\begin{itemize}
\item[0)] If the $k$th eigenvalue is to be obtained, choose a subspace of
$l \geq k$ orthonormal vectors $|v_1 \rangle,\, |v_2 \rangle, \,
\ldots, \, |v_l\rangle$. 
In the following, the matrix $\mathbf{B}$ is the matrix containing
these vectors as columns.
\item[1)] 
\begin{itemize}
\item[i)] Form and save the vectors $H | v_1 \rangle, \, H
|v_2 \rangle, \, \ldots, \, H | v_l \rangle$. 
In the following, the matrix $\mathbf{\tilde{A}}$ is the matrix
containing these vectors as columns.
\item[ii)] Construct the matrix $\mathbf{A} =  \langle v_i | H|
v_j \rangle$ and diagonalize it, obtaining the $k$th eigenvalue
$\lambda_k^{(l)}$ and the corresponding eigenvector
$|\alpha_k^{(l)} \rangle$. 
The upper index $(l)$ denotes that this eigenpair was obtained by
keeping $l$ vectors for building the matrix $\mathbf{A}$. 
\end{itemize}
\item[2)] 
Form the residual vector corresponding to the $k$th eigenvector,
\[
|q_l \rangle = \left( \mathbf{\tilde{A}} - \lambda_k^{(l)}  \mathbf{B}
 \right) | \alpha_k^{(l)} \rangle.
\]
\item[3)] 
Calculate the norm $| \, | q_l \rangle \, |$.  
If $| \, | q_l \rangle \, | < \epsilon$, accept this eigenpair, otherwise
continue. 
\item[4)] Compute the correction vector 
$|w_l \rangle = (\mathbf{D} -  \lambda_k \mathbf{I})^{-1}  | q_l \rangle$,
where $\mathbf{D}$ is a matrix containing the diagonal elements of
$\mathbf{A}$ and $\mathbf{I}$ is the unit matrix.
Orthonormalize $|w_l\rangle$ against $|v_1 \rangle,\, |v_2 \rangle,
\,\ldots, \, |v_l\rangle$ to form $|v_{l+1}\rangle$. 
Expand the the matrix $B$ by adding $|v_{l+1}\rangle$ as an additional
column.
\item[5)] Form and save the vector $H | v_{l+1}
\rangle$. 
Set $l = l+1$ and continue with step 1).
\end{itemize}

Although the Davidson algorithm is somewhat more complicated to
implement than the Lanczos algorithm, its
convergence is usually higher order than the Lanczos method and it
is more stable. 
In particular, spurious ghost eigenvalues do not appear.
The disadvantages compared to Lanczos are that $\langle u_i | H
| u_j \rangle$ is not tridiagonal, and {\it all} the
$| u_i \rangle$ must be kept in order to carry out the explicit
orthogonalization in step 4). 
Similarly to the Lanczos-method, this algorithm can fail for particular
choices of the initial vector, e.g., if 
$|u_0\rangle = | \psi_0 \rangle$. 
In practice, one performs a small number of Davidson iterations and,
if necessary, restarts the procedure using the outcome of the previous
iteration as initial vectors $|v_1 \rangle,\, \ldots, \, |v_l\rangle$. 

A generalization of the original Davidson approach is given by the
Jacobi-Davidson method \cite{vandervorst}. 
In this approach, one approximates the correction vector $| z \rangle$
by 
\begin{equation}
\left( \mathcal{H} - \tilde{\lambda}_k \mathbf{1} \right) | \tilde{z}
\rangle = - | r_k \rangle - \varepsilon | u_k \rangle \; ,
\end{equation}
where $\varepsilon$ is chosen so that that $\langle u_k | \tilde{z}
\rangle = 0$.
Here the preconditioner
$\mathcal{H}$ is an easily invertible approximation to $\mathbf{H}$.
The choice of an optimal preconditioner is non-trivial
and must be tailored to suit the specific problem treated. 
The Jacobi-Davidson method is more general and flexible than the
Davidson method and does not suffer from the lack of convergence when 
$|u_0\rangle = | \psi_0 \rangle$. 
It is (almost) equivalent to the Davidson method when $
\mathcal{H} = \mathbf{D}$.
Therefore, it is now
widely used instead of the Davidson algorithm, especially in quantum
chemistry. 

An additional advantage of the Jacobi-Davidson algorithm over the Davidson
procedure is that it can be applied to generalized eigenvalue
problems, 
\begin{equation}
A \; |x \rangle = \lambda \; B \; | x \rangle,
\end{equation}
where $ A$, $ B$ are general, complex $n \times n$ matrices 
\cite{vandervorst}.

Using these iterative diagonalization algorithms, many different
quantum many-body systems can be treated.
However, the methods outlined thus far are primarily useful for
calculating properties of the ground-state (highest state) and
low-lying excited states.
Many experimentally interesting properties, such as dynamical
correlation functions and other frequency-dependent quantities, or 
finite-temperature properties cannot be directly calculated from 
the ground state and low-lying excited states.
Nevertheless, methods based on iterative diagonalization have been
developed to calculate such properties; these extensions to iterative
diagonalization will be discussed in the remainder of this section.




\subsection{Investigation of Dynamical Properties}
\label{sec:ed_dynamics}
Dynamical quantities in general involve excitations
at a particular frequency $\omega$ and are often additionally resolved
at momentum $\mathbf{q}$.
It is clear that extensions to exact diagonalization must generate a
set of states that are appropriate to describe these energy and
momentum scales.
This can be done by starting with the ground state obtained by a
converged iterative diagonalization procedure and applying an
appropriate operator or sequence of operators.

Frequency-dependent quantities can be defined as the Fourier
transform of time-dependent correlation functions 
\begin{equation}
C(t) = -i\langle \psi_0 | A(t) \; A^\dagger(0) | \psi_0 \rangle \; ,
\label{eqn:t_Green_fctn}
\end{equation}
where the operator $A$ generates the desired correlations.
The Fourier transform to frequency space then yields ($\omega \geq 0$)
\begin{equation}
\tilde{C}(\omega + i\eta) = \langle \psi_0 | \, A \, 
\left ( \omega + i\eta -H + E_0 \right )^{-1}\, A^\dagger \, | \psi_0
\rangle \; ,
\label{eqn:dyn_corr_omega}
\end{equation}
where the inverse operator in the center is termed the resolvent operator.
The spectral function then is given by \cite{mahan}
\begin{equation}
\label{eqn:spectral_function}
I(\omega) = -\frac{1}{\pi}\lim_{ \eta\to 0^+} {\rm Im}\, 
\tilde{C}(\omega + i\eta) \; .
\end{equation}
Dynamical quantities can be probed in scattering experiments
such as photoemission or neutron-scattering.
Examples of choices of $A$ for some typical spectral functions are
displayed in Table~\ref{table::spectralfunctions}. 
\begin{table}
\label{table::spectralfunctions}
\begin{tabular}{|c|c|c|c|}
\hline
name & notation & operators & experiment \\
\hline
single-particle spectral weight & $A({\bf k},\omega)$ &
$A = c^{\phantom{\dagger}}_{{\bf k},\sigma}$ & photoemission \\
structure factor & $ S_{zz}({\bf q},\omega)$ & $ A = S^z_{\bf q}$ & neutron scattering\\
optical conductivity  & $ \sigma_{xx}(\omega)$ & $ A = j_x$ & optics \\
4-spin correlation & $ R(\omega)$ & 
$ A = \sum_{\bf k} R_{\bf k} \, {\bf S}_{\bf k}\cdot {\bf S}_{\bf -k}$
& Raman scattering \\
\hline
\end{tabular}
\caption{}
\end{table}

A spectral function can also be expressed in the Lehmann representation 
\begin{equation}
I(\omega) = \sum_n \; |\langle \psi_n | A^\dagger|\psi_0\rangle|^2
\; \delta(\omega-E_n + E_0) \; , 
\label{eqn:lehmann_repn}
\end{equation}
which involves a sum over the complete eigenspectrum of the system.
Thus, the numerical method must provide the poles $E_n$ and the
weights $\langle \psi_n | A^\dagger|\psi_0\rangle$ for each
eigenstate. 
While it is possible to calculate the spectral function
using this formulation in principle, 
it is generally prohibitive to calculate the entire eigenspectrum of
the system; 
this amounts to a complete diagonalization of the Hamiltonian.

An alternative approach is to apply 
Eq.~(\ref{eqn:spectral_function}) directly. 
In this case, the resolvent operator must either be calculated
directly or approximated.

\subsubsection{Krylov Space (or Continued Fraction) Method }
\label{sec:cont_fraction}
Within the Lanczos method, it is convenient to represent operators in
a Krylov basis.
The frequency-dependent correlation function (\ref{eqn:dyn_corr_omega})
can be expanded in such a basis if a Lanczos 
procedure is carried out starting with the initial vector
\begin{equation}
|u_0 \rangle = 
\frac{1}{\sqrt{\langle \psi_0 | \, A \; A^\dagger \, | \psi_0 \rangle}} \, 
A^\dagger \,| \psi_0 \rangle \; .
\end{equation}
Within the Lanzcos basis generated by 
$H |u_0 \rangle$,  $H^2 |u_0 \rangle, \, \ldots \,$, the resolvent
operator can be expressed in terms of the Lanczos coefficients $a_n$
and $b_n$ as a continued fraction \cite{dagotto}.
The correlation function then has the form
\begin{equation}
\label{eqn:cont_frac}
\tilde{C}(z = \omega + i \eta + E_0) = 
\frac{\langle\psi_0 |\, A \; A^\dagger \,  |\psi_0\rangle }
{z  - a_1 - \frac{b_2^2}{z  - a_2 - \ldots }}
\; .
\end{equation}
Therefore, each set of Lanczos coefficients $a_n$, $b_n$ generates an
additional pole of the frequency-dependent correlation function.
In practice, the procedure is truncated after a number of poles
appropriate to the size of the system and required resolution of the
spectral function has been obtained.
The weight of each pole typically decreases rapidly with number of
steps $n$ so that a moderate number of iterations is usually
sufficient in order to obtain the desired quantities.

The low energy part of the spectrum can also be obtained by directly
using the Lehmann representation of the spectral function
Eq.~(\ref{eqn:lehmann_repn}). 
Truncation after a finite number of terms yields a finite number
of poles. 
However, a fairly large number of eigenstates would have to be
calculated directly, which is awkward within iterative
diagonalization.
Additionally, it is not clear that the weight of each subsequent term would
dimish sufficiently rapidly to justify a truncation after only a small
proportion of the eigenstates are obtained.
The resulting spectral function using this approach is a series of
sharp delta peaks located at the poles.
In order to obtain a continuous spectrum and to be able to compare to
experimental data, it is necessary to introduce a broadening of the
peaks by calculating the convolution, e.g., with a Lorentzian.

Note that the appearance of ghost-eigenvalues is not a problem in the
calculation of the spectral functions, since the matrix elements
associated with the ghost eigenvalues are negligibly small
\cite{mail_lauchli}. 
This turns out to be true when using the Lehmann-representation as
well as when applying the continued fraction approach.


\subsubsection{Correction Vector Method} 
\label{sec::ed_correction_vector}
An alternative approach, suggested by Soos and Ramasesha
\cite{soos_ramasesha},
is to apply the resolvent operator directly.
The vectors 
\begin{equation}
|\phi_0 \rangle = A^\dagger \, |\psi_0\rangle \; , \hspace*{2cm}
|\phi_1 \rangle = \left ( \omega + i\eta -H + E_0 \right )^{-1}\,
|\phi_0\rangle,
\end{equation}
are calculated directly starting with the ground state
$|\psi_0\rangle$ obtained using iterative diagonalization.
The spectral function is then given by
\begin{equation}
I(\omega) = \frac{1}{\pi}\; {\rm Im} \, \langle \phi_0 | \phi_1 \rangle \;.
\end{equation}
The {\it correction vector} $|\phi_1 \rangle$ is
obtained by solving the linear system
\begin{equation}
\label{eqn:solve_correctionvector}
\left ( \omega + i\eta -H + E_0 \right ) | \phi_1 \rangle = | \phi_0
\rangle  
\; .
\end{equation}
One advantage of this approach are that the spectral weight is
calculated exactly for a given frequency $\omega$ rather than
approximated as in the continued fraction approach.
In addition, it is possible to obtain nonlinear spectral functions by
computing higher order correction vectors.
This scheme is naturally used in conjunction with the Davidson
algorithm, in which the correction to the residual vector must be
calculated at each step, whereas the Krylov subspace is not easily available.

The disadvantage of this approach is that it generally is more
expensive in computer time
because the system of equations, Eq.~(\ref{eqn:solve_correctionvector}),
must be solved for each $\omega$ desired.

%
%
%

\subsection{Real Time Evolution}
\label{sec:ed_time_evolution}

The time evolution of a quantum system is governed by the
time-dependent Schr\"odinger equation 
\begin{equation}
i \hbar\frac{\partial}{\partial \, t} | \psi(t)\rangle = H |
\psi(t) \rangle \; .
\end{equation}
Using the Lanczos-vectors, the time evolution through one interval
$U_{dt} | \psi\rangle$ can be approximated  
by \cite{lubich_timeevolve,19dubiousways}   
\begin{equation}
|\psi(t+dt)\rangle = e^{-idt/\hbar \, H} \, |\psi(t)\rangle
\approx  \mathbf{V}_n(t) \; e^{-i dt/\hbar \, \mathbf{T}_n(t)} \;
\mathbf{V}_n^{T}(t) \; |\psi(t)\rangle \; , \label{eqn:lanczos_timeevolve} 
\end{equation}
where $\mathbf{V}_n$ is the matrix containing all the L\'anczos
vectors $|u_j\rangle$.
One finds that for a high accuracy only a very small Krylov space is
needed; $n \leq 20$ is sufficient.  
Therefore, the matrices in Eq.\ (\ref{eqn:lanczos_timeevolve}) are
very small, and the time-evolution can be computed efficiently.  
%
For details see the authors contribution to this topic in this volume,
Ref.~\cite{our_timeevolution}. 

Using this approach, it is possible to study a number of different
non-equilibrium properties of strongly correlated quantum systems.
It can also be used to obtain dynamical properties by computing
the Fourier-transform of a time-dependent correlation
function directly. 
As described in Ref.~\cite{our_timeevolution}, it is possible to
construct a time-evolution scheme for the DMRG based on this approach.  

\subsection{Calculations at Finite Temperature using Exact
Diagonalization} 

All approaches presented so far have been formulated for zero temperature.
The calculation of finite-temperature properties is a more demanding task
because thermodynamic expectation values are given by sums such as
\[
\langle A \rangle = \frac{1}{Z} \sum_n^N \langle n | A e^{-\beta
H}|n\rangle 
\]
with the partition function
\begin{equation}
Z = \sum_n^N \langle n | e^{-\beta H}|n\rangle
\end{equation}
where $ | n \rangle$ is an orthonormal basis.
As in the calculation of dynamical properties, 
it is prohibitively expensive to sum over
all $ | n \rangle$ due to the large
dimension of the basis.  
This difficulty is circumvented by a stochastic
approach (``stochastic sampling of Krylov space'') developed by
Jakli\v{c} and Prelov\v{s}ek \cite{jaklic}.
Their approach enables the calculation of thermodynamic
properties such as the specific heat, the entropy, or the static
susceptibilities of a system, as well as its static and dynamic
correlation functions. 
It is therefore a useful tool for comparing to experiments.

The approach is related to the high temperature series expansion, in
which expectation values are given by
\begin{eqnarray}
\langle A \rangle &=& Z^{-1} \sum\limits_n \sum\limits_{k=0}^{\infty}
\frac{\left( -\beta \right)^k}{k!} \langle n | H^k A | n \rangle 
; , \\
Z  &=& \sum\limits_n \sum\limits_{k=0}^{\infty}
\frac{\left( -\beta \right)^k}{k!} \langle n | H^k | n \rangle \; .
\end{eqnarray}
Expressions such as $\langle n | H^k A | n \rangle$ can be evaluated
using the Lanczos vectors and eigenvalues resulting from a
Lanczos run:
\begin{equation}
\langle n | H^k A | n \rangle = \sum\limits_{i=1}^M \langle n |
\psi_i^M \rangle \langle \psi_i^M | A | n \rangle \left(
\varepsilon_i^{(M)} \right)^k  \; .
\end{equation} 
However, the number of vectors needed is still too large.
To overcome this, in a second step one introduces a stochastic
sampling over different Krylov spaces, i.e., the Lanczos procedure is
performed repeatedly for a variety of random initial vectors, and an
average is taken over the samples at the end. 
This finally leads to the expression
\begin{equation}
\langle A \rangle \approx \frac{1}{Z} \sum_s \frac{N_s}{R} 
\sum_r^R \sum_m^M e^{-\beta \varepsilon^{(r)}_m} 
\langle r | \Psi^{(r)}_m \rangle \langle \Psi^{(r)}_m | A | r \rangle
\end{equation}
where
\begin{equation}
Z \approx \sum_s \frac{N_s}{R} 
\sum_r^R \sum_m^M e^{-\beta \varepsilon^{(r)}_m} 
\left |\langle r | \Psi^{(r)}_m \rangle\right |^2 \; .
\end{equation}
Here $ \sum_ s$ denotes summation over symmetry sectors of dimension $
N_s$, $ \frac{1}{R}\sum^R_r$ denotes the average over $R$ random
starting vectors  
$ |\Psi^{(r)}_0\rangle$, and $\sum_m$ is over the $m^{\rm th}$ Lanczos
propagation of the corresponding random starting vectors $
|\Psi^{(r)}_m\rangle$. 

This approach is useful if convergence with the number of
Lanczos propagations $ M \ll N_s$ and the number of random samples 
$R \ll N_s$ is sufficiently fast. 
Due to the relation to the high-$T$ expansion, the $T \to \infty$ limit
is reproduced correctly.
One obtains high- to medium-temperature properties in the
thermodynamic limit.
For finite systems, the low-temperature limit is reproduced
correctly up to the sampling error \cite{jaklic}.
Aichhorn {\em et al.} \cite{aichhorn} invented an approach which can be used
to reduce the sampling errors at low temperature.
They use the property that a twofold insertion of a Lanczos basis
leads to smaller fluctuations at low $T$.
To do this, the expectation value is expressed as
\begin{equation}
\langle A \rangle = \frac{1}{Z} \sum_n^N \langle n | 
e^{-\beta H/2} \, A \, e^{-\beta H/2}
|n\rangle \; , 
\end{equation}
and the projection to the Lanczos basis is performed.

This method has been used, e.g., for the investigation of the planar
$t-J$ model at finite temperature~\cite{jaklic_PRL}.

%
%
%
%

\subsection{Discussion: Exact Diagonalization}

As we have seen, the exact numerical treatment of quantum many body
systems is hindered by a number of significant obstacles.
However, it is possible to formulate conceptually straightforward
numerically exact methods which allow the treatment of surprisingly
large Hamiltonian matrices, especially when symmetries of the
Hamiltonian are taken into account.
Extensions to the basic method also make it possible to calculate
quantities which, in their simplest formulation, require the 
calculation of the full eigenspectrum of the Hamiltonian, such as
dynamical correlation functions, the time evolution and finite
temperature properties.
Nevertheless, maximum system sizes remain strongly limited because of 
the exponential growth of the many-body Hilbert space with system
size. 
In the following sections, we discuss additional numerically exact
methods which overcome this restriction, especially for
one-dimensional systems, albeit with the introduction of additional
approximations. 
These methods are capable of treating much larger systems,
containing up to as much as several thousand sites, 
allowing a reliable extrapolation to the thermodynamic limit to be
performed.
However, exact diagonalization nevertheless plays the role of a benchmark for
other methods because it provides numerically exact results in most
cases, and is useful for problems for which the other
approaches fail.

\section{Numerical Renormalization Group}

The numerical renormalization (NRG) was developed by Wilson 
as a numerical approach to the single-impurity Kondo and Anderson
problems, which, in conjunction with various analytic methods,
provides an accurate, complete solution of the problems
\cite{wilson_rg}.
The difficulty in these problems lies in the widely varying energy
scales that need to be accurately described.
For example, Kondo screening occurs at an exponentially small energy
scale set by the Kondo temperature.
Perturbation theory, while providing a good description at higher
temperatures, breaks down at the Kondo temperature.

Here we will be concerned primarily with illustrating the general
concepts and features of the method, exploring the conditions under
which it works or does not work, and relating it to other
methods; in particular, to exact diagonalization and to the DMRG.
More complete treatments of the NRG include 
Refs.\ \cite{wilson_rg,hewson_book,costi_dmrg_book}, on which the material
presented here is based.

\subsection{Anderson and Kondo problems}
The single-impurity Anderson model (SIAM) was introduced by Anderson
in 1961 to 
describe a spherically symmetric strongly correlated impurity in an
uncorrelated non-magnetic metal \cite{anderson}.
In lattice form, its Hamiltonian can be written
\begin{equation}
H^{AI} \; = \; \varepsilon_d \, \sum_\sigma n^d_\sigma 
\; + \; U n^d_{\uparrow} n^d_{\downarrow}
\; + \; \sum_{{\bf k},\sigma} \left (  V_{{\bf k},d} \,
c^\dagger_{{\bf k}, \sigma} d^{\phantom{\dagger}}_{{\bf k}, \sigma} +
{\rm H.c.}
\right )
\; + \; \sum_{{\bf k},\sigma} \varepsilon_{{\bf k}}\, 
c^\dagger_{{\bf k},\sigma} c^{\phantom{\dagger}}_{{\bf k},\sigma}
\label{eqn:SIAM_chain}
\end{equation}
where the non-degenerate impurity has energy $\varepsilon_d$ and
$n^d_\sigma = d^\dagger_{\sigma} d^{\phantom{\dagger}}_\sigma$
is its number operator and 
$d^\dagger_{\sigma}$ the corresponding creation operator.
The Coulomb interaction $U$ is local and takes place only on the
impurity, but the hybridization $V_{{\bf k},d}$ in general allows
scattering from all momentum states.
The hybridization function
\begin{equation}
\Delta(\omega) = \pi \sum_{\bf k}|V_{{\bf k},d}|^2 
\delta(\omega - \varepsilon_{\bf k} ) \; ,
\end{equation}
along with the density of states of the conduction band,
$\rho(\omega) =  \sum_{\bf k}\delta(\omega - \varepsilon_{\bf k} )$,
determine the behavior of the system.
Here we follow the usual treatment and consider only the orbitally
symmetric case, i.e., $V_{{\bf k},d} = V_{k d}$ and 
$\varepsilon_{\bf k} = \varepsilon_k$.
This corresponds to taking a completely isotropic electron gas, which
can be realized only approximately in a real solid.
We then need to consider only the s-wave states of the conduction
electrons, so that $c_{\bf k, \sigma}$ can be replaced by  
$c_{k, l=m=0, \sigma} \equiv c_{k, \sigma}$.

A closely related model is the Kondo or s-d exchange model, 
\begin{equation}
H^{K} \, = \, J_K \, {\bf S}_d \cdot {\bf s}_0
\, + \, \sum_{{\bf k},\sigma} \varepsilon_{{\bf k}}\, 
c^\dagger_{{\bf k},\sigma} c^{\phantom{\dagger}}_{{\bf k},\sigma}
\label{eqn:kondo}
\end{equation}
with
$
{\bf s}_0 = f^{\dagger}_{0,\sigma} {\vec{\sigma}}_{\sigma\mu}
f^{\phantom{\dagger}}_{0,\mu} 
$ and localized Wannier state generated by
$f^{\phantom{\dagger}}_{0,\sigma} = 
\sum_k c^{\phantom{\dagger}}_{k,\sigma}$, 
in which the electronic impurity is replaced by a localized spin-1/2
described by the spin operator ${\bf S}_d$.
The Kondo model can be derived as a strong coupling ($U \gg V$)
approximation to the SIAM via the Schrieffer-Wolff 
transformation in the symmetric case ($\varepsilon_d = -U/2$)
\cite{schrieffer-wolff}, leading to the correspondence 
$J_K = 8 V^2/U$. 

In order to bring the SIAM and the Kondo model into a form amenable to
numerical treatment, the models are mapped onto a linear chain model
using a Lanczos tridiagonalization procedure.
For the SIAM with orbital symmetry, the procedure is carried out as
follows.
The object is to transform the hybridization into a local term.
This can be done by defining
the localized Wannier state $|0,\sigma \rangle \equiv
f^\dagger_{0,\sigma} |\psi_{\rm vac} \rangle$ ($|\psi_{\rm vac}
\rangle$ designates the vacuum) with
\begin{equation}
f_{0,\sigma} = \frac{1}{V}\sum_k V_{k d} c_{k,\sigma}
\end{equation}
and the normalization $V \equiv \left (\sum_k |V_{k d}|^2 \right )^{1/2}$.
One then transforms the kinetic energy of the conduction band,
$H_c = \sum_k \varepsilon_k 
c^\dagger_{ k,\sigma} c^{\phantom{\dagger}}_{k,\sigma}$ to the new set
of operators, taking it from diagonal 
to tridiagonal form, by constructing a sequence of orthogonal states
generated by applying $H_c$:
\begin{eqnarray}
|1 \rangle & = & \frac{1}{\lambda_0}
\left [ H_c | 0 \rangle - |0 \rangle \langle 0 | H_c | 0 \rangle
\right ]\nonumber\\
|n + 1 \rangle & = & \frac{1}{\lambda_n}
\left [ H_c | n \rangle - |n \rangle \langle n | H_c | n \rangle
- |n -1 \rangle \langle n-1 | H_c | n \rangle \rangle \right ]
\end{eqnarray}
(where we have dropped the spin index $\sigma$ for compactness).
Note that this is a rather unusual analytic application of what is
essentially the Lanzcos procedure of Section \ref{lanczos_method}.
The result is that the SIAM is transformed to the form of a
semi-infinite tight-binding chain
\begin{eqnarray}
\tilde{H}^{AI} \; = \; \varepsilon_d \, n^d_\ell 
\;& + &\; U n^d_{\ell, \downarrow} n^d_{\ell, \downarrow}
\; + \; V \sum_{{\bf k},\sigma}\, \left (
f^\dagger_{0, \sigma} d^{\phantom{\dagger}}_{\sigma} +
{\rm H.c.} \right ) \nonumber \\
\; & + & \; \sum_{n=0,\sigma}^\infty \left [ \epsilon_{n}\,
f^\dagger_{n,\sigma} f^{\phantom{\dagger}}_{n,\sigma}
\, + \, \lambda_n \left ( 
f^\dagger_{n, \sigma} f^{\phantom{\dagger}}_{n+1,\sigma} +
{\rm H.c.} \right )
\right ]
\; , 
\end{eqnarray}
where, in second quantized notation, the operators
$f^\dagger_{n,\sigma}$ create an electron in state $|n \rangle$.
The Lanczos coefficients, $\lambda_n = \langle n+1 | H_c| n\rangle$
and $\epsilon_n =  \langle n | H_c| n \rangle$, depend
on the dispersion $\varepsilon_k$ and the hybridization function
$\Delta(\omega)$ of the original SIAM, and, in general, can be
determined, in the worst case numerically, during the Lanczos
procedure.
In general, they do not necessarily fall off with $n$, a property which is
crucial for the convergence of the NRG procedure, as we will see
below.
An analogous procedure can be carried out for the Kondo Hamiltonian 
(\ref{eqn:kondo}), starting with the local Wannier state 
$|0, \sigma\rangle$ generated by
$f^{\phantom{\dagger}}_{0,\sigma} = 
\sum_k c^{\phantom{\dagger}}_{k,\sigma}$.

\begin{figure}
\includegraphics[width=0.6\textwidth]{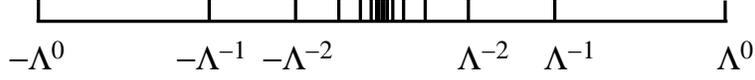}
\caption{logarithmic discretization of the conduction band.}
\label{fig:log_discret}
\end{figure}

In order to ensure that the tridiagonal chain formulations of the
SIAM or the Kondo model lead to a convergent NRG
procedure, additional approximations must be made.
In particular, a logarithmic discretization of the conduction band
leads to coefficients that fall off exponentially with $n$.
Here, we will discuss the Kondo model for concreteness, but extension
to the SIAM is straightforward.
In the first step, we take the density of states of the conduction
band to be a constant, $D$.
As long as there are no divergences at the Fermi level, the low-energy
physics should be dominated by the constant part, as can be justified by
expanding the dispersion $\varepsilon_k$ in a power series about the
Fermi vector $k_F$ \cite{wilson_rg}.
(Generalizations can be made for other densities of states.)
The next is a logarithmic discretization of the conduction
band $[-D < \varepsilon_k < D ]$ in energy, i.e., a division into 
intervals $D^+ = [\Lambda^{-(n+1)},\Lambda^{-n}]$ and 
$D^- = [-\Lambda^{-n},-\Lambda^{-(n+1)}]$ for the positive and negative
parts, respectively (see Fig.\ \ref{fig:log_discret}).
For each interval, we can expand the electron creation operators in a
Fourier series, defining
\begin{equation}
a^\dagger_{n,p,\sigma} \equiv \sum_{[k]_n} e^{i \omega_n p k } 
c^\dagger_{k,\sigma} \; ,
\end{equation}
where $[k]_n$ is the set of momentum points in the interval 
$\Lambda^{-(n+1)} < k < \Lambda^{-n}$, $p$ is an integer, and
$\omega_n = 2 \pi \Lambda^{n+1}/(\Lambda-1)$.
For an interval $\Lambda^{-(n+1)} < k < \Lambda^{-n}$ in the 
negative range, operators $b^\dagger_{n,p,\sigma}$ can be defined
analogously,
It can be shown that the operators $a^\dagger_{n,p,\sigma}$,
$b^\dagger_{n,p,\sigma}$, and their hermitian conjugates obey the usual
anticommutation rules for fermions. 
For $H^K$, Eq.\ (\ref{eqn:kondo}), the localized state can be
expressed as
\begin{equation}
f^\dagger_{0,\sigma} = 
\left ( 1 - \Lambda^{-1} \right )^{1/2} \;
\sum_n \Lambda^{-n/2} \; 
\left ( a^\dagger_{0,p,\sigma} + b^\dagger_{0,p,\sigma} \right )
\end{equation}
At this point, the approximation is made to neglect all higher terms
in the Fourier series, keeping only one electron per logarithmic
energy interval, i.e., 
$a^\dagger_{n,0,\sigma}$ and $b^\dagger_{n,0,\sigma}$.
Since the impurity only couples directly to the localized state,
neglecting these states amounts only to neglecting off-diagonal matrix
elements, which can be shown to be proportional to
$(1-\Lambda^{-1})$. 
Therefore, the approximation becomes valid in the limit $\Lambda\to 1$;
a more detailed analysis of the errors can be found in 
Ref.\ \cite{wilson_rg}.

After applying the tridiagonalization procedure outlined above to the
logarithmically discretized version of the Hamiltonian $H^K$, we
arrive at the effective chain Hamiltonian for the Kondo 
model
\begin{equation}
\tilde{H}^K =  \tilde{D} \, \sum_{n=0, \sigma}^\infty \,
\Lambda^{-n/2} \, \left ( 
f^\dagger_{n, \sigma} f^{\phantom{\dagger}}_{n+1,\sigma} +
{\rm H.c.} \right )
\, + \,
2 J_K \, \sum_{\sigma, \mu} \, f^\dagger_{0, \sigma} {\vec{\sigma}}_{\sigma\mu}
f^{\phantom{\dagger}}_{0,\mu} \; 
\label{eqn:kondo_chain}
\end{equation}
where $\tilde{D} = D (1 + \Lambda^{-1})/2$.
The tight-binding part has the same form as that in 
Eq.\ (\ref{eqn:SIAM_chain}) with 
$\epsilon_n = 0$ and 
$\lambda_n \approx \frac{1}{2} (1+\Lambda^{-1})\Lambda^{-n/2}$ for
large $n$ \cite{wilson_rg}.

This form of the tight-binding Hamiltonian for the Kondo model and an
analogous one for the SIAM are treatable numerically with the NRG.
Crucial for the convergence is that the tight-binding coupling decays
exponentially with the position on the lattice.
Physically, this discretization was carefully thought out by Wilson to
reflect the exponentially small energy scales evident in the behavior
of perturbation theory for the Kondo problem \cite{wilson_rg,hewson_book}.
Note that it is crucial to adjust the discretization parameter
appropriately.
If $\Lambda$ is too close to unity, the NRG will not converge
sufficiently quickly, if $\Lambda$ is chosen to be too large, the error
from the logarithmic discretization becomes too large.
In practice, one chooses a value around 2, but the extrapolation
$\Lambda\to1$ should, in principle, be carried out.

\subsection{Numerical RG for the Kondo Problem}

\subsubsection{Renormalization Group Transformation}

We will now outline the NRG procedure as applied to Hamiltonian
(\ref{eqn:kondo_chain}).
The goal is to investigate the behavior of the system at a given
energy scale by treating a finite system of length $L$ with
Hamiltonian
\begin{equation}
\tilde{H}_L =  \tilde{D} \, \sum_{n=0, \sigma}^{L-1} \,
\Lambda^{-n/2} \, \left ( 
f^\dagger_{n, \sigma} f^{\phantom{\dagger}}_{n+1,\sigma} +
{\rm H.c.} \right )
\, + \,
2 J \, \sum_{\sigma, \mu} \, f^\dagger_{0, \sigma} {\vec{\sigma}}_{\sigma\mu}
f^{\phantom{\dagger}}_{0,\mu} 
\end{equation}
(dropping the superscript/subscript ``K'' for compactness).
Due to the exponentially decaying couplings, a particular system size will
then describe the energy scale set by 
$D_L \equiv \tilde{D}/\Lambda^{(L-1)/2}$.

The idea of Wilson was to examine the behavior (i.e., the renormalization
group flow) of the lower part of the appropriately rescaled eigenvalue
spectrum of the sequence of Hamiltonians 
$\tilde{H}_L$, $\tilde{H}_{L+1}$, $\ldots$ numerically.
It is convenient to rescale the Hamiltonians directly, defining
$H_L \equiv \tilde{H}_L/D_L$.
One can relate $H_L$ to $H_{L+1}$ through the recursion relation
\begin{equation}
H_{L+1} = \Lambda^{1/2} \, H_L + \sum_{\sigma}
\left ( f^\dagger_{L, \sigma} f^{\phantom{\dagger}}_{L+1,\sigma} +
{\rm H.c.} \right )
 \; \equiv {\cal R}[H_L] \; .
\label{eqn:NRG_recursion}
\end{equation}
This defines the RG transformation.
In principle, this transformation is exact (up to the discretization
error associated with $\Lambda$).
However, if one were to treat each subsequent $H_L$ by numerical
diagonalization, the memory and work needed would increase
exponentially because the number of degrees of freedom is multiplied
by four at each step.
As an approximation, Wilson suggested to keep at most a fixed number $m$
of the lowest-lying eigenstates of $H_L$ at each step.
For the Kondo problem, the error made at each step can be shown to be
of order $\Lambda^{-1/2} < 1$ \cite{wilson_rg}.

\subsubsection{Numerical Procedure}

\label{sec:NRG_procedure}

\begin{figure}
\includegraphics[width=0.6\textwidth]{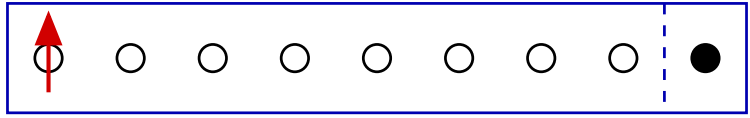}
\caption{Schematic depiction of the NRG procedure.}
\label{fig:NRG}
\end{figure}

The NRG method then proceeds as follows:
\begin{itemize}
\item[1)] Diagonalize $\mathbf{H}_L$ numerically, finding the $m$ lowest
eigenvalues and the corresponding eigenvectors.
\item[2)] Use the undercomplete similarity transformation $\mathbf{O}_L$ formed
by the $m$ eigenstates obtained in 1) to transform
all relevant operators on the $L$-site system to the new basis.
For example, $\bar{\mathbf{H}}_L = \mathbf{O}^\dagger \mathbf{H}_L
\mathbf{O}_L$ is a diagonal matrix of dimension $m$, but other
operators $\bar{\mathbf{A}}_L = \mathbf{O}^\dagger \mathbf{A}_L
\mathbf{O}_L$ 
will not, in general, be diagonal.
\item[3)] Form $\mathbf{H}_{L+1}$ from $\bar{\mathbf{H}}_L$ using the
recursion relation (\ref{eqn:NRG_recursion}), i.e., by adding a site
to the chain and constructing $\mathbf{H}_{L+1}$ in the expanded
product basis. 
\item[4)] Repeat 1)-3), substituting $\mathbf{H}_{L+1}$ for
$\mathbf{H}_L$. 

\end{itemize}
This procedure is illustrated schematically in Fig.\ \ref{fig:NRG}.
The procedure can be started with the purely local term $H_0$
consisting of the impurity coupled to a single site, but in practice,
the relevant first step occurs when the dimension of $\mathbf{H}_L$ is
greater than $m$.
In step 1), one quarter of the eigenstates must be found at a general
step for the Kondo problem, so that typically ``complete
diagonalization'' algorithms as discussed in
Sec.~\ref{sec:ed_complete} 
are used.  
In general, the matrices to be diagonalized are block diagonal with
respect to the conserved quantum numbers of the system, such as number
of conduction electrons $N$ and the projection of the total spin
$S^z$.
As described in Sec.~\ref{sec::representation_manybodystates}, it is
important for numerical efficiency to separate these blocks. 
Once this is done, the matrices to be diagonalized are not
particularly sparse.
In step 3), matrix elements of operators linking sites $L$ and $L+1$ such as 
$f^\dagger_{L,\sigma} f_{L+1,\sigma}$ must be constructed in the
product basis $| i, s_L\rangle \equiv | i \rangle | s_{L+1} \rangle $, 
with the first ket
representing the basis of $\bar{H}_L$, and the second the basis of the
added site (consisting of 4 states for the Kondo problem).
The expression for the matrix elements of $H_{L+1}$ is
\begin{eqnarray}
\langle i, s_L | H_{L+1} | i', s'_L \rangle & = &
\Lambda^{1/2} \; \langle i | i' \rangle \, 
\langle s_{L+1} | s'_{L+1} \rangle \; E^L_i \nonumber \\
& & + \; (-1)^{N_{S_{L+1}}} \; 
\langle i | f^\dagger_{L,\sigma}| i' \rangle  \,
\langle s_{L+1} | f^{\phantom{\dagger}}_{L+1,\sigma}| s'_{L+1}\rangle
 \\ 
& & + \; (-1)^{N_{S'_{L+1}}} \; 
\langle i | f^{\phantom{\dagger}}_{L,\sigma}| i' \rangle  \,
\langle s_{L+1} | f^{\dagger}_{L+1,\sigma}| s'_{L+1}\rangle \nonumber
\end{eqnarray}
where the $E^L_i$ are the eigenvalues of $H_L$, $i$ runs from 1 to
$m$, and $N_{S_L}$ is the number of electrons in state $S_L$.

Note that at a particular point in the procedure, the range of
eigenvalues of $\tilde{H}_{L} = D_L H_L$ which accurately approximates
the spectrum of the infinite system is limited.
In particular, the accuracy breaks down due to the truncation at a
scale that is a multiple $\alpha$ of $D_L$ which depends on the value of
$\Lambda$ and the number of states kept $m$; typically $\alpha$ is of
order 10 when $m$ is of order 1000.
A lower limit is set by the energy scale $D_L$: eigenvalues below
$D_L$ are approximated more accurately at subsequent steps, i.e., for
$L'$ larger than $L$ and smaller $D_{L'}$.
Therefore, the non-rescaled eigenvalues $\tilde{E}_L$ in the range 
$D_L \le \tilde{E}_L \le \alpha D_L$ can be accurately calculated at a
particular step.

\subsubsection{Renormalization Group Flow and Fixed Points}

In order to understand the behavior of a system within the
renormalization group in general, one searches for fixed points of the
renormalization group transformation \cite{wilson_rg}, defined by
\begin{equation}
{\cal R}[H^*] = H^* \; .
\label{eqn:RG_fixed_point}
\end{equation}
In analytic variants of the renormalization group, the behavior of the
Hamiltonian $H$ is
generally parameterized by a small number of coupling constants.
Finding fixed points then amounts to finding stationary points in
the flow equations governing these coupling constants.
In the NRG, identifying fixed points is little more subtle: in 
practice, when the first $p$ rescaled energy levels $E_p^L$
are independent of $L$ for a particular (appropriately chosen) range
of $L$, one identifies a fixed point.
Some insight into the physics of the problem is usually necessary 
to choose appropriate ranges of $p$ and $L$.
Once fixed points are identified, the physicical behavior governed by
the fixed point is determined by the structure of the low-lying
eigenstates.
Note that such behavior is not guaranteed for a more general tight-binding
Hamiltonian.

For the Kondo problem (and the SIAM in appropriate parameter
regimes), two fixed points can be clearly identified;
one associated with the behavior of model (\ref{eqn:kondo_chain})
at $J=0$ and one with its behavior at $J=\infty$.
In both cases, the behavior at the fixed point is easy to understand:
for $J=0$, the impurity is uncoupled to the conduction band and the 
excitation energies are those of the non-interacting tight-binding
band extending from 0 to $L$.
For $J=\infty$, the impurity forms an infinitely tightly bound
singlet with site 0 of the tight-binding chain, effectively removing it
from the system. 
The excitation energies relative to the ground state are therefore
those of a chain extending from 1 to $L$, i.e., the excitation
spectrum for $H_L(J=\infty)$ is the same as that for $H_{L-1}(J=0)$.
An additional complication is that the nature of the spectrum for
$H_L(J=0)$ depends on whether $L$ is even or odd; the asymptotic
values of the scaled excitation energies are different for even and
odd $L$ are different, even in the limit of large $L$.
This can be taken into account, however, by always applying a 
sequence of two renormalization group transformations, i.e., by
replacing $\cal R$ with ${\cal R}^2$ in (\ref{eqn:RG_fixed_point}) to
determine the fixed points and then considering the odd $L$ and even
$L$ cases separately.
In fact, the crossover from the $J=0$ fixed point to the $J=\infty$
fixed point amounts to a reversal of the behavior for odd and even $L$
because the zeroth lattice site is effectively removed from the chain
at the strong coupling fixed point. 
For details on the structure of the excitation energies, see 
Refs.\ \cite{wilson_rg,hewson_book,krishna-murthy}.

\begin{figure}
\includegraphics[width=0.5\textwidth]{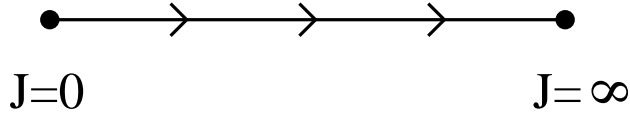}
\caption{Renormalization group flow diagram for the Kondo model.}
\label{fig:RG_flow}
\end{figure}

Numerically, one finds that for small $L$ and small but finite $J$,
the structure of the excitations is that of the $J=0$ fixed point,
whereas for large $L$ the structure is that of the $J=\infty$ fixed
point.
Stability analysis, which can be carried out analytically near the fixed
points, shows that the effective Hamiltonian for the weak-coupling
fixed point has a marginal operator, indicating that it is unstable,
while the effective Hamiltonian for the strong-coupling fixed point has no
relevant operators, indicating that it is stable
\cite{wilson_rg,krishna-murthy,hewson_book}, in agreement with
the behavior observed in the NRG.
The renormalization group flow from the unstable $J=0$ fixed point to
the stable $J=\infty$ fixed point is depicted in 
Fig.\ \ref{fig:RG_flow}.

\subsubsection{Calculation of Thermodynamic Properties}

Thermodynamic quantities such as the specific heat or the impurity
susceptibility can be easily calculated within the NRG if the range of
validity of the excitation spectrum is taken into account.
Generally, one is interested in the impurity contribution to the
thermodynamic quantities, derived from the impurity free energy
$F_{\rm imp}(T) = -k_B T \ln Z / Z_c$, where $Z_c$ is the exactly
calculable partition function for the noninteracting conduction band.
If the entire eigenvalue spectrum were known, the partition function
would be given by $Z(T) = {\rm Tr} \exp (-\tilde{H}^K/k_B T)$.
However, at a particular stage of the NRG, what one can calculate is
the partition function for the truncated lattice
\begin{equation}
Z_L(T) \equiv {\rm Tr} e^{-\tilde{H}_L/k_B T} 
= \sum_n e^{-\tilde{E}^L_n/k_B T}
= \sum_n e^{- D_L E^L_n/k_B T} \; .
\end{equation}
Evidently, $Z_L(T)$ can only be a good approximation for $Z(T)$ when
the temperature for a particular system size $T_L$ is chosen so that
$k_B T_L \ll \alpha D_L$, the largest
energy scale accurately described by $\tilde{H}_L$.
We previously argued that the minimum energy is set by $D_L$; the
error made in substituting $Z_L$ for $Z$ in calculating impurity
properties has been more rigously estimated to be $D_L/\Lambda k_B T$ 
in Ref.\  \cite{krishna-murthy}.
Therefore, the valid temperature range for a given $L$ is set by
\begin{equation}
\Lambda^{-1} \ll \frac{k_B T_L}{D_L} \ll \alpha \; , 
\end{equation}
where $\alpha$ depends on $m$ and $\Lambda$.
In practice, $k_B T_L \approx D_L$ is a reasonable choice.

Experimentally interesting quantities are the impurity
specific heat
\begin{equation}
C_{\rm imp} = - T \frac{\partial^2}{\partial T^2} F_{\rm imp}(T) 
\end{equation}
and the magnetic susceptibility at zero field due to the impurity 
\begin{equation}
\chi_{\rm imp} = \frac{(g \mu_B)^2}{k_B T} \left [
\frac{{\rm Tr} (S_L^z)^2 \, e^{-\tilde{H}/k_B T}}{Z} -
\frac{{\rm Tr} (S_L^{z,c})^2 \, e^{-\tilde{H}_c/k_B T} }{Z_c}
\right ]
\end{equation}
where $S_L^z$ and $S_L^{z,c}$ are the $z$-components of the total spin
on the $L$-site chain with and without the impurity spin,
respectively.
The Hamiltonian $H_c$ is that of the noninteracting conduction band.
The high- and low-temperature limits, as well as the leading behavior
around these limits have been calculated analytically.
These limits can serve as a check of the accuracy of the NRG
calculations.
For a more complete depiction and discussion of results for
thermodynamic properties, see
Refs.\ \cite{wilson_rg,krishna-murthy, hewson_book}.

\subsubsection{Dynamical Observables and Transport Properties}

Dynamical properties, both at zero and at finite temperature, can also
be calculated within the NRG procedure.
To be concrete, we will discuss perhaps the most experimentally
interesting quantity, the impurity spectral function at zero
temperature (for simplicity):
\begin{equation}
A(\omega) = -\frac{1}{\pi} \,{\rm Im} \, G(\omega + i\eta) \; ,
\end{equation}
where 
\begin{equation}
G(t) = 
-i \langle \psi_0 |\, T \, d(t) d^\dagger(0) \, | \psi_0 \rangle
\end{equation}
is the retarded impurity Green function 
[c.f.\ Eqs. (\ref{eqn:t_Green_fctn})-(\ref{eqn:lehmann_repn})].
For finite $L$, it is convenient to calculate the spectral weight
within the Lehmann representation
\begin{equation}
A_L(\omega) = \frac{1}{Z_L} \sum_{p} 
\left | \langle p | d^\dagger_\sigma | 0 \rangle \right |^2
\delta(\omega - E_p + E_0)
\, + \, 
\left | \langle 0 | d^\dagger_\sigma | p \rangle \right |^2
\delta(\omega + E_p - E_0) \; .
\label{eqn:lehmann_impurity}
\end{equation}
Since the excitations out of the ground state for $\tilde{H}_L$ are
well-represented in the energy range $D_L \le \omega \le \alpha D_L $, 
as discussed previously, $A(\omega_L) \approx A(\omega)$ when
$\omega$ is chosen to be within this range.
In practice, a typical choice is $\omega = 2 \omega_L$ where 
$\omega_L \equiv k_B T_L = D_L$.
One usually uses Eq.\ (\ref{eqn:lehmann_impurity}) directly to
calculate the spectrum, rather than the more sophisticated methods
such as the Krylov method outlined in Sec.\ \ref{sec:cont_fraction} or
the correction vector method outlined in 
Sec.\ \ref{sec::ed_correction_vector} because all eigenstates within
the required 
range of excitation energy are available within the NRG procedure; it
is then easy to calculate the poles and matrix elements within this
range.
Note that the result obtained is a set of positions and weights of
$\delta$-functions; in order to compare with continuous experimental
spectra, they must be broadened.
Typical choices for a broadening function are Gaussian or Logarithmic
Gaussian distributions of width $D_L$ \cite{sakai,costi_dmrg_book}.

One interesting application of impurity problems comes about in the
context of the 
Dynamical Mean Field Theory (DMFT), in which a quantum lattice model
such as the Hubbard model is treated in the limit of infinite
dimensions \cite{metzner-vollhardt,jarrell,georges_RMP}.
The problem can be reduced to that of a generalized SIAM interacting
with a bath or host.
The fully frequency-dependent impurity Green function of this
generalized SIAM must then be calculated in order to iterate a set of
self-consistent mean-field equations.
Various methods can be used to solve the impurity problem including
perturbation theory, exact diagonalization, quantum Monte Carlo,
dynamical DMRG (DDMRG), and the NRG \cite{georges_RMP,bulla_DMFT}. 
Dynamical properties at finite temperature can be calculated
using similar considerations, as long as the additional energy scale
$k_B T$ is taken into account 
\cite{costi-hewson91,costi-hewson92,costi-hewson93,costi_dmrg_book}.
In particular, the procedure outlined above can be used as long as
frequencies $\omega > k_B T$ are considered.
For $\omega < k_B T$, additional excitations at higher energies than
$D_L < \omega < \alpha D_L$ become important.
Frequencies in this range can then be handled using a {\sl smaller}
$L$ so that $\omega_L < K_B T$.
Since transport properties such as the resistivity $\rho(T)$ can be
formulated in terms of integrals over frequency of frequency-dependent
dynamical correlation functions \cite{costi-hewson93}, these methods
can also be used to calculate them.

\subsection{Numerical RG for Quantum Lattice Problems}

There are a number of quantum lattice problems that have Hamiltonians
whose structure is formally similar to the tight-binding
Hamiltonian for the Kondo model (\ref{eqn:kondo_chain}) such as the
Hubbard model (\ref{eqn:hubbard}) or the Heisenberg model
(\ref{eqn:heisenberg}) in one dimension.
While one could consider carrying out a variation of the NRG procedure on
these models in order to perform an approximate exact diagonalization
on a finite lattice, the physically interesting case is the one in
which the couplings between nearest-neighbor sites are all equal.
This amounts to setting $\Lambda=1$ in the Kondo case, the point at 
which the convergence of the NRG method breaks down completely because
the identification of the length $L$ with the energy scale is lost.
Nevertheless, adaptations of the NRG procedure were applied to small
Hubbard chains \cite{bray-chui}, obtaining an error
in the ground-state energy of approximately 10\% after 4
renormalization group steps.
Later work on the spin-1 Heisenberg chain obtained an error of
approximately 3\% in the ground-state energy of the $L=18$ chain
\cite{xiang-gehring}.
It is important to note that such calculations are variational so that
quantities which characterize the long distance behavior such as
correlation functions which depend on the wave function can have much
larger errors than the error in the energy.
Finally, an adaptation of the NRG procedure to a two-dimensional
noninteracting electron gas was used to study the Anderson localization
problem in two dimensions \cite{lee}.
The result obtained was that the system undergoes a
localization-delocalization transition as a function of disorder
strength, a result later discovered to be incorrect: there is no
transition because two is the lower critical dimension
\cite{lee_fisher,bigfour}. 

\subsection{Numerical RG for a Noninteracting Particle}
\label{sec::nrg_noninteracting}
More insight into the breakdown of the NRG procedure for
one-dimensional lattice problems with non-decaying couplings can be gained by
applying a variant of the procedure to the problem of a single
particle on a tight-binding chain, Eq.~(\ref{eqn:tb}).
We consider the Hamiltonian in the formulation
\begin{equation}
  H = -\sum_{\ell=1}^{L-1} \left ( ~
  | \ell \rangle \langle \ell+1 | + | \ell+1 \rangle \langle \ell | ~\right )
  + 2 \sum_{\ell=1}^L |\ell \rangle \langle \ell |  \; ,
\label{eqn:nonint_tight_bind}
\end{equation}
where that state $|i\rangle$ represents an orbital localized on site
$\ell$.
In a matrix representation, this Hamiltonian is tridiagonal and
is equivalent to a discretized second derivative operator, 
$-\partial^2/\partial x^2$.
Note that Hamiltonian (\ref{eqn:nonint_tight_bind}) does not include a
nonzero matrix element between sites 1 and $L$, so that fixed boundary
conditions have been applied to the chain, i.e., the wave function is
required to vanish at the ends.
We modify the Wilson NRG procedure slightly to take into account that
we are treating a simpler, noninteracting system.
There are two significant changes: first, we put together 
two equal-sized systems (called ``blocks'') rather than just adding a
site because the dimension of the Hilbert space grows more slowly than
in an interacting system: linearly rather than exponentially with the
length.
Second, the mechanics of putting two blocks together is simpler in the
interacting system.

In terms of the procedure outlined in Sec.\ \ref{sec:NRG_procedure}, the
diagonalization of the Hamiltonian, step 1),  and the transformation
of the relevant operators using $\mathbf{O}_L$, a matrix whose columns
are the $m$ eigenvectors of $\mathbf{H}_L$ with the lowest
eigenvalues, [step 2)] are carried out as before. 
The procedure is somewhat simplified because of the less complicated
structure of  Hamiltonian (\ref{eqn:nonint_tight_bind}); for example,
there are no $S^z$ and $N$ quantum numbers which decouple sectors of
$\mathbf{H}$. 
The operators to be transformed are the Hamiltonian for a block
$\bar{\mathbf{H}}_L = \mathbf{O}^\dagger_L \mathbf{H}_L \mathbf{O}_L$
which is diagonal and $\bar{\mathbf{T}}_L = \mathbf{O}^\dagger_L
\mathbf{H}_L \mathbf{O}_L$, where $\mathbf{T}_L$ represents the
connection between the blocks. 
The procedure is conveniently started at $L=1$ for which $\mathbf{H}_1
= 2$ and $\mathbf{T}_L=-1$ can be represented as $1\times 1$
matrices. 
Matrix representations of a system of size $2L$ are then formed [step 3)]
as
\begin{eqnarray}
\mathbf{H}_{2L}=
\left(
\begin{array}{cc}
\bar{\mathbf{H}}_L & \bar{\mathbf{T}}_L \\
\bar{\mathbf{T}}^\dagger_L & \bar{\mathbf{H}}_L \\
\end{array}
\right)
\label{eqnH_2L}
\end{eqnarray}
and
\begin{eqnarray}
\mathbf{T}_{2L}=
\left(
\begin{array}{cc}
0 & 0 \\
\bar{\mathbf{T}}_L & 0 \\
\end{array}
\right) \; .
\label{eqnT_2L}
\end{eqnarray}
The procedure is then iterated [step 4)], doubling the size of the
system at each step.
The matrix to be diagonalized, $\mathbf{H}_{2L}$, is of size $2m\times
2m$ at a general step, and the truncation is to an $m\times m$ matrix,
i.e., one half of the Hilbert space is discarded. 
Note that the procedure is exact for the first few steps, as long as
$m \le L$.

\begin{table}
\caption{ Lowest energies after 10 blocking
transformations for the  
noninteracting single particle on a 1-D chain with fixed boundary
conditions, keeping up to $m=8$ states.}
\begin{tabular}{|c|c|c|c|}
\hline 
{$\;$} &   {Exact} & {Wilson} &  {Fixed-Free} \\
\hline 
$\ E_0\ $ 
& $\; 2.3508\times 10^{-6}\; $ 
& $\; 1.9207 \times 10^{-2} \; $
& $\; 2.3508 \times 10^{-6} \;$ 
\\
$\ E_1\ $ 
& 9.4032$\; \times 10^{-6}\; $ 
& 1.9209$\; \times 10^{-2}\;$
& $9.4032\times 10^{-6}$ 
\\
$\ E_2\ $ 
& $2.1157\times 10^{-5}$ 
& $1.9214\times 10^{-2}$
& $2.1157\times 10^{-5}$ 
\\
$\ E_3\ $ 
& $3.7613\times 10^{-5}$ 
& $1.9217\times 10^{-2}$
& $3.7613\times 10^{-5}$ 
\\
\hline
\end{tabular}
\label{tab:box_results}
\end{table}

\begin{figure}
  \includegraphics[width=0.5\textwidth]{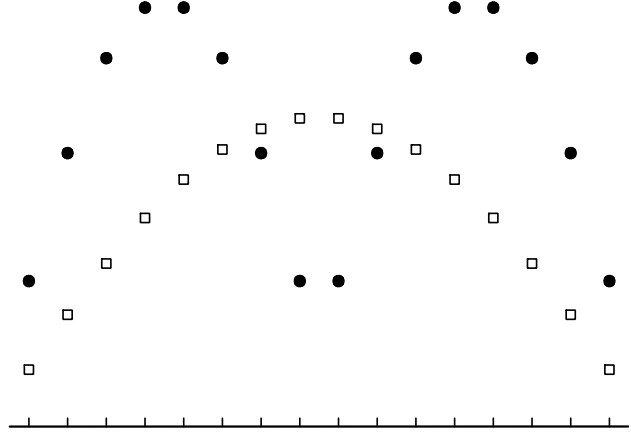}
\caption{The lowest eigenstates of two 8-site blocks (solid circles) and a
16-site block (open squares) for the one-dimensional tight--binding
model with fixed boundary conditions. }
\label{fig:pboxwf}
\end{figure}

As shown in the first two columns of Table \ref{tab:box_results},
the accuracy of the eigenvalues obtained breaks completely after a
moderate number of steps.
The failure of the NRG for this simple problem as well as the reason
for the failure was pointed out by
Wilson in 1986 \cite{wilson_seminar}.
For a particular $L$, the eigenfunctions have the form
\begin{equation}
\psi^L_n(\ell) \propto \sin \left ( 
\frac {n \pi \ell}{L+1} \right ) \; , 
\hspace*{2cm} n=1,\ldots,L
\end{equation}
because fixed boundary conditions have been applied to the system.
In one iteration of the NRG procedure, a linear combination of the
$\psi^L_n(\ell)$ for $n \le m$ are used to form an approximation to
$\psi^{2L}_n(\ell)$.
As depicted in Fig.\ \ref{fig:pboxwf}, the boundary conditions that
are applied to $\mathbf{H}_L$ lead to a non-smooth ``dip'' in the
wave function in the middle of the block of size $2L$ when the NRG
transformation is carried out.
Clearly, this dip can only be removed by forming a linear combination
of almost all of the $\psi^L_n(\ell)$.
The lesson that is learned, then, is that how the 
boundaries of the blocks are treated in the NRG procedure is crucial: 
a more general treatment is necessary to formulate a numerically
accurate real-space NRG procedure for short-ranged quantum lattice
models. 

\section{From the NRG to the Density Matrix Renormalization Group}

\subsection{Better Methods for the Noninteracting Particle}
\label{sec:better_methods}
Perhaps the most obvious way of improving the NRG, at least for the
single tight-binding particle, is to apply a more general set of
boundary conditions to the system to be diagonalized.
For example, one can apply fixed (vanishing wave function in the
continuum limit) or free boundary conditions (vanishing first
derivative) by forming the matrix
\begin{eqnarray}
\mathbf{H}^{bb'}_{2L} =
\left(
\begin{array}{ccc}
\bar{\mathbf{H}}^{b, \rm fixed}_L & \; & \bar{\mathbf{T}}_L \\
\bar{\mathbf{T}}_L^\dagger & \; & \bar{\mathbf{H}}^{\rm fixed, b'}_L \\
\end{array}
\right)
\end{eqnarray}
when putting two blocks together, where the boundary condition $b$ can
be fixed or free \cite{white_and_noack}.
For example,
\begin{eqnarray}
\mathbf{H}^{\rm free, fixed}_{L=2} = 
\left(
\begin{array}{ccc}
1 & \; & -1 \\
-1 & \; & 2 \\
\end{array}
\right) \; .
\end{eqnarray}
The Hamiltonian $\mathbf{H}^{b b'}_{2L}$ can then be diagonalized for all 4
combinations of boundary conditions to obtain a more general set of
basis functions, albeit an overcomplete set which is not
orthogonalized. 
However, if only a small number (e.g., $m/4$) of eigenstates from each set
$\{b,b'\}$ are kept in the transformation matrix $\mathbf{O}_L$, the
columns of $\mathbf{O}_L$ can then be orthogonalized numerically using
the  Gram--Schmidt procedure, resulting in an undercomplete orthogonal
transformation, as before.
The matrices  $\bar{\mathbf{H}}^{b b'}_L = \mathbf{O}_L^\dagger \mathbf{H}^{b b'}_L \mathbf{O}_L$ and 
$\bar{\mathbf{T}}_L = \mathbf{O}_L^\dagger \mathbf{H}^{b b'}_L \mathbf{O}_L$ must then be transformed to
prepare for the next step.
Note that $\bar{\mathbf{H}}^{b b'}_L$ is not diagonal.

As can be seen in Table \ref{tab:box_results}, this ``combination of
boundary conditions'' method works
astoundingly well, producing the first few eigenvalues to almost machine
accuracy for a 2048-site lattice, keeping only $m=8$ states
\cite{white_and_noack}.
However, the crucial question is whether the technique can be generalized
to interacting many-particle systems.
Unfortunately, no good method has been found to extend these
``combination of boundary conditions'' methods.
The reason is that it is not clear that an appropriately general set
of boundary conditions can be found for a many-particle wave function,
which is a complicated function of the coordinates of all the
particles.

However, another method developed in Ref.\ \cite{white_and_noack},
called the {\it superblock} method, can be generalized to
interacting systems.
Rather than applying a general set of boundary conditions to a block
of size $L$, 
this method forms a new basis for $\bar{\mathbf{H}}_{2L}$ and $\bar{\mathbf{T}}_{2L}$
based on the idea that these blocks will eventually make
up part of a larger system.
In order to do this, a ``superblock'' (with periodic boundary
conditions) made up of $p>2$ blocks is formed and diagonalized.
For example, 
\begin{eqnarray}
\mathbf{H}^{p=4}_{2L}=
\left(
\begin{array}{cccc}
\bar{\mathbf{H}}_L & \bar{\mathbf{T}}_L & 0 & \bar{\mathbf{T}}^{\dagger}_L \\
\bar{\mathbf{T}}^{\dagger}_L & \bar{\mathbf{H}}_L & \bar{\mathbf{T}}_L & 0 \\
0  & \bar{\mathbf{T}}^{\dagger}_L & \bar{\mathbf{H}}_L & \bar{\mathbf{T}}_L  \\
\bar{\mathbf{T}}_L & 0 & \bar{\mathbf{T}}^{\dagger}_L & \bar{\mathbf{H}}_L   \\
\end{array}
\right) \; .
\end{eqnarray}
The transformation $\mathbf{O}_{2L}$ is then made up by projecting the $m$
lowest--lying eigenstates of $\mathbf{H}^{p}_{2L}$ onto the coordinates of
the first two blocks, and then orthonormalizing its columns.
In other words, if $u^\alpha_j$ (with $j=1,\ldots,4m$) is an
eigenvector of $\mathbf{H}^{4}_{2L}$, then a nonorthonormalized column vector
of $\mathbf{O}_{2L}$ is composed of the first $2m$ elements, $j=1,\ldots,2m$,
of $u^\alpha_j$, assuming $\bar{\mathbf{H}}_L$ is an $m\times m$ matrix.
This new basis is used to transform 
$\bar{\mathbf{H}}_{2L} = \mathbf{O}^\dagger_{2L} \mathbf{H}_{2L} \mathbf{O}_{2L}$ and 
$\bar{\mathbf{T}}_{2L} = \mathbf{O}^\dagger_{2L} \mathbf{T}_{2L} \mathbf{O}_{2L}$, as defined in 
(\ref{eqnH_2L}) and (\ref{eqnT_2L}).

The superblock method yields good results for the single-particle
tight-binding chain, albeit not as accurate as the fixed-free variant
of the combination of boundary conditions method \cite{white_and_noack}.
As one might expect, the method also becomes more accurate as the
number of blocks making up the superblock, $p$, is increased.
If $p$ can be made arbitrarily large, it can be adjusted so that one
is simply carrying out an exact diagonalization of a system of the
desired size.
Since the additional blocks in the superblock allow fluctuations at
the boundaries of the blocks whose wave functions are used in the
renormalization group blocking step, this method is more promising to
apply to interacting systems: such fluctuations should also treat the
boundaries of blocks of a many-body system in a more general way.
For a single-particle system, the projection of the wave function of
the superblock onto the subsystem of interest is trivial: it is a
one-to-one projection because the Hilbert space of the superblock is a
direct sum of that of its subsystems.
This projection is more complicated for a many-body system: one state
of the superblock can project onto many states of a subsystem because
the Hilbert space grows exponentially, as a direct product of that
of the subsystems.

It is also possible to treat the single particle on a tight-binding
chain, Hamiltonian (\ref{eqn:nonint_tight_bind}), using an algorithm
that is a direct adaptation of the DMRG for interacting systems
\cite{reinhard_dmrg_book}.
This algorithm is linear in the system size $L$
and involves diagonalizing only a $4\times4$ matrix at each DMRG step
\cite{single_particle_dmrg}.
Since the single-particle algorithm contains the essential ingredients 
of the DMRG algorithm for interacting systems, we believe that
understanding the single-particle algorithm is useful in gaining
insight into the algorithm for interacting systems. 
We therefore encourage the reader to work through the example
implementation of this algorithm \cite{reinhard_dmrg_book} and to
examine a similar program 
which is part of the ALPS application library~\cite{alps}.

\subsection{Density Matrix Projection for Interacting Systems}

\label{sec:den_mat_proj}
In the following sections, we will discuss how to generalize the
superblock method to interacting systems. 
This will lead to the density matrix renormalization group method
(DMRG), one of the most efficient numerically exact methods for
one-dimensional strongly correlated quantum systems. 
The reader is encouraged to supplement the discussion given here with
White's original papers \cite{dmrg_prl,dmrg_prb} and other
introductory and review articles on the method
\cite{dmrgbook,hallberg,uli_rmp}. 
In addition, a collection of papers using the DMRG or treating
subjects connected to the DMRG can be found in
Ref.~\cite{nishino_webpage}, a survey is 
presented in Ref.~\cite{eric_webpage}.

As we have seen in Sec.\ \ref{sec:better_methods}, the crucial step of
the superblock method is a projection of the wave function of the
superblock onto a subsystem, consisting of two identical blocks for
the algorithm for the noninteracting particle.
Here we consider how to carry out such a projection for a many-body
wave function in an optimal way.
In order to do this, we briefly review the general quantum mechanical
description of a system divided into two parts. 

The density matrix is the most general description of a quantum
mechanical system because, in contrast to a description in terms of
the wave function, it can 
be used to describe a system in a mixed state as well as in a pure
state \cite{landau,feynman}. 
For a general, mixed state, the density matrix of a system is given by
\begin{equation}
\rho = \sum\limits_{\alpha} \, C_{\alpha} \, | \Psi_{\alpha}
\rangle \langle \Psi_{\alpha} | ,
\label{eqn:densmat}
\end{equation}
where the coefficients $C_{\alpha}$ are the weights of the states in
the mixed ensemble and are normalized so that $\mbox{Tr} \; \rho = 1$.
The density matrix for a system in a pure state would have just one
term in the sum.
Given $\rho$, a subsystem $A$ can be described by {\it
tracing out} the degrees of freedom $| j \rangle$ of the rest of the
system, yielding the {\it reduced density matrix}
\begin{equation}
\rho_A = \mbox{Tr}_{| j \rangle} \rho \; .
\end{equation}
The eigenstates of $\rho_A$ form a complete basis for the subsystem
$A$; its eigenvalues $w_{\alpha}$ give the weight of 
state $\alpha$ in the ensemble and carry the information about the
entanglement of the subsystem with the rest of the system.
The amount of entanglement can be quantified by the mutual quantum
information entropy 
\begin{eqnarray}
S(\rho) &=& - \mbox{Tr}_{|j\rangle} \left ( \rho\log \rho \right )
        \nonumber \\
        &=& - \sum_\alpha w_\alpha \log w_\alpha \; .
\label{eqn:von_neumann}
\end{eqnarray} 
This feature gives an interrelation between the DMRG and quantum 
information theory as discussed in more detail in
Sec.~\ref{sec:dmrg_qit}. 
If a system in a pure state $|\psi\rangle$ is divided into two
parts (A and B), $| \psi\rangle$ can be expressed in terms of the
eigenstates 
of the reduced density matrices of part A, 
$|\phi_\alpha\rangle$,  and part B, $| \chi_\alpha \rangle$
using the {\it Schmidt decomposition}
\cite{schmidt_decomposition,preskill_lecturenotes} 
\begin{equation}
|\psi \rangle = \sum_\alpha \sqrt{w_\alpha} \; | \phi_\alpha \rangle
 \; | \chi_\alpha \rangle \; .
 \label{eqn:schmidt_decomp}
\end{equation}
Here the sum is over the nonzero eigenvalues $w_\alpha$ of the density
matrices of either part, which can be shown to be the same.
The behavior of the eigenvalues $w_\alpha$ with $\alpha$ depends on
the wave function considered and on how the two subsystems are chosen.
If only a small number (i.e., substantially less than the dimension of
the smallest of the two density matrices) of the $w_\alpha$ are
nonzero, $|\psi \rangle$ can be represented exactly by the sum over
the corresponding states.
If the $w_\alpha$ fall off sufficiently
rapidly with $\alpha$, $|\psi\rangle$ can be well approximated
by truncating the sum in Eq.\ (\ref{eqn:schmidt_decomp}) to the $m$ 
eigenstates of the density matrix with the largest eigenvalues.
This is the case for, e.g., one-dimensional quantum many-body
systems with a gapped spectrum, i.e., away from a quantum critical
point, for which $w_{\alpha}$ falls off exponentially
(see the discussion of this topic in Ref.~\cite{uli_rmp}).  
A useful approximation can be achieved if $m$ can be taken to be much
smaller than the dimension of the eigenbasis of the density matrix.
This approximation can be shown to be optimal in the sense of
a least-squares minimization of the differences between 
the exact $|\psi\rangle$ and the approximate one \cite{dmrg_prb} and 
is equivalent to the singular value decomposition
\cite{numerical_recipes}. 
Finally, if the $w_\alpha$ fall off too slowly or not at all, a
truncation of the Schmidt decomposition becomes a bad approximation
for $|\psi\rangle$.
The worst case is when all the $w_\alpha$ are equal, which occurs for
a maximally entangled state.
A representation optimized for multiple states rather than just
$|\psi\rangle$ can be constructed by including additional states
in the density matrix, Eq.\ (\ref{eqn:densmat}) with appropriate
weights $C_\alpha$.
The Schmidt decomposition, Eq.\ (\ref{eqn:schmidt_decomp}) is then no
longer applicable and the approximation to a particular state for a
given $m$ will become less accurate.

A measure for the error of this approximation is the sum over the
weights of the discarded density-matrix eigenstates,
\begin{equation}
\varepsilon_{\rm discard} = \sum_m^N w_\alpha,
\end{equation}
with $N$ the dimension of the system's Hilbert space.
This is called the {\it discarded weight}.
Thus, using the basis of density matrix eigenstates, an ``optimal''
description for a quantum many-body system can be found (for details,
see Ref.~\cite{reinhard_dmrg_book}).
In the usual case, the calculation is performed in order to obtain the
ground state of the system in a particular symmetry sector of the
Hilbert space, thus only one target state is kept and the density
matrix, Eq.~(\ref{eqn:densmat}), has just one term in the sum. 
Here and in the following, we will therefore discuss the procedure for
a calculation with a single target state; the generalization to
multiple target states is straightforward.

The basic procedure to carry out the truncation is then: 
\begin{enumerate}
\item Obtain the ground state $| \psi\rangle$ of a finite lattice
system using an iterative diagonalization procedure (e.g., the Lanczos or
Davidson algorithms). 
\item Divide the system in two and obtain the $m$ most important
eigenstates of the reduced density matrix of one of the subsystems. 
\item Transform the system block into the new (approximate) basis with
only $m$ states. 
\end{enumerate}
The DMRG is a numerically implemented variational method based on this
truncation.
In the following, the implementation of this method is discussed. 
For pedagogical example implementations for the noninteracting case,
we refer  the reader to Refs.~\cite{alps,reinhard_dmrg_book}. 


\subsection{DMRG Algorithms}

The goal in formulating DMRG algorithms is to embed an NRG-like
iterative buildup and truncation of a system, termed ``system block'',
in a larger system, the superblock.
As described in the previous section, an iterative diagonalization is
carried out on the superblock to obtain the ground state and possibly
some other states, and the eigenstates of the corresponding reduced
density matrix with the largest weights are used to form a new
truncated basis for the system block.
In order to construct DMRG algorithms, two elements of the procedure
must still be formulated: how the system block is built up, i.e., how 
degrees of freedom are added, and how the remainder of the superblock,
termed ``environment'' or ``environment block'', is chosen.
In the NRG a single site at a time is added to the system, allowing a
substantial fraction of the energy eigenstates ($1/N_\ell$, where
$N_\ell$ is the number of states on site $\ell$) to be kept at each step.
In the DMRG, it is also important to minimize the number of degrees of
freedom, especially since the system to be diagonalized, the
superblock, contains many additional degrees of freedom in the environment
block. 
Therefore, a single site is added to the system block at each step in
almost all variants of the DMRG algorithm.
There is more freedom to choose how the environment block is
constructed: the algorithms generally fall into two classes, depending
on how the superblock evolves with iteration.
In the infinite system procedure, the size of the superblock increases
at each step in a fashion reminiscent of the NRG, and in the finite
system procedure, the environment block is chosen so that the
size of the superblock remains constant, allowing an iterative
improvement of the wave function or wave functions for one particular
finite system.

\subsubsection{Infinite System Algorithm}

In the infinite system algorithm, the environment block is chosen to be
a reflection of the system block, (usually) including the added site.
Therefore, the superblock grows by two sites at each iteration (as
opposed to one site in the NRG) and the algorithm can be used to scale
towards the infinite-system fixed point, or, as we will
see, can be used to build up an initial approximation to a
system of a particular size, which can then be further improved with
the finite system procedure.
Such additional finite-system iteration is necessary for most systems
because the infinite-system algorithm is not guaranteed to achieve
variational convergence with the number of states kept
\cite{reinhard_dmrg_book,boschi}.


\begin{figure}
\includegraphics[width=9.5cm]{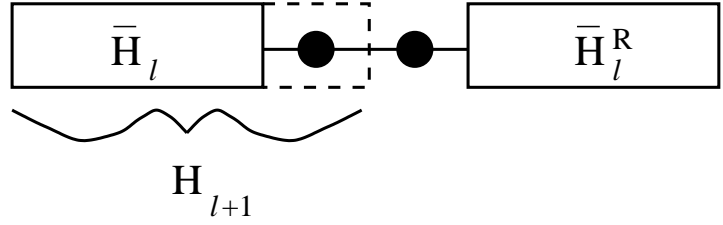}
\caption{Superblock configuration for the infinite system algorithm.}
\end{figure}

The infinite system algorithm for the calculation of the ground state
of a one-dimensional reflection symmetric lattice proceeds as follows.
\begin{enumerate}
  \item Form a superblock containing $L$ sites which is small enough
    to be exactly diagonalized.
   \item Diagonalize the superblock Hamiltonian $H^{\rm super}_L$
    numerically, obtaining {\em only} the ground state eigenvalue and
    eigenvector $\psi$ using the Lanczos or Davidson algorithm.
  \item Form the reduced density matrix $\rho_{i i'}$ for the current 
    system block from $\psi$ using 
    $\rho_{ii'} = \sum_j \psi^*_{ij} \psi_{i'j}$, where $|i\rangle$
    is the basis of the system block of size $\ell+1$, $|j\rangle$ the
    basis of the corresponding environment block, and $\psi_{ij}=
    \langle i | \,\langle j | \psi\rangle$.
    Note that $\ell'=\ell=L/2-1$.
  \item Diagonalize $\rho_{i i'}$ with a dense matrix
    diagonalization routine to obtain the $m$ eigenvectors with the
    largest eigenvalues.
  \item Construct the Hamiltonian matrix $\mathbf{H}_{\ell+1}$ of the
    new system block (i.e. the left block $A+s$) and other operators 
    needed in the course of the iteration (e.g. observables).
    Transform them to the reduced density matrix eigenbasis
    using $\bar{\mathbf{H}}_{\ell+1}^{\alpha} = \mathbf{O}^\dagger_L 
    \mathbf{H}_{\ell+1}^{\alpha} \mathbf{O}_L$,
    $\bar{\mathbf{A}}_{\ell+1} = \mathbf{O}^\dagger_L
    \mathbf{A}_{\ell+1} \mathbf{O}_L$, etc., where the columns of
    $\mathbf{O}_L$ contain the $m$  eigenvectors of $\rho_{i i'}$ with
    the highest eigenvalues, and $\mathbf{A}_{\ell+1}$ is an operator
    in the system block. 
  \item Form a superblock of size $L+2$ using $\bar{H}_{\ell+1}$, 
    two single sites and $\bar{H}^R_{\ell+1}$.
  \item Repeat starting with step 2, substituting 
    $H^{\rm super}_{L+2}$ for $H^{\rm super}_{L}$.
\end{enumerate}
The implementation details for the individual steps will be described in
Sec.~\ref{sec:programming_details}. 
In order to obtain an efficient program, bases and operators should
be decomposed according to the symmetries of the
Hamiltonian whenever possible.
For an introduction to the use of symmetries within the DMRG
framework, see Ref.~\cite{ED_introduction}.

This procedure has been formulated to obtain only the ground state,
but it is easy to extent it to multiple target states by constructing
additional states during step 2, either by continuing the
diagonalization to obtain excited states or by applying operators to
$|\psi\rangle$. 
These additional states are then mixed into the reduced density matrix
in step 3 with appropriate weights, as  discussed in 
Sec.\ \ref{sec:den_mat_proj}. 

Here a reflection-symmetric one dimensional system has been
assumed.
It is possible to generalize the infinite system algorithm to
non-reflection-symmetric lattices by building up the left half and the
right half of the system alternately.
However, such algorithms have not been
particularly important because the finite system algorithm, discussed
below, can treat such cases and 
can be generalized to systems that are not one-dimensional chains.

The results of the procedure are the energies and wave functions
obtained in step 2.
At this point, matrix elements of operators within a state and between
states can be calculated, provided that the operators have been formed
in the appropriate basis, i.e., the same basis in which $|\psi\rangle$
is represented in step 2.



\subsubsection{Finite System Algorithm}
\label{sec:finite_system_alg}

The infinite system method has the weakness that the wave function
targeted at each step is different because the lattice size is
different.
This can lead to poor convergence or complete lack of convergence with
$m$ if the wave function changes qualitatively between steps.
This can occur for states with some incommensuration with the lattice
such as, for example, fermions with a nonintegral filling or excited
states characterized by a particular wavevector.

Therefore, an algorithm in which the same finite system is treated at
each step is very useful.
Instead of convergence to an infinite-system fixed point with
iteration, there is variational convergence to the wave function or
set of wave functions for a particular finite system.
Such an algorithm can be formulated by choosing a block of appropriate
size from a previous step as the environment block.



\begin{figure}
\includegraphics[width=10cm]{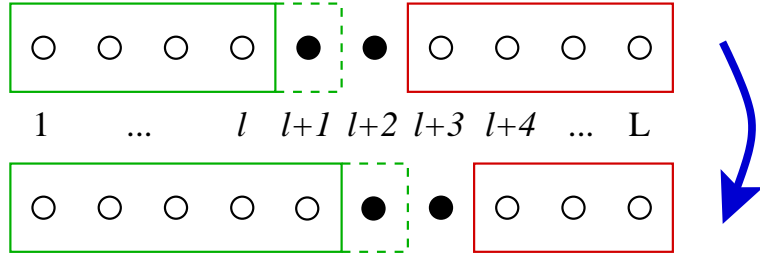}
\caption{A schematic depiction of a step in the finite-system algorithm.}
\label{fig:finite_system_alg}
\end{figure}

The finite system algorithm for finding the ground state on a
one-dimensional lattice proceeds as follows:
\begin{enumerate}
  \setcounter{enumi}{-1}
  \item Carry out the infinite system algorithm until the superblock
    reaches size $L$, storing $\bar{H}_\ell$
    and the operators needed to connect the blocks
    at each step.
  \item Carry out steps 3-5 of the infinite system algorithm to obtain 
    $\bar{H}_{\ell+1}$.  Store it. (Now $\ell\ne \ell'$.)
  \item Form a superblock of size $L$ using $\bar{H}_{\ell+1}$, 
    two single sites and $\bar{H}^R_{\ell'-1}$.
    The superblock configuration is shown in
    Fig.~\ref{fig:finite_system_alg} where $\ell' = L - \ell - 2$. 
  \item Repeat steps 1-2 until $\ell = L-3$ (i.e. $\ell' = 1$).
    This is the {\em left to right} phase of the algorithm.
  \item Carry out steps 3-5 of the infinite system algorithm, reversing
    the roles of $\bar{H}_{\ell}$ and
  $\bar{H}^R_{\ell'}$, i.e. switch 
    directions to build up the right block and obtain 
    $\bar{H}^R_{\ell'+1}$ using the stored $\bar{H}_{\ell}$ as the
    environment. 
    Store $\bar{H}^R_{\ell'+1}$. 
  \item Form a superblock of size $L$ using $\bar{H}_{\ell-1}$, 
    two single sites and $\bar{H}^R_{\ell'+1}$.
  \item Repeat steps 4-5 until $\ell=1$.
    This is the {\em right to left} phase of the algorithm.
  \item Repeat starting with step 1.
\end{enumerate}
This procedure is illustrated schematically in 
Fig.\ \ref{fig:finite_system_alg}.

One iteration of the outermost loop is usually called a finite-system
iteration or finite-system sweep.
If the lattice is reflection symmetric, the procedure can be shortened
by reversing direction at the reflection symmetric point, 
$\ell = L/2-1$, i.e., by using the reflection symmetry to interchange
the role of the left and right blocks at this point.
In this formulation, we have assumed that the infinite system
algorithm can be carried out to build up the lattice to the desired
size and to generate an initial set of environment blocks.
If this is not the case, the finite system method can still be applied
if a reasonable approximation is used for the environment
block in the first finite-system sweep.
The simplest such approximation is to use a null environment block,
which is equivalent to the Wilson NRG procedure;
a better one is to use a few exactly treated sites as the environment.
As long as this initial procedure does not lead to a system block
basis that is too bad, convergence is reached after a relatively small
number of finite system iterations for most systems, typically between
2 and 10 for one-dimensional systems.

Note that, as in the infinite system algorithm, it is also
possible to obtain results for several target states by mixing
additional states into the density matrix in step 3.

\begin{figure}
\includegraphics[width=0.65\textwidth]{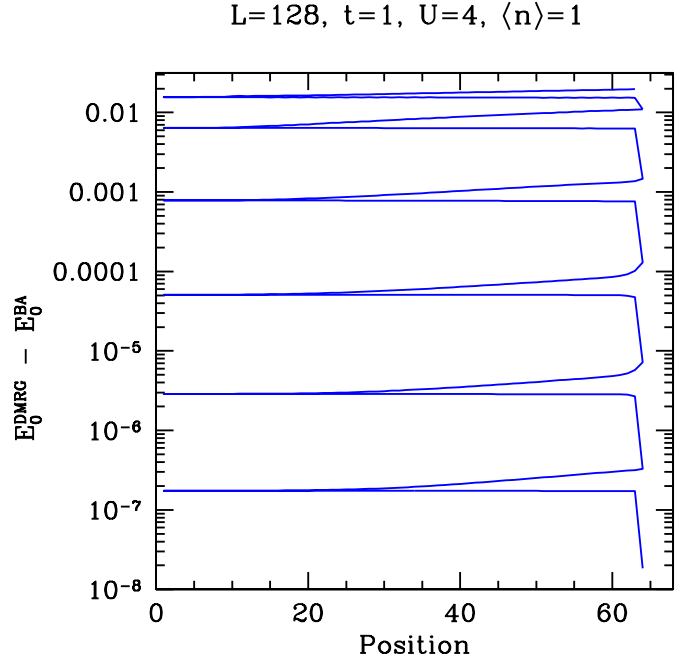}
\caption{Ground-state energy $E_0^{\rm DMRG}$ obtained by the DMRG
algorithm compared with the exact ground state energy from the Bethe
Ansatz $E_0^{\rm BA}$ for the one-dimensional Hubbard model of length
$L=128$ with $U/t = 4$ at half-filling.
``Position'' refers to the position at which a site is added.
The infinite system algorithm (not shown) followed by six sweeps of the
reflection-symmetric version of the finite system algorithm are
carried out.
\label{fig:DMRG_convergence}
} 
\end{figure}

We show an example of the behavior of the ground-state energy in the
course of a DMRG run for the half-filled one-dimensional Hubbard model
in Fig.\ \ref{fig:DMRG_convergence}.
Note the convergence is clearly variational and that there is a
significant downwards jump in the energy when the direction is changed
at the middle of the chain in the finite system algorithm.
This is due to a qualitative improvement in the representation of the
system as the reflected system block can also be used as the
environment block.

Since its introduction thirteen years ago, the DMRG has become a
standard method for obtaining the ground-state properties of
one-dimensional short-range quantum lattice models.
Among the first and most extensive areas of application have been 
spin chains with half-integral and integral spin
\cite{dmrg_prl,dmrg_prb,white-huse}, as well as frustrated spin chains 
\cite{white_affleck}. 
One-dimensional fermionic systems
\cite{noack_georgia,hubbard_chain_dmrg_book,manmana:155115} have been 
treated in many variations.
Since the amount of work done on such models is quite large, a 
comprehensive survey is beyond the scope of this pedagogical
introduction; we refer the reader to 
Refs.\ \cite{dmrgbook,hallberg,uli_rmp} for further references.

\section{The DMRG in Detail}

\subsection{Programming Details}
\label{sec:programming_details}
In this section, we discuss the details of implementing the DMRG
algorithm, paying particular regard to efficiency.
We follow Ref.~\cite{dmrgbook} in that we take the one-dimensional Heisenberg
model with  nearest-neighbor exchange term
\begin{equation}
{\bf S_\ell} \cdot {\bf S_{\ell+1}} = S^z_\ell S^z_{\ell+1} + \frac{1}{2} 
\left ( S^+_\ell S^-_{\ell+1}  + S^-_\ell S^+_{\ell+1} \right )
\end{equation}
as an example. 
Any DMRG program will contain the following three crucial elements:
\begin{enumerate}
\item The addition of two blocks (usually the system block and a site).
\item The multiplication $H^{\rm super} |\psi \rangle$. 
\item The basis transformation of relevant operators on a block to the
truncated basis of density matrix eigenstates.
\end{enumerate}

When two blocks are added, the part of the Hamiltonian internal to
each block as well as all terms connecting the two blocks must be
constructed.
For the example, the terms that must be constructed are
\begin{eqnarray}
\left [H_{12}\right]_{ii' ; jj'} =&& \left[ H_1\right ]_{i i'} \delta_{jj'} 
\; + \; \delta_{i i'} \left[ H_2\right ]_{jj'} 
\; + \;  \left[ S^z_\ell\right ]_{i i'} \left[ S^z_{\ell+1}\right ]_{j
j'} \nonumber \\ 
&& \; + \; \frac{1}{2}\left ( \left[ S^+_\ell\right ]_{i i'} \left[
S^-_{\ell+1}\right ]_{j j'} +  \left[ S^-_\ell\right ]_{i i'} \left[
S^+_{\ell+1}\right ]_{j j'} \right ) \; . 
\label{eqn:adding_two_blocks}
\end{eqnarray}
Note that the matrix representation of each operator 
appearing in Eq.~(\ref{eqn:adding_two_blocks}) 
must be available in the current basis; such operators must be
constructed and transformed in previous steps.

\begin{figure}
\label{fig::b1b2}
\includegraphics[width=0.75\textwidth]{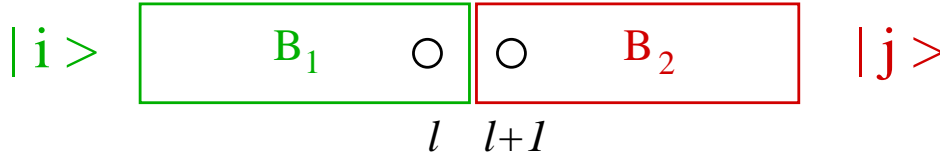}
\caption{A schematic depiction of the addition of two blocks.}
\end{figure}

The multiplication $H^{\rm super} | \psi\rangle$ is the most time-consuming
operation in the DMRG algorithm.
The following is an efficient procedure to perform this operation
within the framework of the DMRG; it should be considered to be the standard
method.
Assuming that the system is composed of two parts, the
superblock Hamiltonian can be constructed using 
\begin{equation}
\left[H^{\rm super}\right]_{ij;i^{\prime}j^{\prime}} =
\sum\limits_{\alpha} A_{ii^{\prime}}^{\alpha}
B_{jj^{\prime}}^{\alpha} 
\; , 
\end{equation}
where $\mathbf{A}^{\alpha}$ are the matrix representations of the
appropriate operators on the left block, while the matrices
$\mathbf{B}^{\alpha}$ are those on the right block.
The index $\alpha$ iterates over all pairs of operators that are
needed to construct the superblock Hamiltonian, as given in
Eq.~(\ref{eqn:adding_two_blocks}) for the example system. 
The product with the wave function in the appropriate superblock
configuration then is
\begin{equation}
\sum\limits_{i^{\prime}j^{\prime}} \left[H^{\rm
super}\right]_{ij;i^{\prime}j^{\prime}} \psi_{i^{\prime} j^{\prime}} =
\sum\limits_{\alpha} \sum\limits_{i^{\prime}} A_{ii^{\prime}}^{\alpha}
\sum\limits_{j^{\prime}} B_{jj^{\prime}}^{\alpha} \psi_{i^{\prime}
j^{\prime}}  \; .
\end{equation}
For each $\alpha$, this expression is equivalent to the multiplication
of three matrices
\begin{equation}
H^{\rm super} \psi = \sum\limits_{\alpha} \mathbf{A}^{\alpha}
\left(\mathbf{B}^{\alpha} \psi^T \right)^T \;, 
\label{eqn:dmrg_Hpsi}
\end{equation}
which can be carried out in order $m^3$ operations.

In the course of the DMRG procedure, all of the above matrices as well
as the operators required to calculate desired observables must be
transformed into the new truncated basis given by the $m$
density-matrix eigenvectors with largest weight.
If the transformation matrix $O_{ i j;\alpha}$ is composed of $m$ basis
vectors  $u^\alpha_{i j}$, the operator $A_{i j ; i' j'}$ is
transformed as 
\begin{equation}
\label{eqn:operator_transformation}
A_{\alpha\alpha'} = \sum_{i,j,i',j'} O_{i j;\alpha}
A_{i j;i' j'} O_{i' j';\alpha'} \; ,
\end{equation}
leading to a reduction of the dimension of the matrix representing
$A$ from $(m_1 m_2) \times (m_1 m_2)$ to $m \times m$. 

In order to make the above procedures as efficient as possible,
it is necessary to exploit the system's Abelian symmetries.
The sums over sets of states can be divided up into sectors
corresponding to different quantum numbers.
Once this is done, only the nonzero parts of the matrix
representations of the operators which connect particular quantum
numbers must be stored and the multiplications can be decomposed into
sums over multiplications of these non-vanishing pieces.
Use of non-Abelian symmetries is also possible, but is substantially
more difficult 
\cite{mcculloch-gulasci2000,mcculloch-gulasci2001,mcculloch_et_al_2001}.
In order to treat large systems and make $m$ as large as possible, 
it is also important to minimize the use of main memory.
This can be done by storing the matrix representations of operators
not currently needed on secondary storage.
In particular, relevant operators for blocks that are not 
needed in the superblock configuration at a given step of the finite
system procedure, but which will be
needed in a subsequent step, need not be retained in primary storage.


\subsection{Measurements}

What are termed ``measurements'' in the DMRG framework are expectation
values of operators calculated within a state or between states of the
superblock, which are obtained in the iterative diagonalization step.
The procedure is straightforward provided that the necessary operators 
are available in the appropriate basis.
Given a state $\psi_{i j}$ of the two-block system, the single-site
expectation value $\langle \psi | S^z_\ell | \psi \rangle$ is given by 
\begin{equation}
  \langle \psi | S^z_\ell| \psi \rangle = \sum_{i, i', j} 
  \psi^*_{i j} \, [S^z_\ell]_{i i'} \, \psi_{i' j} \; .
\end{equation}
The matrix representation $[S^z_\ell]_{i i'}$ is constructed when the
site $\ell$ is added to the system block, and must be transformed at
each subsequent step so that it is available in the basis
$|i\rangle$. 


For expectation values of operators on two different sites such as 
the correlation function
$\langle \psi | S_\ell^z S_m^z | \psi \rangle$, how the operators are
constructed depends on whether the two sites $\ell$
and $m$ are on the same or on different blocks.
If they are located on different blocks, then the expectation value
can be formed using 
\begin{equation}
    \langle \psi | S^z_\ell S^z_m| \psi \rangle = \sum_{i, i', j, j'} 
    \psi^*_{i j} \, [S^z_\ell]_{i i'} [ S^z_m]_{j j'}\, \psi_{i' j'}
    \; ,
\end{equation}
where $[S^z_\ell]_{i i'}$ and $[S^z_m]_{j j'}$ are the individual
single-site operators.
However, if $\ell$ and $m$ are on the {\sl same} block, the expression 
\begin{equation}
    \langle \psi | S^z_\ell S^z_m| \psi \rangle \approx \sum_{i, i',i'', j} 
    \psi^*_{i j} \, [S^z_\ell]_{i i'} [ S^z_m]_{i' i''}\, \psi_{i'' j} 
\end{equation}
is {\sl incorrect} within the approximatation for $\psi_{ij}$ because
the sum over $i'$ extends over the current
truncated basis rather than a complete set of states.
The correct way to calculate the two-site expectation value is to use the
relation 
\begin{equation}
    \langle \psi | S^z_\ell S^z_m| \psi \rangle = \sum_{i, i', j} 
    \psi^*_{i j} \, [S^z_\ell S^z_m]_{i i'} \, \psi_{i' j},
\end{equation}
where the operator $[S^z_\ell S^z_m]_{i i'}$ has been calculated at
the appropriate step; this is correct within the variational
approximation for $\psi_{ij}$.
The general rule is, that compound operators internal to a block must
be accumulated as the calculation proceeds. 
In this way, almost all correlation
functions as well as more complicated equal-time expectation values
can be calculated. 
The calculation of dynamical correlation functions and time-dependent
quantities have been made possible by recent developments and will be
discussed in Sec.\ \ref{sec:DDMRG} and 
Sec.\ \ref{sec:DMRG_time_evolution}, respectively. 

\subsection{Wave Function Transformations}
\label{sec:wavefunctrans}

The most time-consuming part of the DMRG algorithm is the
iterative diagonalization of the superblock Hamiltonian. 
Here we discuss how to optimize this procedure significantly by
reducing the number of steps in the iterative diagonalization, in some
cases by up to an order of magnitude.

As discussed in Sec.~\ref{sec:iterative_diag}, the key operation and
most time-consuming part of any iterative diagonalization procedure is
the multiplication of the Hamiltonian and an arbitrary wave function, 
i.e., the operation 
$H^{\rm super} \psi$ in the DMRG.
Typically, of the order of 40-100 such multiplications are required to
reach convergence.
Reducing this number would thus directly lead to a proportional
speedup of the diagonalization. 
If the Davidson or Lanczos procedure is started with a wave vector
that is a good approximation to the desired wave vector,
much fewer iterations will be required.
Since an approximation to the same system is treated at each step of
the finite system algorithm, an obvious starting point is  the result
$|\psi^\ell_0\rangle$ of a previous finite system step.
However, this wave function is not in an appropriate basis for 
$H^{\rm super}$ because it was obtained using a different superblock
configuration. 
In order to be able to perform the multiplication 
$H^{\rm super}_{\ell+1}\, |\psi_0^{\ell}\rangle$, the wave function must
be transformed from the basis at step $\ell$ to a basis suitable to
describe the system configuration at step $\ell+1$.
\begin{figure}
\includegraphics[width=9.5cm]{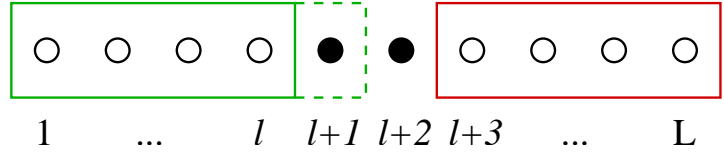}
\caption{Superblock configuration relevant for the wave function
transformation.}
\label{fig:WF_transform}
\end{figure}

At step $\ell$, a state in the superblock basis is given by 
\begin{equation}
|\alpha_{\ell} \, s_{\ell+1} \, s_{\ell+2}\, \beta_{\ell+3}\rangle =
|\alpha_{\ell}\rangle \otimes |s_{\ell+1}\rangle \otimes 
|s_{\ell+2}\rangle \otimes |\beta_{\ell+3}\rangle \; , 
\end{equation}
where $|\alpha_{\ell}\rangle$ is the basis of the left block
containing sites $1, \ldots , \ell$,
$|s_{\ell+1}\rangle$ and $|s_{\ell+2}\rangle$ are the bases of the single
sites, $\ell+1$ and $\ell+2$, and $|\beta_{\ell+3}\rangle$ is the basis
of the right block in the 4-block description of the superblock, as
depicted in Fig.\ \ref{fig:WF_transform}.
Assuming the algorithm is building up the system block from left to
right, these states must be transformed to the configuration of the
superblock at step $\ell+1$ with basis
\begin{equation}
|\alpha_{\ell+1} \, s_{\ell+2} \, s_{\ell+3} \, \beta_{\ell+4}\rangle \; .
\end{equation}
This transformation is performed in two steps.
The left block is transformed 
from the original
product basis $\{| \alpha_{\ell} \rangle \otimes | s_{\ell+1} \rangle \}$
to the effective density matrix basis obtained at the end of the
previous DMRG step, $\{ | \alpha_{\ell + 1} \rangle \}$ using
\begin{equation}
|\alpha_{\ell+1}\rangle = \sum_{s_{\ell+1},\alpha_{\ell}} 
L^{\ell+1}[s_{\ell+1}]_{\alpha_{\ell+1},\alpha_{\ell}} 
|\alpha_{\ell}\rangle \otimes |s_{\ell+1}\rangle \; ,
\end{equation}
where the transformation matrix
$L^{l+1}[s_{l+1}]_{\alpha_{l+1},\alpha_{\ell}}$ contains the
density matrix eigenvectors 
$u^{\alpha_{\ell+1}}_{s_{\ell+1} \, \alpha_\ell}$ and is a simple
rearrangement of the matrix elements of the transformation matrix
$O_{s_{\ell+1} \, \alpha_\ell \, ; \, \alpha_{\ell+1}}$ used in
Eq.~(\ref{eqn:operator_transformation}) for the transformation of
operators to the new (truncated) density matrix basis. 
Similarly, for the right basis one defines
\begin{equation}
|\beta_{\ell+3}\rangle = \sum_{s_{\ell+3},\beta_{l+4}} 
R^{\ell+3}[s_{\ell+3}]_{\beta_{\ell+3},\beta_{\ell+4}}
|s_{\ell+3}\rangle \otimes |\beta_{\ell+4}\rangle \; .
\end{equation}
Transformation matrices similar to 
$L^{l+1}[s_{l+1}]_{\alpha_{l+1},\alpha_{\ell}}$ and
$R^{\ell+3}[s_{\ell+3}]_{\beta_{\ell+3},\beta_{\ell+4}}$ were
introduced by \"Ostlund and Rommer in Ref.~\cite{rommer}, where they
show that the resultant wave function 
is a {\it matrix product state}.
Such states will be discussed in more detail in Sec.~\ref{sec:MPS}.

To perform the wave function transformation needed for
the left-to-right part of the DMRG, we expand the superblock wave
function at step $\ell$ as 
\begin{equation}
|\psi\rangle = \sum_{\alpha_{\ell},s_{\ell+1},s_{\ell+2},\beta_{\ell+3}}
\psi(\alpha_{\ell},s_{\ell+1},s_{\ell+2},\beta_{\ell+3}) \;
|\alpha_{\ell} \, s_{\ell+1} \, s_{\ell+2} \, \beta_{\ell+3}\rangle 
\; .
\end{equation}
The basis transformation is formally performed by inserting
$\sum_{\alpha_{\ell+1}} |\alpha_{\ell+1} \rangle \langle
\alpha_{\ell+1} |$. 
Since there is a truncation, this is only an approximation,
\begin{equation}
\sum_{\alpha_{\ell+1}} |\alpha_{\ell+1} \rangle \langle
\alpha_{\ell+1} | \approx 1 \; . 
\end{equation}
The coefficients of the wave function in the new basis therefore become
\begin{eqnarray*}
\lefteqn{\psi(\alpha_{\ell+1},s_{\ell+2},s_{\ell+3},\beta_{\ell+4}) \approx }\\
& & \sum_{\alpha_{\ell},s_{\ell+1},\beta_{\ell+3}} 
L^{\ell+1}[s_{\ell+1}]_{\alpha_{\ell+1},\alpha_{\ell}}
\; \psi(\alpha_{\ell},s_{\ell+1},s_{\ell+2},\beta_{\ell+3}) \;
R^{\ell+3}[s_{\ell+3}]_{\beta_{\ell+3},\beta_{\ell+4}} \; .
\end{eqnarray*}
It is convenient to perform the procedure in two steps:
\begin{itemize}
\item[1.] Form the intermediate result
\begin{equation}
\psi(\alpha_{\ell+1},s_{\ell+2},\beta_{\ell+3}) =
\sum_{\alpha_{\ell},s_{\ell+1}} 
L^{\ell+1}[s_{\ell+1}]_{\alpha_{\ell+1},\alpha_{\ell}}
\psi(\alpha_{\ell},s_{\ell+1},s_{\ell+2},\beta_{\ell+3}) \; .
\end{equation}
\item[2.] Then form
\begin{equation}
\psi(\alpha_{\ell+1},s_{\ell+2},s_{\ell+3},\beta_{\ell+4}) =
\sum_{\beta_{\ell+3}} 
\psi(\alpha_{\ell+1},s_{\ell+2},\beta_{\ell+3})
R^{\ell+3}[s_{\ell+3}]_{\beta_{\ell+3},\beta_{\ell+4}} \; .
\end{equation}
\end{itemize}
An analogous transformation is used for a step in the right to left
sweep.

The wave function transformations are relatively inexpensive in CPU
time and in memory compared to other steps of the DMRG procedure.
In addition to leading to faster convergence in the iterative
diagonalization due to the good starting vector, wave function
transformations also allow the convergence criterium of the Davidson- or
Lanczos-algorithm to be relaxed so that the variational error in the
iterative diaogonalization is comparable to the variational error of
the DMRG truncation, saving additional iterations.
Normally such a relaxation would lead to the strong possibility of
convergence to a qualitatively incorrect state.
However, since the initial state comes from a diagonalization of the
same finite lattice, there is little danger of this occuring.

Implementing this transformation requires saving the
transformation matrices $\mathbf{L}[s_\ell]$ and $\mathbf{R}[s_\ell]$ 
at every finite-system step, which was not necessary in the original
formulation of the algorithm.
In an efficient implementation of the DMRG algorithm, the
transformation matrices (as well as the matrices needed to describe a
block) are stored on hard disk.
This additionally makes it possible to reconstruct all operators 
{\sl after} the final ground state wave function $|\psi_0\rangle$ has been
obtained, which saves memory during the DMRG run when many
measurements are made.

\subsection{Extensions to Higher Dimension}

\subsubsection{Real-Space Algorithm in Two Dimensions}
\label{sec:2D_ext}
\begin{figure}
\includegraphics[width=0.6\textwidth]{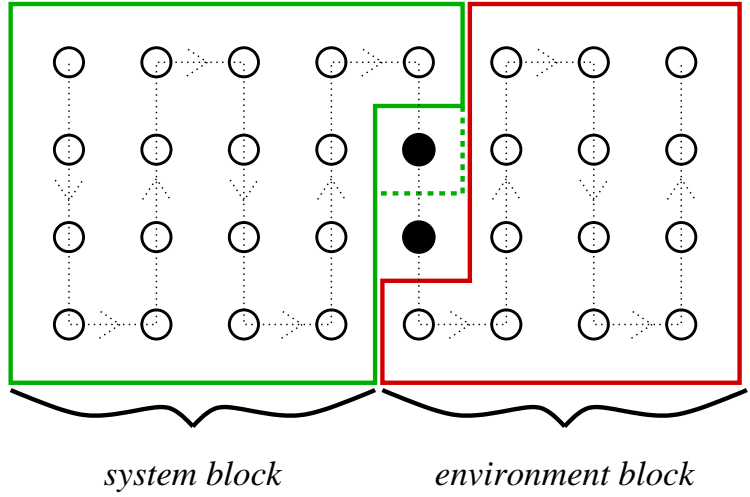}
\caption{Superblock scheme for a two-dimensional lattice. }
\label{fig:2D_superblock}
\end{figure}

The extension of the DMRG algorithm to higher dimensional systems,
i.e., to two or three-dimensional quantum systems or to
three-dimensional classical systems is a difficult problem.
The most straightforward, and, until now, most extensively used,  way
of extending the DMRG algorithm to, for
example, systems of coupled chains, is to simply fold the
one-dimensional algorithm into the two-dimensional lattice, as
depicted in Fig.\ \ref{fig:2D_superblock}
\cite{noack_georgia,liang-pang}. 
In effect, one is treating a one-dimensional lattice with longer-range
interactions generated by the coupling between rows.
Note that one site is added to the system block at each step as in the
one-dimensional algorithm.
One could consider instead adding portion of the lattice that is more
appropriate for preserving the two-dimensional symmetry such as a row
of sites \cite{henelius}; however, this leads to a prohibitive
explosion in the number of states in the superblock basis unless
additional approximations are made.
The infinite-system algorithm outlined above cannot be
straightforwardly extended to the two-dimensional lattice because
there is no reflection symmetry between the system and environment
blocks at a particular point.
However, once the system has been built up so that a set of system
blocks has been generated, the finite-system algorithm can be used
with no problems, even taking into account reflection 
symmetry, if it is present.
As long as this set of system blocks is a reasonable approximation, the
convergence in the number of finite-system sweeps is rapid for most
systems.

In order to carry out the buildup and form the first generation of
system blocks, different schemes can be used.
The simplest is to not include the environment at all, which amounts
to carrying out the Wilson procedure.
However, this algorithm has serious drawbacks for quantum lattice
problems, as discussed above.
In addition, for systems of fermionic or bosonic particles at
fillings with a non-integral number of particles per site, a chemical
potential must be introduced and carefully adjusted so that the
average particle density is appropriate.
An improvement on this scheme is to take the environment block
to be a small number of exactly treated sites, typically 3 to 8, depending
on the system.
This scheme avoids the problems in the Wilson procedure, is flexible
and easy to implement, and generates sufficiently good system blocks
in most cases.
Another possibility is to use a hybrid scheme in which the
finite-system algorithm is run on a lattice smaller than the target
size, and the system size is increased a row at a time when
an appropriately sized block are available \cite{noack_georgia}.
(This scheme seems to have been rediscovered in Ref.\ \cite{farnell}.)
Additional improvements to two-dimensional algorithms include adding
symmetry-adapted bands rather than sites \cite{deJongh} and a scheme
in which a the choice of particular diagonal path through a
two-dimensional lattice makes it possible to formulate both
infinite-system and finite-system 
algorithms that add only one site at a time and break up the lattice
more symmetrically \cite{xiang_2D}.

However, all of these schemes suffer from the fundamental limitation
that the boundary between the system and environment blocks is
relatively long (proportional to the linear dimension in two
dimensions) with many interaction terms along the boundary.
In practice, this leads to a large entanglement between the system and
environment blocks and therefore a slower fall-off of the eigenvalues
of the density matrix, requiring many more states to be kept to attain
a given accuracy.
This scaling of the required number of states has been studied
systematically for the case of two-dimensional noninteracting spinless
fermions, where it is found that the number scales {\it exponentially} with
the linear dimension of the lattice \cite{liang-pang,chung-peschel2001}.
It is not completely clear that this exponential scaling is
fundamental for all systems; in fact, noninteracting particles are
probably the worst case because of the absence of length scales in the
wave functions and because of the highly degenerate gapless excitation
spectrum.
For two-dimensional gapped systems, Chung and Peschel
\cite{chung-peschel2000} have shown that the scale of exponential decay of the
density-matrix eigenvalues for a system of interacting harmonic
oscillators diverges linearly with system width; an argument exists
that this behavior is generic \cite{deJongh}.  
In practice, multichain systems up to a particular width that depends
on the system studied can be treated with sufficient accuracy to
obtain well-controlled results.

The DMRG has been used to study a number of multi-chain
and two-dimensional systems, including  the multichain Heisenberg
model \cite{white_heisenberg_chains}, the frustrated Heisenberg model
in one dimension \cite{white_affleck} and in two dimensions
\cite{deJongh}, the two-leg Hubbard ladder
\cite{noack_hubbard_ladder}, CaV$_4$O$_9$ \cite{white_CaVO}, and the
multichain and two-dimensional $t$--$J$ model \cite{white_tJ_1998a,
white_tJ_1998b,white_tJ_2000,white_tJ_2003}.

Recently, a variational scheme related to the DMRG and based on a
tensor-based generalization of matrix product states has been argued
to provide a better \cite{verstraete-cirac_2D} representation of
two-dimensional wave functions.
While the prelimary calculations look promising, 
the usefulness of this scheme in practice relative to the existent
DMRG algorithms has not yet been explored.

\subsubsection{Momentum-Space DMRG}

Another approach to treating higher dimensional systems has been to
work with a momentum-space formulation of a local Hamiltonian such as
the Hubbard model.
The Hamiltonian of the Hubbard model in momentum space reads
\begin{equation}
H = \sum_{k \sigma} \varepsilon_k \; c^\dagger_{k \sigma} 
c^{\phantom \dagger}_{k \sigma}
+ \frac{U}{N} \sum_{p k q} 
c^\dagger_{p-q \uparrow} c^\dagger_{k+q \downarrow} 
c^{\phantom \dagger}_{k \downarrow} c^{\phantom \dagger}_{p \uparrow} \;,
\label{eqn:hubbard_kspace}
\end{equation}
where 
$c^\dagger_{k \sigma} = 
1/\sqrt{N}\; \sum_{\ell} e^{ik\ell} c^\dagger_{\ell \sigma}$ is the
Fourier-transformed electron creation operator and $N$ is the number
of lattice sites.
DMRG methods were first applied to this Hamiltonian by White in 1993
in unpublished work \cite{white_kspace} and subsequently
independently implemented by Xiang with some additional algorithmic
improvements \cite{xiang_kspace}.

The basic idea is to carry out the standard finite-system DMRG
algorithm (see Sec.\ \ref{sec:finite_system_alg}) using the
single-particle momenta as lattice points.
The motivation for treating the momentum-space model is that all
information about the dimensionality and lattice structure in 
Eq.\ (\ref{eqn:hubbard_kspace}) is contained in the first term which is
diagonal. 
Therefore, the hope is to avoid the drastic loss in accuracy with
dimension present in the real-space DMRG.
The momentum-space basis is also a natural starting point for treating
low-energy excitations relative to the filled Fermi sea.
In addition, unlike in the real-space algorithm, it is easy to
preserve and take advantage of the translational symmetry of the
system, which translates to the conservation of total momentum in
Hamiltonian (\ref{eqn:hubbard_kspace}).
This can be used both to reduce the dimension of the Hilbert space in
the diagonalization step and to calculate momentum-dependent
observables.
The disadvantage of the momentum-space formulation lies in the fact
that the interaction
term, the second term in Eq.\ (\ref{eqn:hubbard_kspace}), is highly
non-local, which leads to both technical and fundamental problems.

There are three significant technical problems: the first is that,
unlike for a short-range interactions like those in the Hubbard model
in real space, there are formally a large number of terms in the
interaction; in this case $N^3$.
The impact of this problem can be reduced by factorizing the
interacting into sums of terms that are internal to one block.
As shown in Ref.\ \cite{xiang_kspace}, this reduces the number of
terms linking  the parts of a two-part system to $6N$.
Second, as in the two-dimensional
algorithm described above, it is not possible to carry out the
infinite-system algorithm to build up the lattice due to the lack of
reflection symmetry at each step.
However, a similar problem is present in the simplest variant of the
real-space DMRG for two dimensions, as discussed in 
Sec.\ \ref{sec:2D_ext}; some of the same techniques can be used here, such as
taking a null environment block (equivalent to the Wilson procedure)
or taking a small number of exactly treated sites as the environment.
Third, it is not immediately clear how to order the momentum sites
optimally in the DMRG algorithm because the interaction term links all
sites with the same amplitude.
Orderings in which momenta that are a similar distance from the Fermi
surface are grouped together seem to be favorable
\cite{nishimoto_kspace,legeza_DSS}, but the issue is still being
explored \cite{legeza_solyom_QIE}.
This problem also occurs in the DMRG treatment of other nonlocal
Hamiltonians such as those obtained in quantum chemistry; see 
Sec.\ \ref{sec:quantum_chem}.

The crucial issue in the usability of the momentum-space DMRG is the
convergence of the method, especially as compared to the real-space
DMRG applied to similar systems.
In particular, it is clear that the momentum-space DMRG becomes exact
in the limit of weak interaction $U/t$ because Hamiltonian
(\ref{eqn:hubbard_kspace}) becomes diagonal \cite{xiang_kspace}.
In contrast, the real-space DMRG becomes exact for the Hubbard model
in the atomic limit $t\to 0$, but not directly in the strong-coupling
limit $U/t\to \infty$.
Therefore, for a given dimensionality, there is some interaction
strength at which the momentum-space DMRG becomes more accurate than
the real-space DMRG for the same lattice with the same boundary
conditions.
This occurs at approximately $U/t=8$ for a $4\times 4$ lattice
\cite{xiang_kspace,nishimoto_kspace}.
A comparison of the convergence for one- and two-dimensional lattices
shows that comparable accuracy is attained when the ratio of the
interaction to the bandwidth $U/W$ is approximately the 
same, with some additional small loss of accuracy with
dimension \cite{nishimoto_kspace}.

\section{Recent Developments in the DMRG}

\subsection{Classical Transfer Matrices}

In contrast to interacting quantum systems, the ground state of a
discrete classical model such as the Ising model can be trivially
obtained: it is the behavior at finite temperature, characterized for
instance by continuous phase transitions, that is
interesting.
The DMRG algorithm is powerful technique for obtaining
approximate but very accurate information about the extremal
eigenstates of an operator that operates on many-body Hilbert space
composed of the direct products of the Hilbert spaces of its component
systems.
Transfer matrices for a classical systems have such
properties and can be used to obtain thermodynamic
properties.
Therefore, a DMRG treatment of suitable transfer matrices, called the
Transfer Matrix Renormalization Group (TMRG) can be used
to calculate the thermodynamic properties of a classical system
\cite{nishino95}.
In addition, there is a well-known correspondence between a
$d$-dimensional quantum and a $(d+1)$-dimensional classical system
which makes it possible to calculate the thermodynamic properties of
some quantum system, as will be discussed in 
Sec.\ \ref{sec:quantum_transfer}.

\begin{figure}
\includegraphics[width=0.5\textwidth]{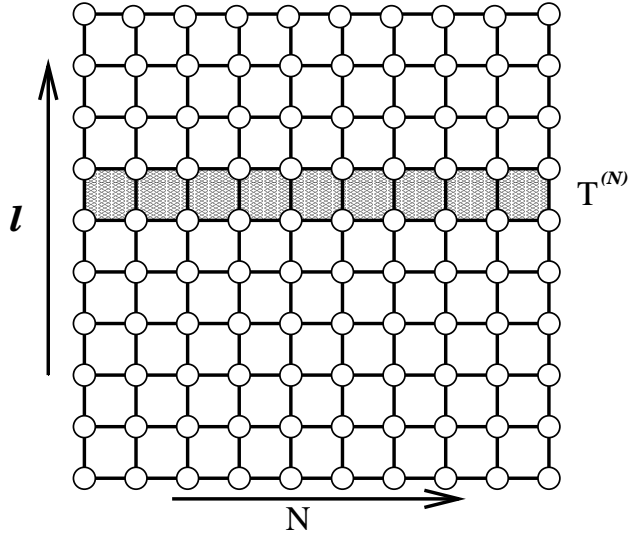}
\caption{Schematic depiction of the finite row transfer matrix $T^{(N)}$ for
the two-dimensional Ising model.
Open boundary condition are applied in the $N$ direction and periodic
boundary conditions in the $\ell$ direction, which is taken to become
infinite. } 
\label{fig:row_transfer}
\end{figure}

Consider the two-dimensional Ising model with cylinder geometry so
that $\ell$ ranges from 1 to $L$ with PBC and $n$ ranges from 1 to $N$
with OBC (see Fig.\ \ref{fig:row_transfer}).
The energy between nearest neighbors $i$ and  $j$ is  $-J s_i s_j$
where $s_i = \pm 1$ is an Ising spin variable.
The symmetrical row transfer matrix in the $\ell$ direction can then
be written as  
\begin{equation}
 T^{(N)}({\bf s}'|{\bf s}) = 
\exp\left( \frac{K}{2} s_1 s'_1 \right )
\left \{ 
\prod_{n=1}^{N-1} W(s'_i s'_{n+1} | s_i s_{n+1} )
\right \}
\exp\left( \frac{K}{2} s_N s'_N \right )
\end{equation}
where $\bf s = (s_1, \ldots, s_N)$ designates a vector of Ising spins, 
$K = J/k_B T$ is the dimensionless coupling, and 
\begin{equation}
W(s'_i s'_{n+1} | s_n s_{n+1} ) = \exp \left [
\frac{K}{2} (s'_n s_n + s_n s_{n+1} + s'_{n+1} s'_{n} + s_{n+1} s'_{n+1} ) 
\right ]
\end{equation}
is the Boltzmann weight connecting sites around a plaquette of four
nearest neighbors.
This matrix has dimension $2^N$ and is the object that will be treated
with DMRG methods. 

The thermodynamic properties can be derived from the partition
function $Z$, which can be written in terms of the row transfer matrix
as 
\begin{equation}
Z = {\rm Tr} \; \rho = {\rm Tr} \; (T^{(N)})^L = 
\sum_\alpha (\lambda_\alpha)^L \; 
\end{equation}
where the $\lambda_\alpha$ are the eigenvalues of $T^{(N)}$.
(We assume that $T^{(N)}$ is diagonalizable and has positive
semi-definite eigenvalues.)
The trick in calculating $Z$ with a method that can accurately
approximate the extremal eigenvalues of a finite sparse matrix
such as the DMRG or iterative diagonalization is the take the limit
$L\to \infty$, keeping $N$ finite, i.e., to make the lattice
infinite in one direction only.
The partition function can then be approximated as
$Z \simeq \lambda_1^L$, where $\lambda_1$ is the {\it largest}
eigenvalue of $T^{(N)}$.
The (unnormalized) density matrix
is given by $\rho = \lambda^L \; | \lambda_1 \rangle \langle \lambda_1|$ 
where $|\lambda_1 \rangle$ is the corresponding eigenvector.

In order to apply the DMRG method, the system, i.e., $T^{(N)}$, must
be divided into two parts.
If this is done at site $M$, $T^{(N)}$ can be written
\begin{equation}
T^{(N)}({\bf s'}|{\bf s}) = 
T_L({\bf s'_L}|{\bf s_L}) 
W(s_M' s_{M+1}'|s_M s_{M+1})
T_R({\bf s'_R}|{\bf s_R}) \; 
\end{equation}
where
\begin{equation}
T_L({\bf s'_L}|{\bf s_L}) = \exp \left( \frac{K}{2} s'_1 s_1 \right )
\prod_{n=1}^{M-1} W(s'_i s'_{n+1} | s_n s_{n+1} ) 
\end{equation}
is the transfer matrix for the left half-row 
${\bf s_L}=(s1,\ldots,s_M)$ and $T_R({\bf s'_R}|{\bf s_R})$ is
defined analogously for the right half-row.
The equivalent of the reduced density matrix used in the original DMRG
is the left density submatrix defined as
\begin{equation}
\rho_L({\bf s}'_L|{\bf s}_L) = \sum_{{\bf s}_R} 
\rho({\bf s}'_L \, {\bf s}_R|\, {\bf s}_L \,{\bf s}_R)
\end{equation}
(or the right density submatrix $\rho_R$, defined analogously).
More than one eigenvalue of $\rho_L$ is important in the $L\to\infty$
limit, unlike the eigenvalues of $\rho$, but the eigenvalues generally
decay rapidly, almost exponentially 
\cite{baxter_book,peschel_kaulke_legeza} when the
correlation length is finite.
The partition function 
\begin{equation}
\tilde{Z} = \sum_{i=1}^m \omega_i^2 \; , 
\end{equation}
where the $\omega_i^2$ are the eigenvalues of $\rho_L$,
for the partial system is bounded from above by the full partition
function $Z$, and the difference becomes small when $m$ becomes large,
or when the the eigenvalues $\omega_i^2$ fall off sufficiently
rapidly.

The DMRG procedure is analogous to that for the quantum
system, using iterative diagonalization to obtain the largest
eigenstate of $T^{(N)}$ rather than the ground
state of $H$ as for quantum systems.
At each step, the local transfer
matrix for a single plaquette $T^{(N)}$ is added to the finite lattice
instead of a quantum site.
As for the quantum system, the system can be built up to a given row
length using the infinite-system algorithm, and then 
finite-system sweeps can be carried out until convergence is reached.
One major difference in terms of efficiency compared with the
calculation for the quantum algorithm is that there are
no conserved quantum numbers, reducing the maximum number of states
$m$ that can be kept for a given amount of numerical effort.

While all thermodynamic quantities can, in principle, be calculated
by taking numerical derivatives of the free energy per site 
$f = -k_B T \ln Z / N \approx k_B T \ln \tilde{\lambda}_1$ (where 
$\tilde{\lambda}_1$ is the variational approximation to $\lambda_1$
obtained from the DMRG), second derivatives needed to obtain the
susceptibility or specific heat are numerically unstable and it is
better to use the first derivative of the energy per site, which can
be calculated using matrix elements such as
$\langle \lambda_1 | s_i s_j | \lambda \rangle$, which, of course,
also yield the short-range spin correlations.
The correlation length $\xi$  can be calculated from the short-range spin
correlations if they are small, or directly using 
$\xi = 1/ \ln {\rm Re} (\lambda_2/\lambda_1)$, where $\lambda_2$ is
the second-largest eigenvalue.

Applications of the TMRG for classical row transfer matrices include
the two-dimensional Ising model \cite{nishino95}, the spin-3/2 Ising
model \cite{tsushima}, wetting phenomena
\cite{carlon-drzewinski97,carlon-drzewinski98}, 
the $q$-state Potts model \cite{igloi-carlon}, the $q$-state Potts model
with random exchange \cite{carlon99}, 
the 19-vertex model \cite{honda-horiguchi}, and the three-state chiral
clock model \cite{sato-sasaki}.
For a more extensive recent survey, see Ref.\ \cite{uli_rmp}.

It is also possible to treat Baxter's corner transfer matrices
\cite{baxter68,baxter_book} using
the DMRG, leading to the so-called Corner Transfer Matrix
Renormalization Group (CTMRG) \cite{nishino-okunishi97}.
Since corner transfer matrices break up a two-dimensional square
lattice into quadrants, the partition function is given by the fourth
power of the corner transfer matrix
\begin{equation}
Z^{(2N-1)} = {\rm Tr} \, \rho_c 
\approx {\rm Tr} \left ( C^{(N)} \right )^4
= \sum_{\nu = 1}^m \alpha_\nu^4 \; , 
\end{equation}
and converges variationally to the exact partition function in the
thermodynamic limit and as the number of states kept $m$ becomes
sufficiently large.
Here the corner transfer matrix is defined as
\begin{equation}
C^{(N)} = \sum_{\{s\}} \prod_{\langle ijkl \rangle} 
W(s_i \, s_j \, | \, s_k \, s_l) 
\end{equation}
and $\alpha_v$ are its (largest) eigenvalues.
Baxter calculated the free energy per site in the $N\rightarrow
\infty$ limit variationally \cite{baxter68}.
Okunishi pointed out that Baxter's treatment is essentially the
infinite-system method of the DMRG \cite{okunishi}; Nishino and
Okunishi formulated a numerical DMRG algorithm based on these ideas,
the CTMRG \cite{nishino-okunishi97}.
It has the advantage that the eigenvalues of $C^{(N)}$ can be found
without diagonalizing a large sparse matrix.
The CTMRG has been used to find the critical exponents of the
two-dimensional Ising model to high precision
\cite{nishino-okunishi97}, to investigate the spin-3/2 Ising model 
\cite{tsushima-horiguchi98} and a 7-configuration vertex model
\cite{takasaki}, and to study models of two-dimensional self-avoiding
walks \cite{foster-pinettes1,foster-pinettes2}.
Efforts are underway to generalize CTMRG-like algorithms to
three-dimensional lattices \cite{nishino-okunishi98}.

\subsection{Finite Temperature}

\label{sec:quantum_transfer}

\subsubsection{Low Temperature Method}

\begin{figure}
\includegraphics[width=0.6\textwidth]{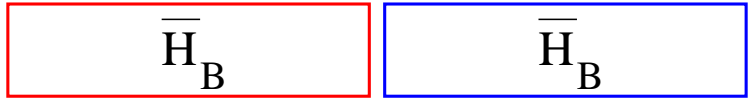}
\caption{Superblock configuration consisting of two truncated blocks.}
\label{fig:truncated_superblock}
\end{figure}

In the original formulation of the DMRG \cite{dmrg_prb}, it was clear
that a system in a mixed state could the treated within the DMRG
simply by obtaining excited states of the superblock in addition to
the ground state and by including them in the system block density matrix:
\begin{equation}
\rho_{i i'} = \sum_\alpha P_\alpha \sum_j 
\psi^\alpha_{ij} \psi^{\alpha \; *}_{i'j} \; ,
\end{equation}
where the  weight $P_\alpha \ge 0$, $\sum_\alpha P_\alpha =1$, and the
$|\psi^\alpha\rangle$ are target states. 
For a system at finite temperature, the natural choice of the weight
$P_\alpha$ might seem to be the Boltzmann weight
$P_\alpha = e^{-\beta E_\alpha}$.
Moukouri and Caron \cite{moukouri-caron} pointed out, however, that
$P_\alpha$ is only used to form the basis of the system block and does
not enter into the calculation of expectation values directly.
They found that it is more efficient to target a moderate number $M$
(typically 10 to 30) of superblock states and weight them equally,
i.e., choose $P_\alpha =1/M$.
In order to calculate thermodynamic properties, they form the
superblock configuration 
shown in Fig.\ \ref{fig:truncated_superblock} in which the number of
states kept in each block is sufficiently small so that a complete
diagonalization can be carried out.
The thermodynamical expectation value of an operator $A$ is then
approximated by the Boltzmann sum
\begin{equation}
\langle A \rangle = \sum_\gamma e^{-\beta E_\gamma} 
\langle \psi^\gamma_{BB} | A | \psi^\gamma_{BB} \rangle \; ,
\end{equation}
where the set of eigenstates $E_\gamma$ is approximate and incomplete.
The assumption is that the truncated basis of $\bar{H}_B$ gives an
adequate description of the states in the Boltzmann sum even
if they are not included as target states.
This approximation is clearly better in the low-temperature regime and 
most likely breaks down at moderately high temperature.
However, it has the advantage that it is a straightforward extension
of the ground-state DMRG and therefore can treat the same systems,
albeit with a somewhat higher computational cost.
This method has been applied to $S=1/2$ and $S=3/2$ Heisenberg chains
\cite{moukouri-caron} and to an $S=1$ model for Y$_2$BaNiO$_5$
\cite{batista98,hallberg99}.

\subsubsection{Transfer Matrix Method}

\begin{figure}
\includegraphics[width=0.5\textwidth]{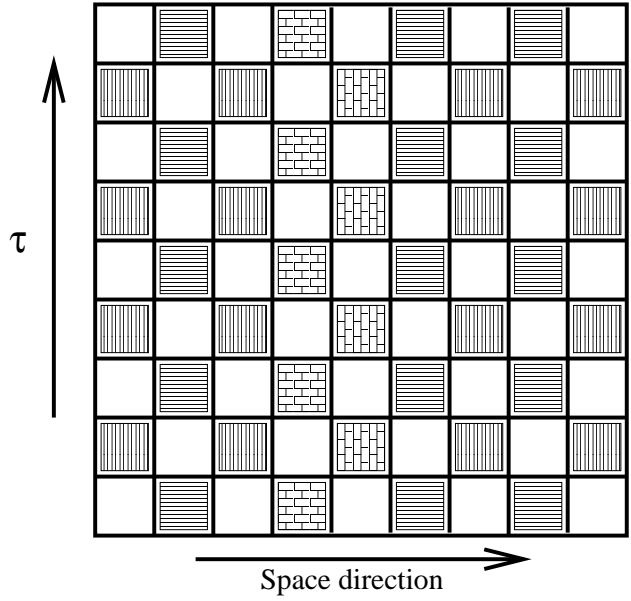}
\caption{Checkerboard decomposition of the partition function.
The quantum transfer matrix in the spacial direction is indicated by
the region of differently marked squares. }
\label{fig:checkerboard}
\end{figure}

The correspondence between two-dimensional classical systems and
one-dimensional quantum systems at finite temperature has led to the
application of TMRG techniques to quantum transfer matrices
\cite{bursill96,wang-xiang,shibata1,shibata2}.
In particular, taking the transfer matrix in the spatial
direction leads to a method in which the system is at finite
temperature, but in the thermodynamic limit.
In order to formulate the quantum system as a two-dimensional
classical lattice model, the partition function is discretized in the
imaginary time or temperature direction using a
Trotter-Suzuki (checkerboard) decomposition
\begin{equation}
Z = {\rm Tr} \; e^{-\beta H} = {\rm Tr} 
(e^{-\Delta \tau H_{\rm odd}} e^{-\Delta \tau H_{\rm even}} )^{M/2}
+ {\cal O}(\Delta \tau^2) \; ,
\end{equation}
where $\Delta \tau = \beta/2 M$ is the imaginary time step, 
\[
H_{\rm odd} = \sum_{\ell =1}^{L-1} h_{2 \ell-1,2 \ell}
\ \ \ \mbox{and} \ \ \  
H_{\rm even}= \sum_{\ell = 1}^{L-1} h_{2 \ell,2 \ell+1}
\] 
with $h_{\ell,\ell+1}$ being the Hamiltonian terms linking site $\ell$
with site $\ell+1$.
(We assume that $H = H_{\rm odd} + H_{\rm even}$ contains only
nearest-neighbor interactions.)
The error is due to neglecting the commutator in factorizing the
exponential.
At this point a complete set of states $|\sigma^j_\ell\rangle$ 
is inserted at every point in
space $\ell$ and imaginary time $j$ so that the partition function can
be written as a discretized path integral
\[ 
\displaystyle
Z_M  = 
{\rm Tr}_{\{\sigma^j_\ell \} } \, 
\prod_{\ell=1}^{L/2} \, \prod_{j=1}^{M} \;
\langle 
\sigma^{2j+1}_{2\ell-1} \;  \sigma^{2j+1}_{2\ell} 
|e^{-\Delta \tau h_{2\ell-1,2 \ell}} | 
\sigma^{2j}_{2\ell-1} \;  \sigma^{2j}_{2\ell} 
\; \rangle \nonumber  
\] 
\begin{equation}
\hspace*{3.8cm} 
\langle 
\sigma^{2j}_{2\ell} \;  \sigma^{2j}_{2\ell+1} 
|e^{-\Delta \tau h_{2\ell,2 \ell+1}} | 
\sigma^{2j-1}_{2\ell} \;  \sigma^{2j-1}_{2\ell+1} 
\; \rangle \; .
\end{equation}
The resulting lattice in space and imaginary time is shown
schematically in Fig.\ \ref{fig:checkerboard}.
Note that the finite lattice treated with the TMRG is in the
imaginary-time direction where the boundary conditions are periodic.
By defining local transfer matrices on a plaquette as 
\begin{equation}
\tau(\sigma^{}_{} \sigma^{}_{} | \sigma^{}_{} \sigma^{}_{} ) \equiv 
\langle | 
e^{-\Delta \tau h_{\ell, \ell+1}} 
| \rangle  \; , 
\end{equation}
the transfer matrices in the spatial direction for odd and even sites
can be written as
\[
T_{\rm odd} = \tau_{1,2} \tau_{3,4} \cdots \tau_{2M-1,2M}
\]
and
\[
T_{\rm even} = \tau_{2,3} \tau_{4,5} \cdots \tau_{2M,1} \; , 
\]
and the partition function for a finite discretization as
\begin{equation}
Z^\infty_M = 
\lim_{L \rightarrow \infty} {\rm Tr} 
\left ( T_{\rm odd} T_{\rm even} \right )^{L/2} \; .
\end{equation}

As in the row-transfer TMRG for classical systems, the free energy per
site is related to the largest eigenvalue of the transfer matrix,
$\lambda_1$,
\begin{equation}
f_M = -\frac{k_B T}{2} \ln \lambda_1 \;,
\end{equation}
DMRG methods are used to obtain $\lambda_1$.
The procedure is a direct adaptation of that for the classical case,
except that:
\begin{itemize}
\item[(i)] The detailed expressions for the transfer matrix are
complicated by the checkerboard decomposition -- the local transfer
matrices only operate on the filled squares of the checkerboard, and
the transfer matrix repeats itself after two rows rather than every
row.
\item[(ii)] Since the total transfer matrix is given by 
$T^{(M)} = T_{\rm odd} T_{\rm even}$ and since 
the commutator $[T_{\rm odd},T_{\rm even}] \ne 0$ in general, 
$T^{(M)}$ is not necessarily symmetric.
However, the largest eigenvalue is real for physical cases, and a
stable iterative diagonalization is possible, for example, with the
Arnoldi or unsymmetrical Lanczos methods 
(see e.g. Ref.~\cite{templatized_evproblems})  
with additional re(bi)orthogonalization. 
The left and right eigenvectors $\langle \psi_L|$ and $|\psi_R\rangle $ can
thus be calculated and are most conveniently chosen to be biorthogonal 
$\langle \psi_L | \psi_R \rangle = 0$.
\item[(iii)] Conserved quantum numbers in the system Hamiltonian lead 
to corresponding conserved quantities in the transfer matrix, which
can be used to reduce the dimension of the space treated in the
iterative diagonalization.
\item[(iv)] In order to perform basis reduction, the reduced density
matrix $\rho = {\rm Tr}_e |\psi_R\rangle \langle \psi_L|$, where the
trace is over the basis of the environment block, is used. 
This in general unsymmetric matrix must be carefully diagonalized,
taking into account numerical problems such as the appearance of
spurious complex conjugate pairs of eigenvalues and the loss of
biorthogonality \cite{wang-xiang}.
\end{itemize}

As in the classical case, the infinite-system algorithm followed by a
number of finite-system sweeps can be carried out until convergence is
attained.
Since we are treating the infinite-system limit of the quantum
transfer matrix in the spatial direction, the lattice size corresponds
to the number of imaginary time steps taken, and the temperature 
$\beta = 1/k_B T = M \Delta \tau$.
Therefore, it takes more numerical effort to reach a lower temperature
and the method should be numerically exact at high temperature.
Since there is a systematic error associated with the finite imaginary
time step $\Delta \tau$, the extrapolation $\Delta \tau\to 0$ should
be carried out to obtain (numerically) exact results.
Thermodynamic quantities can be calculated as numerical derivatives of
the free energy, or can be formed a calculation of the local energy,
as in the classical case.
Dynamical quantities for quantum systems, as discussed in 
Sec.\ \ref{sec:ed_dynamics} can be calculated at finite temperature
with two caveats. 
First, only spatially local or at most short-range
expectation values can be easily calculated because the transfer
matrix only contains spatial information at a range of two sites.
Second, the dynamical information in the
imaginary time direction can be obtained directly from the calculation
of the appropriate correlation functions on the imaginary-time
lattice.
However, the experimentally relevant dependence is on real frequency. 
The analytic continuation of imaginary-time data to read frequencies occurs
via an inverse Laplace transform, which is an ill-posed problem.
This issue of how to perform such an analytic continuation numerically
and what the limitations are
has been extensively investigated in the context of quantum
Monte Carlo calculations; the Maximum Entropy
method is the most used technique \cite{maxent}. 

The quantum TMRG has been applied to calculate the thermodynamic
properties of a number of systems, including
spin chins \cite{shibata97,wang-xiang,xiang98}, spin chains with frustration
\cite{maisinger98,kluemper99,kluemper00}, the
one-dimensional $t$-$J$ model
\cite{ammon,sirker-kluemper1,sirker-kluemper2}, the one-dimensional
Kondo lattice model 
\cite{shibata98,shibata-tsunetsugu_B}, and a spin-orbit model
\cite{sirker-khaliullin}.
Dynamical quantities have been calculated for anisotropic spin chains
\cite{naef99}, the one-dimensional Kondo model
\cite{mutou,shibata-tsunetsugu_A}, and the nuclear spin relaxation rate
in two-leg spin ladders \cite{naef-wang}.

\subsection{Dynamics}
\label{sec:DDMRG}

As discussed in Sec.~\ref{sec:ed_dynamics}, the calculation of
dynamical quantities is a demanding task.
In order to obtain the dynamical correlation function
\begin{equation}
G({\bf k},\omega) = \langle \psi_0 | 
A^\dagger_{\bf k} (\omega + i\eta - H)^{-1} A_{\bf k}| \psi_0 \rangle
\; ,
\end{equation}
a ground state calculation is not sufficient.
It is necessary to obtain additional states, which in the DMRG must be
included in the density matrix as target states.
The methods vary in which states are targeted, but 
the state $A_{\bf k} |\psi_0 \rangle$ must be targeted in addition to
the ground state in all methods.

The first approach to the calculation of dynamical properties using
the DMRG that was developed is the {\it Lanczos vector method}
\cite{hallberg_dynamics}.
In this approach, the sequence of Lanczos vectors which span the
Krylov subspace
\begin{equation}
| \psi_0 \rangle, \  A_{\bf k}| \psi_0 \rangle, \  H A_{\bf k}| \psi_0
  \rangle, \  H^2 A_{\bf k}| \psi_0 \rangle, \; \ldots
\end{equation}
are targeted.
In this way, the Lanczos coefficients can be obtained
and the spectral function can be calculated using the
continued-fraction representation, Eq.~(\ref{eqn:cont_frac}).  
This provides the spectral function for the full range of
$\omega$ in one run.
Usually, one restricts the target vectors to approximately the first
10 vectors of the Krylov subspace; otherwise, the DMRG run 
would become too expensive because the number of states kept would
have to be increased in order to obtain the low-lying states
sufficiently accurately.
However, this number of states is generally not sufficient to reach
convergence in the continued fraction expansion.
In order to overcome this problem, the Lanczos recursion is
continued using vectors that are not targeted in the density matrix.
However, the additional Lanczos vectors are represented in a basis of
density-matrix eigenvectors which have not been adapted for these
vectors.
This could lead to uncontrolled errors.

An alternative approach which overcomes this problem is 
the {\it correction vector method} \cite{kuehner_white}.
It is based on the correction vector method for iterative
diagonalization \cite{soos_ramasesha} 
discussed in Sec.~\ref{sec::ed_correction_vector} and was further
developed and successfully applied in Ref.\ \cite{raas:041102}.  
In this approach, the target vectors 
\begin{equation}
| \psi_0 \rangle, \  A_{\bf k}| \psi_0 \rangle, \  (\omega + i\eta -
  H)^{-1} A_{\bf k}| \psi_0 \rangle 
\end{equation}
are included in the density matrix.
The method is more accurate but more costly than the Lanczos
vector method, since for each value of $\omega$ a separate run must
be performed.

The direct inversion of the resolvent operator can be carried out with
a procedure such as the conjugate gradient method, but this procedure
is computationally costly and can be unstable.
(These problems can be minimized by a better choice of methods for the
inversion \cite{raas:041102}.)
A variant of this
method was introduced by Jeckelmann \cite{jeckelmann_dynamics} 
which leads to a more efficient and stable method.
The basic observation is that the correction vector minimizes the
functional 
\begin{equation}
W_{A,\eta}(\omega, \psi) =  
\langle \psi |( E_0 + \omega - H)^2 + \eta^2) | \psi \rangle
\, + \, \eta \langle \psi_0 |A| \psi \rangle \, + \, \eta \langle \psi | A
| \psi_0 \rangle \; . 
\end{equation}
The spectral weight is then given by
\begin{equation}
W_{A_{\bf k},\eta}(\omega, \psi_{\rm min}) = - \pi\eta \, {\rm Im} \,
G({\bf k},\omega) \; .
\end{equation}
In the calculation, the correction vector is split up into imaginary
and real parts
\begin{equation}
(\omega + i\eta - H)^{-1} A_{\bf k}| \psi_0 \rangle = |X_A(\omega +
i\eta) \rangle + i | Y_A (\omega + i\eta) \rangle \; .
\end{equation}
The target-states used in this approach are
\begin{equation}
|\psi_0 \rangle, \, A_{\bf k} | \psi_0 \rangle, \, |X_A(\omega +
 i\eta) \rangle, \, | Y_A (\omega + i\eta) \rangle.
\end{equation}
The DMRG procedure can be formulated so that
it simultaneously minimizes the ground-state energy and provides
the minimum of the functional $W_{A,\eta}(\omega, \psi)$
\cite{jeckelmann_dynamics,holger_diss}.



Using this approach, several investigations have been performed. 
In the following we will outline the procedure by outlining the
calculation of the single-particle spectral weight for the Hubbard
chain \cite{benthien:256401}.
The 1D Hubbard model with open boundary conditions has the Hamiltonian
\begin{equation}
H = -t \sum^{L-1}_{\ell=1, \sigma} 
\left( c^\dagger_{\ell, \sigma} c^{\phantom{\dagger}}_{\ell+1, \sigma} 
+ c^\dagger_{\ell+1, \sigma} c^{\phantom{\dagger}}_{\ell, \sigma} \right) 
+ U \sum^L_{\ell=1} n_{\ell, \uparrow} n_{\ell, \downarrow} \; .
\end{equation}
The single-particle spectral weight for holes (as obtained in
photoemission experiments) is defined by
\begin{equation}
A(k,\omega) = \frac{1}{\pi} \, {\rm Im} \,
\langle \psi_0 | \; c^{\dagger}_{k, \sigma}
\; \frac{1}{H + \omega - E_0 + i \eta } \;
  c^{\phantom{\dagger}}_{k, \sigma} \; | \psi_0 \rangle
\end{equation}
where
\begin{equation}
c^{\phantom{\dagger}}_{k, \sigma} \, = \,
\frac{1}{\sqrt{L}}\sum_\ell e^{i k \ell}\, 
c^{\phantom{\dagger}}_{\ell, \sigma} \; .
\end{equation}
This definition of $c^{\phantom{\dagger}}_{k, \sigma}$ is problematic
because the Fourier transform used is 
only defined for {\it periodic} boundary conditions, while a system
with open boundary conditions is generally treated by the DMRG for
reasons of efficiency (see Sec.\ \ref{sec:den_mat_proj}). 
The solution is to carry out the transform using ``particle-in-a-box''
eigenstates,
\begin{equation}
c^{\phantom{\dagger}}_{k, \sigma} =
\sqrt{\frac{2}{L+1}}\sum_\ell \; 
\sin (k \ell) \; c^{\phantom{\dagger}}_{\ell, \sigma} \; .
\end{equation}
A comparison of the results obtained with the analytic and exact
Bethe-Ansatz method shows excellent agreement.
Thus, it is possible to obtain momentum-dependent dynamical quantities
using open boundary conditions, for which the DMRG works best.
For this model, relatively large
system sizes can be reached (about 200 sites for a Hubbard chain on
present-day desktop PCs).

%
%
%
%
%
%
%
%

\subsection{Quantum Chemistry}

\label{sec:quantum_chem}
Quantum chemical or {\it ab initio} treatments of a solid can be cast
into the form of a lattice Hamiltonian once a suitable single-particle
basis of $N$ orbitals has been chosen.
The generic form of the Hamiltonian is then
\begin{equation}
    H = \sum_{i,j,\sigma} t_{ij} c_{i\sigma}^\dag 
   c^{\phantom \dagger}_{j\sigma}
 + \frac{1}{2} \sum_{i,j,k,l} G_{ijkl} \sum_{\sigma , \sigma'}
  c_{i\sigma}^\dag c_{j\sigma'}^\dag 
   c^{\phantom \dagger}_{k\sigma'} c^{\phantom \dagger}_{l\sigma} .
\label{eqn:quant_chem}
\end{equation}
where $c^{\dagger}_{i\sigma}$ creates an electron in molecular orbital $i$,
$t_{ij}$ is the single-electron integral of molecular orbitals $i$ and $j$,
and $G_{ijkl}$ is the two-electron integral 
(Coulomb repulsion) \cite{white_martin}.
This Hamiltonian can be treated using the iterative
diagonalization methods described in Sec.\ \ref{sec:iterative_diag};
in quantum chemistry nomenclature, this is termed a ``full
Configuration-Interaction'' (full-CI) calculation.
Since the Hamiltonian (\ref{eqn:quant_chem}) has a form that is similar to
but more general than the Hubbard model in momentum-space 
(\ref{eqn:hubbard_kspace}), DMRG methods that are similar to those used
in the momentum-space DMRG can be applied  and
similar technical problems crop up.
In particular, the molecular orbitals take the role of the
single-particle momenta with the difference that there is no
momentum conservation.
A one-dimensional path is chosen in this space of molecular orbitals
and similar problems occur in the initial buildup of the lattice.
Both the single-electron term and the two-electron term are
non-diagonal in general.
The two-electron term can be factorized into sums of terms internal to
a block as in the momentum-space algorithm; however, the minimum
number of terms scales as $N^2$ rather than $N$ because of the absence
of momentum conservation \cite{white_martin}.
The computational cost of the algorithm scales as $(N^4 m^2+N^3m^3)$, 
where the first term comes from the formation of the necessary
operators as a site is added to a block
and the second is due to multiplication of the ${\cal O} (N^2)$
representations of operators on a block, which comes about either when 
performing a change of basis on the representation of an operator, or 
performing the multiplications necessary to form the product 
$H|\psi \rangle$ needed for the iterative 
diagonalization \cite{white_martin}.
(Each operation is repeated ${\cal O}(N)$ times in the finite-system
algorithm.)

As in the momentum-space algorithm, which ordering of the sites, i.e.,
the molecular orbitals, is optimal is not clear.
It seems that the ordering can strongly effect both the performance
(i.e., how many finite-system steps are required to reach convergence)
and the ultimate accuracy of the result
\cite{legeza:195116,rissler_talk}. 
In particular, some orderings can lead to a convergence to
higher-lying metastable states.
Unfortunately, a general method to find a globally optimal ordering
has not yet been found.
The simplest possibility is to use information from the Hartree-Fock
energy and occupation of the orbitals to choose the ordering,
typically keeping partially occupied orbitals near each other in the
center of the chain
\cite{white_martin,daul_quant_chem,mitrushenkov_2001,mitrushenkov_2003}.
Another approach is to use information from the single-particle matrix
elements $t_{ij}$.
In Ref.\ \cite{chan_head-gordon} the symmetric reverse Cuthill-Mckee
reordering, which attempts to make $t_{ij}$ as band-diagonal as
possible, was shown to lead to faster convergence in $m$ and the
number of finite-system sweeps as compared to Hartree-Fock ordering.
Improvement was also obtained by optimizing the ordering using the Fock term
which contains contributions from both $t_{ij}$ and parts of the
interaction $V_{ijkl}$ \cite{legeza_quant_chem_II}.
Ideas from Quantum Information Theory also seem to be important:
Legeza and S\'olyom \cite{legeza_solyom_QIE} have shown that the
profile along the chain of the entanglement entropy between single
sites and the rest of the system is important.
In particular, orderings which lead to fast convergence tend to have
sites with high entanglement entropy grouped together near the center
of the chain.

Another possibility to improve the quantum chemical algorithm is to
change the choice of single-particle orbitals either by choosing a
different set of single-particle orbitals at the outset or by
performing a canonical transformation on the single-particle basis
\[
\tilde{\phi}_i = \sum_j U_{ij} \phi_j
\]
where $U_{ij}$ is unitary and $\phi_i$ and $\tilde{\phi}_i$ represent
single-particle bases.
Since such a transformation can change the range and topology of
connections of both $t_{ij}$ and $V_{ijkl}$, the performance of the
DMRG algorithm can be strongly affected.
Daul {\it et al.} \cite{daul_quant_chem} applied a 
procedure in which orbitals unoccupied at the Hartree-Fock level were
localized, but did not find a 
significant improvement in the convergence as compared to standard
Hartree-Fock orbitals for the HHeH molecule.
White \cite{white_canon_trans} applied a continuous sequence of
canonical transformations related to the flow equation 
method \cite{wegner,glazek_wilson} in order to reduce non-diagonal
matrix elements.
Calculations for the stretched water molecule as compared to full-CI
calculations show promising results \cite{white_canon_trans}; this
method remains to be further explored.

Thus far, the DMRG for quantum chemistry has been primarily applied to
relatively small molecules in order to test and develop the method.
Full-CI results carried out with the same basis set as the DMRG
calculations have provided a useful benchmark for the DMRG method.
In particular, the water molecule has been treated in the DZP basis with
8 electrons on 25 molecular orbitals which is small enough so that
full-CI results are available
\cite{white_martin,daul_quant_chem,chan_head-gordon,legeza_DSS}.
Agreement with full-CI to within 0.004 mH (Hartree) is possible,
keeping up to $m=900$ states \cite{chan_head-gordon,legeza_DSS}.
A TZ2P basis with 10 electrons on 41 orbitals, which is too large to
treat with full-CI methods has also been treated with the DMRG,
yielding an estimated accuracy of 0.005mH in the ground-state energy
for $m=5000$, approximately a factor of three lower than the best
coupled cluster method \cite{chan_head-gordon_II}.
The energy dissociation curves for N$_2$ \cite{mitrushenkov_2001} and
H$_2$O \cite{mitrushenkov_2003}, and 
excited states of the HHeH molecule \cite{daul_quant_chem} and CH$_2$
\cite{legeza_DSS} have
been calculated with good accuracy.
For LiF, the ground and excited states and the dipole moment were
calculated to relative accuracies of $10^{-9}$, $10^{-7}$, and
$10^{-3}$, respectively, as compared to
full-CI results near the avoided crossing associated with the ionic-neutral
crossover of the bond \cite{legeza_quant_chem_II}.

\subsection{Time Evolution}

\label{sec:DMRG_time_evolution}

As discussed in Sec.\ \ref{sec:ed_time_evolution}, the time evolution
of a state in quantum mechanics is governed by the 
time-dependent Schr\"odinger-equation,
Eq.\ (\ref{eqn:timedependent_schroedinger}), with the formal solution 
\begin{equation}
|\psi(t) \rangle = e^{-iHt} \; |\psi(0)\rangle \; .
\end{equation}
Typically, one is interested in the behavior of the system after a
perturbation
\[
H = H_0 +  H_1 \; \Theta(t)
\]
is switched on at $t=0$
or after an external operator
\[
A^\dagger \; | \psi_0 \rangle.
\]
is applied.
In order to calculate the time evolution of the system, one either has to
integrate Eq.\ (\ref{eqn:timedependent_schroedinger}) directly or one
has to find a suitable approximation for the time-evolution operator
$\exp(-iHt)$. 
In the context of the DMRG, both approaches have been utilized.
Recent developments are recapitulated in
Ref.~\cite{uli_timeevolution_review}. 

In the direct integration approach, a
Runge-Kutta integration \cite{numerical_recipes} of the time evolution
was carried out \cite{cazalilla:256403}. 
As pointed out in Ref.~\cite{luo:049701}, the main difficulty in
calculating the time evolution using the DMRG is that the effective
basis determined at the beginning of the time evolution is not able,
in general, to represent the state well at later times because it 
covers a subspace of the system's total Hilbert space which is not
appropriate to properly represent the state at the next time step.
As discussed in Refs. \cite{our_timeevolution,luo:049701}, it is
possible that the representation of the time-dependent wave function
very soon becomes quite bad.
It is therefore necessary either to mix {\it all} time steps
$|\psi(t_i)\rangle$ into the density-matrix
\cite{luo:049701,schmitteckert:121302}, or to {\it adapt} the density
matrix during propagation. 
An approach for adaptive time evolution basing on the Trotter-Suzuki
decomposition of the time-evolution operator was developed in
Refs. \cite{vidal:040502,white:076401, Uli_timeevolve}.  
In this approach, the time-evolution operator is divided into two-site
terms which can be applied to the two sites included exactly in a typical
superblock-configuration.
However, the method is restricted to systems with nearest-neighbor
terms in the Hamiltonian only. 
A more general scheme to carry out the evolution through one time step 
that is closely related to exact techniques and is 
combined with an adaption scheme recently proposed by White
\cite{white_timeevolve_newpaper} is presented in a second contribution
of the authors in this volume, Ref.~\cite{our_timeevolution}. 
In this approach, the time-evolution operator is expanded in a
Lanczos basis in a similar manner to the time evolution approach used in the
exact diagonalization scheme, as discussed in
Sec.~\ref{sec:ed_time_evolution}.
By using the multi-target method discussed in
\cite{white_timeevolve_newpaper,our_timeevolution}, it is possible to
efficiently calculate the time evolution of systems which can be
treated with the DMRG, even if the Hamiltonian contains terms 
connecting sites that are not nearest neighbors, e.g., on ladder
systems. 
However, the how to carry out the targeting optimally is still under
investigation and will be presented elsewhere \cite{our_work}. 
 
Since it is possible to treat fairly large systems with the
time-dependent DMRG, it is well-suited to investigate
strongly correlated quantum systems out of equilibrium; such a tool
was missing until now.
The application of the DMRG to
such systems opens up new possible  applications to a variety of
systems in the near future.
For recent investigations using this method, see
Refs. \cite{uli_timeevolution_review,our_timeevolution} and references
therein.
Note that it is also possible to use the time-dependent DMRG to 
investigate spectral quantities by directly calculating the
Fourier transform of time-dependent correlation functions.
This method should be faster than the DDMRG presented in
Sec.~\ref{sec:DDMRG} when a complete spectrum is required
because information for a wide range of frequencies can be obtained in
one DMRG run.
However, it is likely that the DDMRG will still provide more accurate
results for a given frequency $\omega$.

\subsection{Matrix Product States}
\label{sec:MPS}

In the DMRG procedure, the state of the system block
is changed in two steps:
First, degrees of freedom are added to the system by adding a site (or
other subsystem) to the system block, increasing the size of the basis
\begin{equation}
|\alpha\rangle_\ell \otimes | s_\ell \rangle \to |
 \beta'_{\ell+1}\rangle \; .
\end{equation}
Second, this expanded system is transformed to a new truncated basis
(i.e., the eigenbasis of the density matrix with the largest
eigenvalues).
\begin{equation}
|\beta_{\ell+1}\rangle = u^{\beta}_{\beta'} |\beta'_{\ell+1} \rangle
\end{equation}
In fact, both steps can be combined into one transformation, as was
done in the wave function transformation discussed in 
Sec.\ \ref{sec:wavefunctrans}
\begin{equation}
|\beta_{\ell+1} \rangle = \sum_{s_\ell, \alpha}A^{s}_{\beta \alpha} 
| \alpha_\ell\rangle 
\otimes | s_\ell \rangle \; .
\end{equation}
For each state of the added site $s$, the matrix 
$A^s_{\beta \alpha}$ maps from the basis $\alpha_\ell$ describing a
system block with $\ell$ sites to $\beta_{\ell+1}$, the truncated
basis for a system block with $\ell+1$ sites; typically these bases
are the same size.
Since the other half of the system, i.e., the environment, is built up in a
similar fashion (either symmetrically in the infinite-system algorithm
or in a previous finite-system sweep), its basis can be described as a
similar series of transformations.
The DMRG wave function can then, in general, be written as a sum of such
products, leading to the state \cite{rommer}
\begin{equation}
|\psi_{\rm MP}\rangle = \sum_{\{s_\ell\}}  {\rm Tr}
\left \{ \; A^{s_1} \, A^{s_2} \, A^{s_3} \ldots A^{s_L} \right \}
| s_1, s_2, \ldots, s_L \rangle  \; .
\end{equation}
Here the $A^{s_\ell}$ are $m\times m$ matrices, 
except for $A_1$ and $A_L$ which are $m$-element {\em vectors}.
(We assume that the same number $m$ states are
retained at each step; generalization to a variable number of states
is straightforward.)
The trace is carried out over the set of coordinates of the added
sites.
Generally, a normalization
\begin{equation}
\sum_{\{s_\ell\}} A^{s_\ell} (A^{s_\ell})^\dagger  = \openone
\end{equation}
is chosen for the $A$-matrices.
This is a special case of a {\em matrix product state}.
The DMRG can be viewed as a particular variational calculation in the
space of such states 
\cite{rommer,verstraete-cirac_periodic}.
Note that a matrix product state for periodic
boundary conditions can be formed if the vectors $A_1$ and $A_L$ are
replaced by matrices.
A translationally invariant state would then have all 
$A^{s_\ell} = A^s$ equal \cite{verstraete-cirac_periodic}.

It is useful to consider matrix product states and their properties in
more generality.
In particular, any state can be described as a matrix product state if
the dimension of the $A^{s_\ell}$ at each step is large enough.
However, this is not useful in itself: the required dimension can, in
principle, grow exponentially with the system size.
Nonetheless, some states are very compactly described:
the 
Ne\'el state $| \uparrow \downarrow \uparrow \downarrow
\uparrow \ldots \rangle  $ can be trivially described using 
$1\times 1$ matrices
and the 
Bell state $|\uparrow \uparrow \uparrow \ldots \rangle
\pm |\downarrow \downarrow \downarrow \ldots \rangle $ using simple 
$2\times 2$ matrices.
The valence-bond solid state for the spin-1 chain, which provides a
qualitatively useful description of the ground state for the
Heisenberg Hamiltonian and is the exact ground state of the
Affleck-Kennedy-Lieb-Tasaki (AKLT) Hamiltonian \cite{AKLT1,AKLT2}, can be
described as a matrix-product state with $m=2$.

For a matrix product state with a particular structure, the
elements of the matrices play the role of variational parameters.
The value of the parameters can, in general, be variationally
optimized using a variety of methods.
For example, the 
power method (see Sec.\ \ref{sec:iterative_diag}), in which the
Hamiltonian is repeatedly applied to the variational state,
$H^n | \psi_{\rm MP} \rangle$, 
imaginary time evolution,  $e^{\Theta H} | \psi_{\rm MP} \rangle$,
or iterative diagonalization methods such as the Lanczos method, all
can be used.
The DMRG algorithm is somewhat more complicated to formulate in detail
in terms of matrix product states; see 
Ref.\ \cite{verstraete-cirac_periodic}.
Expectation values of a set of  
arbitrary operators residing on different sites can be evaluated
directly in terms of matrix products if the relevant operators are
expressed as $m^2$-dimensional operators and are inserted into the
matrix product at the appropriate points \cite{andersson}.

This leads to the question of whether other algorithms based on matrix
product states are better than the DMRG for particular systems.
Recently, a number of such algorithms have been formulated, especially
for systems and situations in which the DMRG either does not perform
well or cannot be applied.
In particular, the DMRG does not produce translationally invariant
states, leading to poor convergence for one-dimensional systems with
periodic boundary conditions.
Verstraete, Porras, and Cirac \cite{verstraete-cirac_periodic} have
formulated a variational algorithm that is closely related to the DMRG
based on a translationally invariant matrix product state.
In contrast to the matrix product state generated by the DMRG, their
state has 
same degree of entanglement between all sites, including those
spanning the boundary.
The diagonalization of the superblock Hamiltonian used in the DMRG
algorithm is replaced by the minimization of a more general functional
and the sweeping back and forth in the finite system algorithm is
replaced by sweeping in the positive or negative direction around the
ring-like lattice.
This algorithm apparently yields a convergence with the number of states
kept $m$ for the Heisenberg chain with periodic boundary conditions
that is comparable to that of the DMRG for a chain of the same length
with open boundary conditions.
However, the computational cost of the new algorithm is larger than
the DMRG, order $m^5$ rather than order $m^3$ because sparseness of
the matrix representations is lost, and implementations of
comparable efficiency to current DMRG codes do not yet exist, so that
the utility of the method for realistic systems remains to be
determined.

Generalizations of matrix product states have been formulated which
incorporate ancilla subspaces so that mixed states can be represented 
\cite{verstraete_finiteTemperature,zwolak_finiteTemperature}.
Such states can be used to represent dissipative quantum or
quantum systems at finite temperature.
Starting with a completely mixed state corresponding to infinite
temperature, imaginary time evolution with a Trotter-decomposed
operator can be carried out to obtain a
finite-temperature state
\cite{verstraete_finiteTemperature,zwolak_finiteTemperature}.
Such a state can then be further evolved in real time using the
methods discussed in Sec.\ \ref{sec:DMRG_time_evolution}, yielding the
time-dependent behavior of finite-temperature or dissipative systems.
Since these methods have been proposed quite recently, they still
await application to realistic systems.

Generalizations of matrix product states that are better suited to
two-dimensional systems have also been proposed \cite{verstraete-cirac_2D}.
For a two-dimensional square lattice, such a state must represent the
entanglement of a site to its four nearest neighbors.
This can be done by inserting an ancilla space for each bond direction,
leading to to a state that is represented by a product of rank-4
tensors.
(Such tensor product states are related to corner transfer tensor
states that have been considered for three-dimensional classical systems
\cite{nishino-okunishi98}.) 
In Ref.\ \cite{verstraete-cirac_2D}, such a state is variationally
optimized for the two-dimensional Heisenberg model with and without
frustration using imaginary-time evolution, yielding promising
variational results for bond dimensions (analogous to the number of
states kept, $m$) of up to 4.
However, since the algorithm seems to scale with a quite high power of
$m$, its utility for realistic system remains to be explored.

As we have seen, recent developments in understanding the correspondence
between the DMRG and matrix product states and in formulating
algorithms based on matrix product states and their generalizations
have opened up many exciting new possibilities in developing new
numerical methods for interacting quantum systems.

\subsection{Quantum Information and the DMRG}
\label{sec:dmrg_qit}

As discussed in Sec.\ \ref{sec:den_mat_proj}, the basic approximation
used to truncate the Hilbert space of the superblock in the DMRG is based
on dividing it in two parts given 
that the superblock is in a pure state (in the simplest case).
The accuracy of the approximation is intimately connected to the
degree of entanglement between the two parts of the bipartite system,
i.e., between the system block and the environment block.
Such entanglement is in turn intimately
connected with the mutual quantum information in the two subsystems.
Recently, progress has been made in utilizing this connection between
the DMRG and quantum information theory to improve the understanding
of the behavior of errors within the DMRG, to minimize the error in
the DMRG algorithm for a given computational cost, to improve
extrapolation methods for calculated quantities, and to formulate
improved variants of the DMRG algorithm.

A measure of mutual entanglement is the von Neumann or mutual quantum
information entropy $S$, defined in Eq.\ (\ref{eqn:von_neumann}).
Since this quantity can be written as 
$- \sum_\alpha w_\alpha \log w_\alpha$, its behavior is a
measure of the behavior of the eigenvalues of the density matrix
$w_\alpha$.
The von Neumann entropy and related quantities from quantum
information theory
can be used to analyze the behavior of the DMRG algorithms in a
quantitative way and to reformulate and optimize the algorithms.
In particular, Legeza {\it et al.} \cite{legeza_DSS} have formulated a
scheme to vary the number of states kept $m$ dynamically according to
a criterion based on keeping the truncation error, as measured by the
weight of the discarded density-matrix eigenvalues, constant.
The relationship between the truncation error, the von Neumann
entropy, and the size of the complete Hilbert space was explored in
Refs.\ \cite{legeza_solyom_QIE} and \cite{legeza_QDC} for the Hubbard
model in momentum space and for quantum chemistry.
The Kholevo bound for the accessible information in a quantum channel, 
\begin{equation}
\chi = S(\rho) - p_{\rm typ} \,  S(\rho_{\rm typ}) 
- (1-p_{\rm typ}) \,  S(\rho_{\rm atyp}) \; ,
\end{equation}
where $\rho_{\rm typ}$ is the density matrix and $p_{\rm typ}$ the
weight of the typical subspace, corresponding to the portion of states
kept, and $\rho_{\rm atyp}$ the density matrix of the atypical
subspace, corresponding to discarded states, 
was argued to be a good measure of the information loss due to
truncation, and was used as a criterion for adjusting $m$ and as a
means of extrapolating energies \cite{legeza_QDC}.
Considerations based on the von Neumann block entropy and the
Kullback-Leibler entropy \cite{kullback_leibler,kullback_book} have
been used to investigate criteria for the optimal ordering of sites
and to formulate a more efficient, dynamical procedure to build up the
lattice in the quantum chemistry and momentum-space DMRG
\cite{legeza_solyom_QIE}.

Another interesting issue is how the behavior of the von Neumann
entropy depends on whether or not a system is critical. 
This has been investigated for one-dimensional XY, XXZ, and
Heisenberg spin chains in Ref.\ \cite{latorre}.
The results are that when the system is not at a quantum critical
point, the von Neumann entropy $S_L$ for a particular chain length $L$
saturates when $L$ is of the size of the correlation length.
For critical systems, $S_L$ is shown to grow logarithmically with $L$,
an unbounded behavior, but with a growth that is much slower than the
upper bound on the entropy, $S_L \sim L$.
The authors argue that the prefactor of this logarithmic divergence
depends only on the universality class of the critical point.

The connection between the DMRG and quantum information theory
suggests that the DMRG algorithm could possibly be applicable
to problems in quantum information theory such as how decoherence
occurs in detail or what the bounds on the information transmitted 
in a noisy channel are, either through the direct simulation of model
systems or through the use of insights gained from the analysis of the
DMRG algorithm and its generalization using concepts from quantum
information theory.

\section{Conclusion and Outlook}

In this tutorial, we have discussed a number of closely related
numerical methods for investigating strongly interacting quantum
many-body systems based on numerical diagonalization.
The aim is to treat quantum lattice models, systems composed of sites,
usually on a regular lattice, with a finite number of quantum
mechanical states.
The size of the Hilbert space of such a model grows exponentially
with the number of sites, making treatment of sufficiently large
lattices to scale to the thermodynamic limit difficult.
The prototypical models considered here, such as the quantum
Heisenberg model or the Hubbard model, originate as effective models
for interacting electrons in a solid, but quantum lattice models can
also come about in other physical systems such as atomic nuclei,
arrays of trapped atoms, or, in general, through the discretization of
a quantum field theory.

All methods discussed here are based on numerical
diagonalization, with the NRG and the DMRG having an additional
variational approximation in which the dimension of the Hilbert space
treated numerically is systematically reduced.
``Exact diagonalization'' methods treat the
entire Hilbert space of the finite lattice Hamiltonian, which can be
restricted to particular symmetry sectors.
``Complete diagonalization'' uses standard algorithms whose cost
scales as the cube of the dimension to obtain all eigenstates.
Since at most Hilbert spaces dimension of order $10^3$, corresponding
to very small lattice size, can be treated,
the method is useful primarily for testing and for benchmarks for
other methods.
``Iterative diagonalization'' can calculate extremal eigenstates for
much larger Hilbert spaces, up to of order $10^9$,
corresponding to system sizes of up to approximately 40 sites for the
simpler quantum lattice models. 
This is done by using an iterative procedure to project the Hamiltonian
onto a cleverly chosen subspace.
While the method is formally variational,
convergence to almost machine precision is attained for typical
quantum lattice models.
The methods has also been extended to efficiently calculate dynamical
correlation functions, finite-temperature properties, and time evolution of a
quantum system, properties that, in principle, depend on excited
states.
The ability of iterative diagonalization to provide results that are
essentially exact for numerical purposes for a wide variety of
properties with few restrictions on the type of Hamiltonian or
lattice makes it an important standard tool.
However, the restriction to moderate lattice sizes due to the
exponential growth of the Hilbert space of quantum lattice models with
system size limits its usefulness in extrapolating to the
thermodynamic limit.
The method is more useful for low-dimensional systems simply because
the linear dimension is often the relevant parameter in such an
extrapolation.

The numerical renormalization group is an iterative procedure in which
a (``complete'') numerical diagonalization is combined with
a truncation of the Hilbert space to lowest energy eigenstates and the
addition of degrees of freedom by adding lattice sites.
The numerical effort is linear in the system size when the number of
states kept is held fixed.
However, the approximation is only well-controlled for one-dimensional
quantum lattice models with coupling between sites which fall off
exponentially with position.
Thus, the NRG is limited to systems that can be mapped onto such a
model; these are primarily single-impurity models such as the Kondo
model and the single-impurity Anderson model, including its
generalization for the dynamical mean-field theory.
The NRG is an extremely powerful method for such systems and can be
used to calculate renormalization group flows, finite-temperature
properties and dynamic quantities such as the impurity density of
states.

The density matrix renormalization group generalizes
the NRG by circumventing problems with the treatment of the wavefunction
at the boundaries present in the NRG.
This is done by embedding a NRG procedure in a larger lattice in which
the diagonalization is carried out.
The reduced density matrix of the renormalized subsystem corresponding
to one or more states of the system is used to
carry out the truncation of the basis in a way that can be shown to be
optimal.
The ability to choose the state or states of the system to which the
truncated basis is adapted allows specialization of the algorithm 
for specific purposes, such as calculation of the ground state only or 
the ground state and low-lying excited states or other additional states.
The freedom of choosing how to divide the system at each step leads to
two classes of algorithms: the infinite-system algorithm in which the
system grows by two sites at each step, and the finite-system
algorithm in which the size of the system is kept fixed.
While the infinite-system algorithm is useful in building up the
system at the beginning of the procedure or in scaling directly to the
infinite-system limit, use of the finite-system procedure leads to more
accurate results in almost all cases.

Crucial to the applicability of the DMRG is how quickly the
eigenvalues of the density matrix fall off; this determines the
accuracy of the variational approximation and is intimately related to
the entanglement of the parts of the system.
For determination of ground-state properties, one-dimensional lattices
with open boundary conditions can be most accurately treated.
It is also favorable for the system to have a gap in one or more
symmetry sectors of the spectrum, leading to a finite correlation
length in the associated correlation function.
Critical one-dimensional systems have behavior that is less favorable,
in particular as the lattice size is increased, and two-dimensional
systems require exponentially many states to be kept with lattice size
in the worst case.
Therefore, the DMRG has become an extremely powerful standard tool for
the calculation of low-energy properties of short-range quantum
lattice models in one-dimension and on coupled chains.

Adaptations of the finite-system algorithm to systems with long-range
interactions such as the Hubbard model in momentum space and 
quantum chemical systems have been formulated.
The momentum-space algorithm becomes more accurate as the coupling
becomes weaker and its convergence only weakly dependent on dimension;
however, it is not competitive with the real-space method for
one-dimensional and coupled chain systems at moderate to large
interaction strength.
Application to models with longer-range interactions in real space
corresponding to shorter-range interactions in momentum space should be
promising to explore.
The quantum chemical DMRG is competitive with other quantum chemical
methods for small molecules, but more development to improve the
performance of the algorithm must be done 
for it to be useful for a wider variety of quantum
chemical problems.

Calculation of dynamical correlation functions using the DMRG can be
carried out with adaptations of both methods used in iterative
diagonalization, the continued fraction (Lanczos basis) method and the
correction vector method.
The latter method provides
substantially more accurate information at a given frequency at
the price of having to carry out a separate calculation at each
frequency to be obtained.
While some optimization of algorithm is probably possible, these
techniques seem to be relatively mature, although they remain to be
applied widely to one-dimensional quantum lattice models.

Transfer matrices rather than Hamiltonians can
also be treated using DMRG methods.
For classical systems, determining the largest eigenvalue of the
transfer matrix allows thermodynamic properties to be calculated
for two-dimensional discrete systems with an infinite extent in one
direction (row transfer matrices) or in both (corner transfer
matrices) such as Ising or Potts models.
Extending the techniques to three dimensions remains a major
challenge. 
Treatment of the quantum transfer matrix in the spatial direction for
one-dimensional quantum lattice models allows thermodynamic quantities
and local dynamical quantities to be calculated for infinite lattice
sizes.
This method is accurate in the high-temperature limit and becomes
computationally more demanding as the temperature is lowered, but is
nevertheless applicable to a wide variety of models over a wide
temperature range.

Recent progress in understanding the connection between the DMRG and
quantum information theory has opened up many new possibilities for
the improvement and development of numerical methods.
Quantum information can be used to optimize the DMRG by providing a
criterion for varying the number of states kept to attain a specified
error goal and to provide insight into optimal ordering of sites for
nonlocal models.
Efficient calculation of the time evolution of quantum states now
seems to be possible, a development which has been spurred by insights
from quantum information theory.
Matrix and tensor product states form the basis for possible
new algorithms which generalize the  DMRG and which could potentially
attain much better performance for translationally invariant systems
and for two-dimensional systems and allow the treatment of dissipative
systems and better treatment of systems at finite temperature.
In addition, there is the possibility to apply the DMRG as a numerical
technique to improve the understanding of various aspect of quantum
information theory.


\begin{theacknowledgments}
We would like to thank the organizers A.\ Avella and F.\ Mancini for
inviting us to 
the IX Training Course in Vietri in October 2004 and to
acknowledge useful discussions with H.~Benthien, A.~L\"auchli,
D.~Poilblanc, and M.~Troyer. 
SRM acknowledges financial support by the European Graduate College
``Electron-Electron Interactions in Solids'' at the Universit\"at
Marburg. 
We also thank D.~Ruhl, F.~Gebhard, H.~Gimperlein, and L.~Tincani for a critical
reading of the manuscript.
\end{theacknowledgments}


\bibliographystyle{aipproc}   

\bibliography{vietri_proc}

\IfFileExists{\jobname.bbl}{}
 {\typeout{}
  \typeout{******************************************}
  \typeout{** Please run "bibtex \jobname" to obtain}
  \typeout{** the bibliography and then re-run LaTeX}
  \typeout{** twice to fix the references!}
  \typeout{******************************************}
  \typeout{}
 }

\end{document}